\documentclass[aps,floats,amssymb,amsmath,prb,nofootinbib,twocolumn,superscriptaddress]{revtex4-1}

\usepackage{booktabs}
\usepackage{calc}
\usepackage{psfrag}
\usepackage{graphicx}
\usepackage{color}
\usepackage{units}
\usepackage[dvipsnames]{xcolor}
\usepackage[unicode=true,pdfusetitle,
 bookmarks=true,bookmarksnumbered=false,bookmarksopen=false,
 breaklinks=false,pdfborder={0 0 0},backref=false,colorlinks=true]
 {hyperref}
\usepackage{geometry}
\pdfpageattr {/Group << /S /Transparency /I true /CS /DeviceRGB>>} 

\newcommand{\up}{\uparrow}
\newcommand{\dn}{\downarrow}

\usepackage{pdfpages}

\makeatletter
\AtBeginDocument{\let\LS@rot\@undefined}
\makeatother

\begin{document}

\title{Charge fluctuations, hydrodynamics and transport in the square-lattice Hubbard model} 
\author{J. Vu\v ci\v cevi\'c}
\affiliation{Scientific Computing Laboratory, Center for the Study of Complex Systems, Institute of Physics Belgrade,
University of Belgrade, Pregrevica 118, 11080 Belgrade, Serbia}
\author{S. Predin}
\affiliation{Scientific Computing Laboratory, Center for the Study of Complex Systems, Institute of Physics Belgrade,
University of Belgrade, Pregrevica 118, 11080 Belgrade, Serbia}
\author{M. Ferrero}
\affiliation{CPHT, CNRS, Ecole Polytechnique, Institut Polytechnique de Paris, Route de Saclay, 91128 Palaiseau, France}
\affiliation{Coll\`ege de France, 11 place Marcelin Berthelot, 75005 Paris, France}

\begin{abstract}
Recent experimental results suggest that a particular hydrodynamic theory describes charge fluctuations at long wavelengths in the square-lattice Hubbard model. 
Due to the continuity equation, the correlation functions for the charge and the current are directly connected: the parameters of the effective hydrodynamic model thus determine the optical conductivity.
Here we investigate the validity of the proposed hydrodynamic theory in the full range of parameters of the Hubbard model. 
In the non-interacting case, there is no effective hydrodynamics, and the charge fluctuations present a rich variety of non-universal behaviors. 
At weak coupling, the optical conductivity is consistent with the hydrodynamic theory: at low frequency one observes a Lorentzian-shaped Drude peak, but the high-frequency asymptotics is necessarily different; the high-temperature limit for the product of the two hydrodynamic model parameters is also in agreement with numerical data.
At strong coupling, we find that a generalization of the proposed hydrodynamic law is consistent with our quantum Monte Carlo, as well as the finite-temperature Lanczos results from literature.
Most importantly, the temperature dependence of the hydrodynamic parameters as well as the dc resistivity are found to be very similar in the weak and the strong-coupling regimes.
\end{abstract}

\maketitle

\section{Introduction}

Strange metallic behavior is one of the central subjects for the theory of strong electronic correlations\cite{Keimer2015}.
It appears to be a universal phenomenon, observed in many different systems, often in close proximity to a superconducting phase\cite{Cooper2009,Legros2018,Cao2020,Ayres2021}, or quantum critical points\cite{Grigera2001,Licciardello2019,Cha2020}. 
In this regime, the dc resistivity is linear in temperature, in a very broad range of temperature. 
The origin of this behavior is unclear, but numerical investigations converge to the conclusion that the Hubbard model captures the underlying mechanisms\cite{Deng2014,VucicevicPRL2015,Perepelitsky2016,Kokalj2017,Huang2019,VucicevicPRL2019,VranicPRB2020}.
Very recently, a variational solution of the semi-classical Boltzmann equation revealed a $T$-linear dc resistivity regime at high-temperature, extending towards zero temperature as half-filling is approached\cite{Kiely2021}. 
This finding of Kiely and Mueller bares an important implication that strange metallicity is not necessarily a strong correlation phenomenon, even in cases when it extends to very low temperature (see also Ref.~\onlinecite{Herman2019} highlighting the role of van Hove singularities at the Fermi level).

In the high temperature limit, a simple and quite universal understanding of the linear resistivity was proposed in terms of the effective hydrodynamics that is expected to arise at long wavelengths in interacting systems. 
Diffusive transport should under very general circumstances lead to $\sigma_\mathrm{dc}=\chi_c D$, where $\sigma_\mathrm{dc}$ is the dc conductivity, $\chi_c$ is the charge compressibility and $D$ is the diffusion constant. 
In Refs.~\onlinecite{Perepelitsky2016} and \onlinecite{Kokalj2017}, it was argued that, at high temperature, $D$ approaches a constant, while quite generally $\chi_c\sim 1/T$, which thus leads to $\rho_\mathrm{dc}\sim T$. 
In a subsequent optical lattice simulation of the Hubbard model by P.T.Brown et al.~\onlinecite{Brown2019}, the assumption of hydrodynamic behavior was exploited to extract values for the dc resistivity and the width of the Drude peak.
In almost quantitative agreement with the best available numerical method (finite temperature Lanczos, FTLM\cite{Jaklic2000,Kokalj2013}), the experiment found linear resistivity in a broad range of temperature. However, the width of the Drude peak $\Gamma$ was greatly overestimated in the experiment, which brings into question the quality of the $\rho_\mathrm{dc}$ estimates and the underlying assumptions.
The interpretation of experimental results relied on a specific hydrodynamic ansatz for the charge-charge correlation function, proposed to be valid in the long wavelength limit.
On the other hand, the fits to the direct measurement data were performed at relatively short wavelengths:
Any discrepancy between the ansatz and the actual behavior at these wavelengths (and correspondingly higher frequencies) could have led to the apparent bias in the estimates of $\Gamma$, but perhaps even in the estimates of $\rho_\mathrm{dc}$.

In this paper we investigate the validity of the hydrodynamic theory proposed in Ref.~\onlinecite{Brown2019}.
We first discuss its analytical properties and find that the high-frequency asymptotics is manifestly non-physical.
We propose a modified hydrodynamic law, which corrects the high-frequency behavior, and ultimately allows for a comparison with Matsubara-axis data we obtain from quantum Monte Carlo. We derive the equation of motion for the current, which must present a microscopic basis for the hydrodynamic theory.
We are unable to rigorously connect the hydrodynamic parameters to Hubbard model parameters, but we find evidence that $D\Gamma\approx 2t^2$ (here $t$ is the hopping amplitude), which is consistent with numerical results at both weak and strong coupling.
Moreover, at weak coupling, $D\Gamma= 2t^2$ can be derived rigorously as the high-temperature limit of the hydrodynamic theory, only based on the knowledge of the exact asymptotics of the charge-charge correlation function. 

We perform numerical calculations for the square-lattice Hubbard model and cover a wide range of parameters.
We start with the non-interacting limit where the hydrodynamic theory is not expected to hold and find multiple interesting examples of charge-fluctuation spectra.
At weak coupling we use second order perturbation theory for the self-energy, and compute optical conductivity and the charge-charge correlation function from the bubble approximation.
We confirm the recent findings of Kiely and Mueller\cite{Kiely2021} that the dc resistivity is linear at half-filling and, more generally, at high-temperature.
At stronger couplings, we use the numerically exact continuous-time quantum Monte Carlo (CTINT\cite{Rubtsov2005,GullRMP2011}) on a finite $10\times 10$ lattice, and control for the lattice size.
We show that a modified hydrodynamic law is consistent with the Matsubara-axis results for the charge-charge and current-current correlation functions,
as well as with the FTLM result for optical conductivity.
The hydrodynamic model parameters extracted from FTLM at strong coupling display strikingly similar behavior to what we find from the bubble approximation at weak coupling.

The paper is organized as follows.
In Section \ref{sec:models} we introduce the 2D Hubbard model and the hydrodynamic theory proposed to govern its charge and current fluctuations at long wavelengths.
In Section \ref{sec:results} we show our numerical results, separated in three subsections based on the coupling strength.
In Section \ref{sec:discussion} we discuss our findings in the context of existing literature and give concluding remarks in Section \ref{sec:conclusions}.
In Appendicies \ref{app:continuity}-\ref{app:optmal_selfenergy} we give detailed derivations of equations used in this paper, and outline the fast algorithm we used for computing the second-order self-energy.
In Appendix \ref{app:chic} we show and discuss static charge susceptibility data in the non-interacting limit.

\section{Models}
\label{sec:models}

\subsection{The square-lattice Hubbard model}
\label{sec:hubbard_model}

We solve the Hubbard model given by the Hamiltonian
\begin{equation} \label{eq:hubbard_hamiltonian}
 H = -t\sum_{\sigma,\langle i,j\rangle } c^\dagger_{\sigma,i}c_{\sigma,j} +  U\sum_i n_{\up,i} n_{\dn,i} - \mu \sum_{\sigma,i} n_{\sigma,i}
\end{equation}
where $\sigma\in\{\up,\dn\}$, $i,j$ enumerate lattice sites, $t$ is the hopping amplitude between the nearest-neighbor sites $\langle i,j\rangle$, $U$ is the onsite coupling constant, and $\mu$ is the chemical potential. We absorb the chemical potential in the bare dispersion, which is thus given by
\begin{equation}
 \varepsilon_\mathbf{k} = -2t(\cos k_x + \cos k_y)-\mu .
\end{equation}
We will switch between the site-notation and real-space notation whenever convenient ($A_i \equiv A_\mathbf{r}$, with $\mathbf{r}=\mathbf{r}_i$, which is the real-space position of the site $i$). The density operator is denoted $n_{\sigma,i} = c^\dagger_{\sigma,i}c_{\sigma,i}$.
Througout the paper we use the half-bandwidth $4t$ as the unit of energy. 
We only consider paramagnetic solutions. In equillibrium we assume full lattice symmetry.

\subsection{The hydrodynamic model}
\label{sec:hydrodynamic_model}

In Ref.~\onlinecite{Brown2019} it was proposed that a hydrodynamic model describes the fluctuations of current and charge at long wavelengths in the Hubbard model. The hydrodynamic model reads
\begin{eqnarray} \label{eq:continuity}
 \partial_t n &=& -\nabla\cdot \mathbf{j} \\ \label{eq:constitutive}
 \partial_t \mathbf{j} &=& -\Gamma( D\nabla n + \mathbf{j} )
\end{eqnarray}
where $n$ and $\mathbf{j}$ are scalar and vector fields, respectively, dependent on time and space. The parameters of the model are the momentum-relaxation rate $\Gamma$ and the diffusion constant $D$. 

The first equation (Eq.\ref{eq:continuity}) is the continuity equation, and it is certainly valid in the Hubbard model for the time-dependent operators in the Heisenberg picture. However, on the square lattice, the spatial derivatives must be discretized, and the actual continuity equation reads
\begin{eqnarray} \label{eq:lattice_continuity}
 \partial_t n_\mathbf{r} = -\sum_{\eta\in\{x,y\}} (j^\eta_\mathbf{r}  - j^\eta_{\mathbf{r}-\mathbf{e}_\eta})
\end{eqnarray}
which simply means that any increase in the particle density at a site $\mathbf{r}$ must be due to a disbalance between the currents entering and exiting the given site. The current operator is given by 
\begin{equation}
j^{\eta}_{\mathbf{r}} = it\sum_\sigma (c^\dagger_{\mathbf{r}+\mathbf{e}_\eta,\sigma}c_{\mathbf{r},\sigma} - c^\dagger_{\mathbf{r},\sigma}c_{\mathbf{r}+\mathbf{e}_\eta,\sigma}) .
\end{equation}
where $\mathbf{e}_\eta$ denotes the lattice vector in the direction $\eta$.
A derivation of Eq.\ref{eq:lattice_continuity} is presented in Appendix~\ref{app:continuity}, but can be found elsewhere\cite{Hafermann2014}.
Since the operators are connected instantaneously, the charge and current fluctuate synchronously. The corresponding charge-charge and current-current correlation functions must be connected directly, as well. Following the derivation presented in Appendix~\ref{app:chi_vs_Lambda}, one obtains in the entire complex plane

\begin{eqnarray} \label{eq:chi_vs_Lambda_lattice}
 &&z^2\chi_\mathbf{q}(z)  \\ \nonumber
 &&= \sum_{\eta\in\{x,y\}}\sum_\mathbf{k}\Phi^\eta_{\mathbf{k},\mathbf{q}} (\langle n_{\mathbf{k}+\mathbf{q}}\rangle  - \langle n_{\mathbf{k}}\rangle) \\ \nonumber
 &&\;\;\;+\sum_{\eta,\eta'\in\{x,y\}} \Big( 1 - e^{iq_\eta} - e^{-iq_{\eta'}} + e^{i(q_\eta-q_{\eta'})} \Big) \Lambda^{\eta,\eta'}_\mathbf{q}(z)
\end{eqnarray}

where $\Phi^\eta_{\mathbf{k},\mathbf{q}} = -2t(\cos(k_\eta+q_\eta)-\cos k_\eta )$, and $n_\mathbf{k}=\sum_\sigma c^\dagger_{\sigma,\mathbf{k}}c_{\sigma,\mathbf{k}}$.
We define in imaginary-time/site-space
\begin{equation}
\chi_{ij}(\tau)=\langle n_i(\tau)n_j(0)\rangle - \langle n \rangle^2
\end{equation}
and
\begin{equation}
\Lambda^{\eta\eta'}_{ij}(\tau)=\langle j_i^\eta(\tau)j_j^{\eta'}(0)\rangle 
\end{equation}
the charge-charge and current-current correlation functions, respectively, calculated in thermodynamic equilibrium.
The standard Fourier transform to Matsubara frequencies gives $\chi(z)$ and $\Lambda(z)$ at a discrete set of points along the imaginary axis; a spatial Fourier transform gives the corresponding quantities in reciprocal space.
We are ultimately interested in retarded quantities which correspond to taking $z=\nu+i0^+$ and we denote them as $\chi_\mathbf{q}(\nu)$ and $\Lambda^{\eta\eta'}_\mathbf{q}(\nu)$.
Here, $\nu$ is real frequency and $\mathbf{q}$ is momentum. 
We have checked numerically that Eq.\ref{eq:chi_vs_Lambda_lattice} holds in the non-interacting limit, for any $\mathbf{q}$ and in any parameter regime of the model (data not shown). In general, the transversal components $\Lambda^{\eta,\eta'\neq\eta}$ play a role in Eq.\ref{eq:chi_vs_Lambda_lattice}. However, the expression greatly simplifies for the imaginary part on the real-axis at small $\mathbf{q}$ in the $x$-direction
\begin{equation}\label{eq:chi_vs_Lambda_continuum}
\lim_{q\rightarrow 0} \mathrm{Im} \chi_{\mathbf{q}=(q,0)}(\nu) = \frac{q^2}{\nu^2}\mathrm{Im} \Lambda^{xx}_{\mathbf{q}=(q,0)}(\nu)
\end{equation}
which is the same expression one obtains in the continuum limit, directly from Eq.\ref{eq:continuity}.

The second equation (Eq.\ref{eq:constitutive}) is the so called constitutive equation\cite{Forster2018} of a hydrodynamic theory, needed to close the system of equations, as the continuity equation itself does not fully fix $n$ and $\mathbf{j}$. In the stationary regime, Eq.\ref{eq:constitutive} reduces to Fick's law of diffusion $\mathbf{j}=-D \nabla n$. Eq.\ref{eq:constitutive} is not necessarily satisfied in the Hubbard model, and is an underlying assumption of the work presented in Ref.~\onlinecite{Brown2019}. 
It is precisely the aim of this work to investigate whether the hydrodynamics encoded in Eq.\ref{eq:constitutive} truly emerges in the Hubbard model at the longest wavelenths, and if yes, under which conditions.

\subsubsection{Microscopic constitutive equation for the hydrodynamics in the Hubbard model}

We start by deriving a microscopic expression for the time-derivative of the current operator. The derivation presented in Appendix~\ref{app:constitutive} yields
\begin{widetext}
\begin{eqnarray} \label{eq:microscopic_constitutive}
\partial_t j^\eta_\mathbf{r}
 &&=  -t^2  \sum_{\sigma} 
    \Bigg \{
         2 n_{\sigma,\mathbf{r}+\mathbf{e}_\eta}
      -  2 n_{\sigma,\mathbf{r}}
    +\sum_{\mathbf{u}\in\{-\mathbf{e}_{\eta},\mathbf{e}_{\bar\eta},-\mathbf{e}_{\bar\eta}\}}  \Bigg(     
         c^\dagger_{\sigma,\mathbf{r}+\mathbf{u}} c_{\sigma,\mathbf{r}+\mathbf{e}_\eta} 
    -     c^\dagger_{\sigma,\mathbf{r}}  c_{\sigma,\mathbf{r}+\mathbf{e}_\eta-\mathbf{u}}     
    + \mathrm{H.c.} \Bigg) \Bigg\}   \\ \nonumber
 &&\;\;\;\;\;
 -t U\sum_\sigma
     (n_{\bar\sigma,\mathbf{r}+\mathbf{e}_\eta}-n_{\bar\sigma,\mathbf{r}})\big( 
              c^\dagger_{\sigma,\mathbf{r}} c_{\sigma,\mathbf{r}+\mathbf{e}_\eta}             
             +   c^\dagger_{\sigma,\mathbf{r}+\mathbf{e}_\eta} c_{\sigma,\mathbf{r}} 
     \big) . 
\end{eqnarray}
\end{widetext}
where we used $\bar\sigma=\uparrow$ if $\sigma=\downarrow$, and vice versa; similarly $\bar\eta=y$ if $\eta=x$ etc.
We immediately recognize the lattice version of the local gradient of charge in the direction of the current $n_{\sigma,\mathbf{r}+\mathbf{e}_\eta}-n_{\sigma,\mathbf{r}}$. 
If we are interested in the time-dependent averages, we can split the terms in the second row in the disconnected and connected parts, $\langle n c^\dagger c\rangle = \langle n\rangle \langle c^\dagger c\rangle + \langle n c^\dagger c\rangle^\mathrm{conn}$. Assuming that we are close to and approaching equilibrium, one can further split the averages in the equilibrium value and the time dependent part. Taking into account the lattice symmetries satisfied in equilibrium, the constant appearing in front of the gradient of charge has the following terms $2t^2 + 2tU \langle c^\dagger_{\sigma,\mathbf{r}+\mathbf{e}_\eta} c_{\sigma,\mathbf{r}} \rangle_\mathrm{eq}$, and will therefore decay towards $2t^2$ as $T\rightarrow \infty$ or $U\rightarrow 0$. Identifying this with the term $D\Gamma \nabla n$ in the hydrodynamic theory (Eq.\ref{eq:constitutive}), one could expect that at high temperature $D\Gamma\approx 2t^2$. In Section~\ref{sec:weak_coupling}, we analyze numerical data and indeed find such behavior. However, in Eq.~\ref{eq:microscopic_constitutive} there are also time-dependent factors that multiply the gradient of charge, and other terms which correspond to neither $\nabla n$ or $\mathbf{j}$. It is unclear under which conditions the remaining terms conspire to give rise to the effective Eq.\ref{eq:constitutive}, even if only in the long-wavelength, low-frequency and linear-response limit. In Appendix~\ref{app:constitutive}, we present Eq.~\ref{eq:microscopic_constitutive} also in momentum space, but find no clear simplifications in the $\mathbf{q}\rightarrow 0$ limit (the Fourier transform to the frequency domain would be analogous).

\subsubsection{Experimental quench setup and CDW amplitude evolution: the ballistic and diffusive regimes}

One can combine the Eq.\ref{eq:continuity} and Eq.\ref{eq:constitutive} to obtain a differential equation governing the time-evolution of the amplitude of a charge-density wave $n_\mathbf{q}$ (on the lattice, this field corresponds to the operator $n_\mathbf{q}=\sum_{\sigma,\mathbf{k}} c^\dagger_{\sigma,\mathbf{k}+\mathbf{q}}c_\mathbf{k}+\mathrm{H.c.}$)
\begin{equation} \label{eq:cdw_amp_time_evolution}
 \partial^2_t n_\mathbf{q} + \Gamma\partial_t n_\mathbf{q} + \Gamma D q^2 n_\mathbf{q} = 0.
\end{equation}
Consider a setup where a charge density wave was first thermalized by applying an external density-modulating field $V$
($
 H \rightarrow H + V \int \mathrm{d}\mathbf{r} \sin x n(\mathbf{r})\
$)
for a long time, and was then let to evolve after abruptly switching off $V$. This time-evolution is the solution of Eq.\ref{eq:cdw_amp_time_evolution} with the boundary condition $\partial_t n_\mathbf{q}(t=0) = 0$, $n_\mathbf{q}(t=0) = n_0$. If $n_0$ is small, this behavior should also be described by the linear-response theory 
\begin{eqnarray}\nonumber
n_\mathbf{q}(t) &=& \int_{-\infty}^{t} \chi_\mathbf{q}(t-t') \theta(-t') \mathrm{d}t' \\ \label{eq:linear_response_theory}
                &=& \int_{t}^{\infty} \chi_\mathbf{q}(t') \mathrm{d}t',
\end{eqnarray}                    
assuming the knowledge of the charge-charge correlation function in real-time, obtained as the Fourier-transform from the retarded $\chi_\mathbf{q}(\nu)$ as $\chi_\mathbf{q}(t)=\int \mathrm{d}\nu e^{-it\nu}\chi_\mathbf{q}(\nu)$. One can show\cite{Brown2019} that the solution of Eq.\ref{eq:cdw_amp_time_evolution} is equal to Eq.\ref{eq:linear_response_theory} with the retarded charge-charge correlation function of the form
\begin{eqnarray}\label{eq:chi_hyd}
 \chi_\mathbf{q}(\nu) &=& \frac{\chi_\mathrm{c}}{1-\frac{i\nu}{q^2D}-\frac{\nu^2}{q^2D\Gamma}} 
\end{eqnarray}
where $\chi_\mathrm{c}$ is the charge compressibility, which connects $n_0$ with the strength of the density modulating field at $t<0$, but does not affect the dynamics of Eq.\ref{eq:cdw_amp_time_evolution}. This correlation function has the important property $\chi_{q\rightarrow 0}(\nu\neq 0)=0$. This indicates the conservation of the total number of particles, which is a prerequisite for the continuity equation. This is easy to understand as $n_{\mathbf{q}=0}$ equals the total number of particles $N_\mathrm{tot}$, and therefore $\mathrm{Im}\chi_{q\rightarrow 0}(\nu)$ describes the fluctuations of $N_\mathrm{tot}$.

\begin{figure}
 \includegraphics[width=2.5in,trim=0cm 0cm 0cm 0cm]{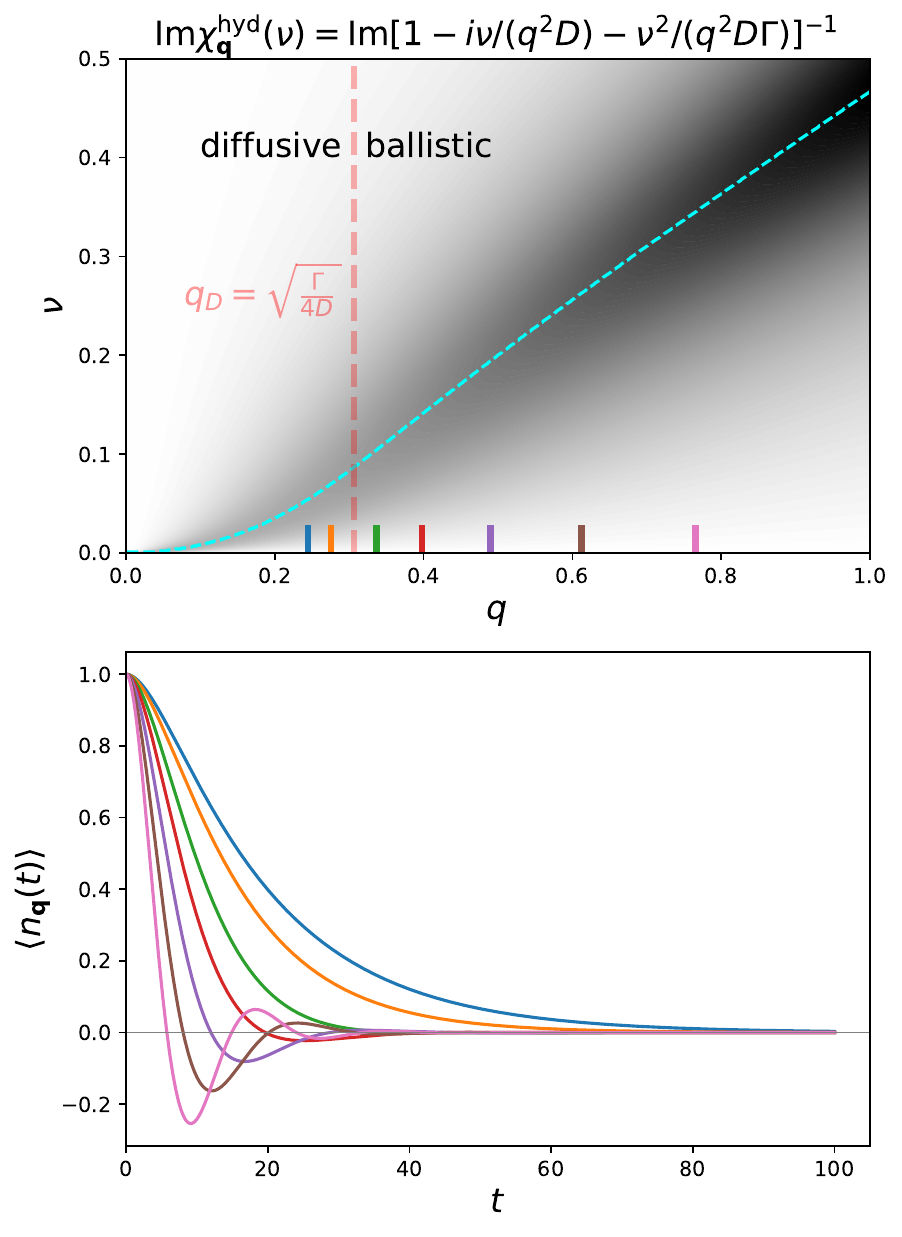}
 \includegraphics[width=2.5in,trim=0cm 0cm 0cm 0cm]{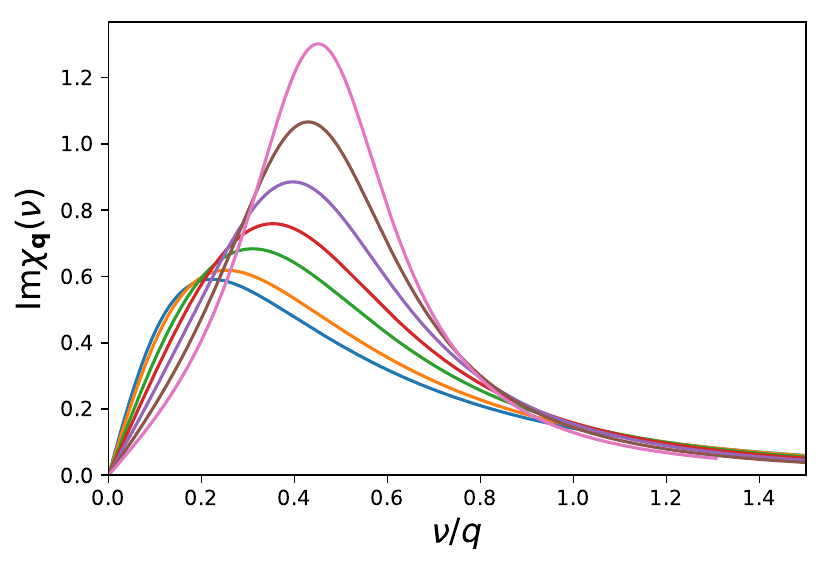}
 \caption{ Illustration of the hydrodynamic theory from Ref.\onlinecite{Brown2019}, defined by Eq.~\ref{eq:continuity} and Eq.~\ref{eq:constitutive}, with parameters taken to be $D=0.8$, $\Gamma=0.3$, $\chi_c=1$. Top: the imaginary part of the charge-charge correlation function as a function of momentum and frequency (Eq.~\ref{eq:chi_hyd}); Middle: the time-evolution of relaxing charge density waves at wave-vectors denoted by matching-color $x$-ticks in the top panel; $\langle n_\mathbf{q}(t) \rangle$ (computed through Eq.~\ref{eq:nqt_hyd}) is normalized to the initial amplitude $\langle n_\mathbf{q}(t=0) \rangle$. Bottom: frequency dependence of the imaginary part of the charge-charge correlation function at the same wave-vectors.
}
 \label{fig:hyd_illustration}
\end{figure}

At any given $\mathbf{q}$, one can rewrite the frequency dependent part of Eq.\ref{eq:chi_hyd} in a more revealing way. In the upper half-plane, the dynamic charge susceptibility Eq.\ref{eq:chi_hyd} can be represented as a sum of two poles in the lower half-plane:
\begin{equation}\label{eq:chitp_z}
 \chi^{\mathrm{tp}}(z^+) = A\left[ \frac{1}{z^+-z_1} - \frac{1}{z^+-z_2}\right]
\end{equation}
with $A = -\chi_\mathrm{c}/r$, $r = \sqrt{4b-a^2}$, $a = \frac{1}{q^2D}$, $b = \frac{1}{q^2D\Gamma}$, $z_1 = \frac{r-ia}{2b}$, $z_2 = \frac{-r-ia}{2b}$.
It is clear that there are two distinct regimes: one where $r$ is purely real, hence the two poles appear at $\mathrm{Re}z_1=-\mathrm{Re}z_2$ and $\mathrm{Im}z_1=\mathrm{Im}z_2=-a$; the other one is when $r$ is purely imaginary, and the two poles appear at $\mathrm{Re}z_1=\mathrm{Re}z_2=0$, $\mathrm{Im}z_1=\mathrm{Im}z_2+2\mathrm{Im}r$. The latter is the ``diffusive regime'', which is realized whenever $4b<a^2$, i.e.
\begin{equation}
 q < q_D \equiv \sqrt{\frac{\Gamma}{4D}}
\end{equation}
To understand why $4b<a^2$ represents the diffusive behavior, and $4b>a^2$ ballistic behavior, we investigate the corresponding solutions of Eq.\ref{eq:cdw_amp_time_evolution}.  The linear response theory Eq.\ref{eq:linear_response_theory} can be solved analytically in the case when $\chi = \chi^{\mathrm{tp}}$. One has
\begin{eqnarray}\label{eq:chitp_t}
 \chi^{\mathrm{tp}}(t) &\sim&  e^{-itz_1}-e^{-itz_2}
\end{eqnarray}
and therefore, under the assumption that neither $z_1$ or $z_2$ are purely real, one gets
\begin{eqnarray} \label{eq:nqt_hyd}
 n_\mathbf{q}(t) &\sim& \frac{e^{-itz_1}}{z_1}-\frac{e^{-itz_2}}{z_2}.
\end{eqnarray}
We see that $n_\mathbf{q}(t)$ will be zero whenever
\begin{equation}\label{eq:condition_to_be_zero}
 \frac{z_2}{z_1} = e^{-it(z_2-z_1)}.
\end{equation}
In the ballistic regime, $z_1=E-i\eta$ and $z_2=-E-i\eta$, and the condition Eq.~\ref{eq:condition_to_be_zero} means
\begin{equation}
 t = \frac{1}{2iE} \log\left( \frac{-E-i\eta}{E-i\eta}\right).
\end{equation}
At a fixed $\eta$ and a finite $E$, there are infinitely many solutions to the above equation: the amplitude of the CDW presents damped oscillations after turning off the external field $V$. In the other case ($4b<a^2$), the poles are placed along the imaginary axis, say $z_1=-i\eta_1$, and $z_2=-i\eta_2$, $\eta_2>\eta_1$ and $r$ is purely imaginary.
One thus has $ n_\mathbf{q}(t) \sim \frac{e^{-t\eta_1}}{\eta_1}-\frac{e^{-t\eta_2}}{\eta_2} $ which can never be zero if $\eta_1\neq\eta_2$. This means that the amplitude of the CDW will ``crawl'' towards zero, signaling an overdamped, or diffusive regime. The correlator $\chi_\mathbf{q}(\nu)$ (Eq.\ref{eq:chi_hyd}) and the corresponding solutions for $n_\mathbf{q}(t)$ (Eq.~\ref{eq:cdw_amp_time_evolution}, or, equivalently, Eq.\ref{eq:linear_response_theory}) are illustrated in Fig.~\ref{fig:hyd_illustration}.

%

\subsubsection{Asymptotic behavior and the connection between hydrodynamics and transport}

The hydrodynamic form for the charge-charge correlation function (Eq.\ref{eq:chi_hyd}) directly implies the form of the current-current correlation function. Inverting Eq.\ref{eq:chi_vs_Lambda_continuum} (which is a direct consequence of the continuity equation) one obtains
\begin{eqnarray} \label{eq:Lambda_hyd}
 \mathrm{Im}\Lambda^{xx}_{\mathbf{q}=(q,0)}(\nu) &=& \frac{\chi_\mathrm{c} D}{\frac{q^4 D^2}{\nu^3}+\frac{1}{\nu}\big(1-2q^2\frac{D}{\Gamma}\big)+\frac{\nu}{\Gamma^2}} 
\end{eqnarray}
At any finite $q$, the behavior at small $\nu$ goes as $\sim\nu^3$. At precisely $q=0$ one gets
\begin{eqnarray} \label{eq:Lambdaq0_hyd}
 \mathrm{Im}\Lambda^{xx}_{\mathbf{q}=0}(\nu) &=& \frac{\chi_\mathrm{c} D}{\frac{1}{\nu}+\frac{\nu}{\Gamma^2}} 
\end{eqnarray}
which at small $\nu$ goes as $\sim\nu$.
Having in mind that the conductivity is obtained as\cite{Coleman2015,VucicevicPRL2019}
\begin{equation}
 \sigma^{\eta\eta'}_\mathbf{q}(\nu) = \frac{1}{\nu}\mathrm{Im} \Lambda^{\eta\eta'}_\mathbf{q}(\nu)
\end{equation}
this model clearly predicts that $\sigma^{xx}_{\mathrm{dc},\mathbf{q}}=0$ for any finite $\mathbf{q}$ in the $x$-direction, which is precisely what is expected on physical grounds. At $\mathbf{q}=0$, which is the most relevant case, one gets a Lorentzian-shaped Drude peak 
\begin{equation} \label{eq:sigmaxx_hyd}
 \sigma^{xx}_{\mathbf{q}=0}(\nu) = \frac{\chi_c D}{1+\big(\frac{\nu}{\Gamma}\big)^2}
\end{equation}
indicating $\sigma^{xx}_{\mathrm{dc},\mathbf{q}=0}\equiv \sigma^{xx}_{\mathbf{q}=0}(\nu=0)= \chi_\mathrm{c} D$, which is the well known Nernst-Einstein equation.

It is important to note that, a priori, the forms Eq.~\ref{eq:Lambda_hyd}, Eq.~\ref{eq:Lambdaq0_hyd} and Eq.~\ref{eq:sigmaxx_hyd} of $\Lambda$ and $\sigma$ are unphysical. The scaling with high frequency
\begin{equation}
\mathrm{Im}\Lambda_{\mathbf{q}=0}(\nu\rightarrow\infty)\sim \frac{1}{\nu} 
\end{equation}
cannot be obtained from a correlation function in imaginary time $\Lambda_{\mathbf{q}=0}(\tau)$ that has the correct symmetries. The Lorentzian Drude peak $\sigma \sim \frac{1}{\nu^2}$ must be restricted to some finite frequency. In general, one expects that at high-frequency $\mathrm{Im}\chi(\nu)$ and $\mathrm{Im}\Lambda(\nu)$ decay exponentially. In the non-interacting case, there is even a sharp cutoff - both charge and current fluctuations are bounded in frequency from above, with a bound that depends on $\mathbf{q}$ (see Section \ref{sec:noninteracting}). In any case, on the Matsubara axis one must have $\chi(i\nu\rightarrow i\infty)\sim 1/\nu^2$ and $\Lambda(i\nu\rightarrow i\infty)\sim 1/\nu^2$. The hydrodynamic ansatz for the charge fluctuations Eq.\ref{eq:chi_hyd} does not violate this, as on the upper half of the Matsubara axis
\begin{equation}\label{eq:chi_hyd_imag_axis}
\chi_\mathbf{q}(i\nu)=\frac{\chi_\mathrm{c}}{1+\frac{\nu}{q^2D}+\frac{\nu^2}{q^2D\Gamma}},
\end{equation}
but, through the continuity equation, it does imply a non-physical asymptotic behavior $\Lambda_{\mathbf{q}=0}(i\nu\rightarrow i\infty)\sim \frac{1}{\nu}$. To be able to compare the hydrodynamic theory with Matsubara-frequency results for the charge-charge and current-current correlation functions, we thus propose a modified hydrodynamic form. The details are given in Section \ref{sec:strong_coupling}.

\begin{figure*}[!ht]
 \includegraphics[width=6.0in,trim=0cm 0cm 0cm 0cm]{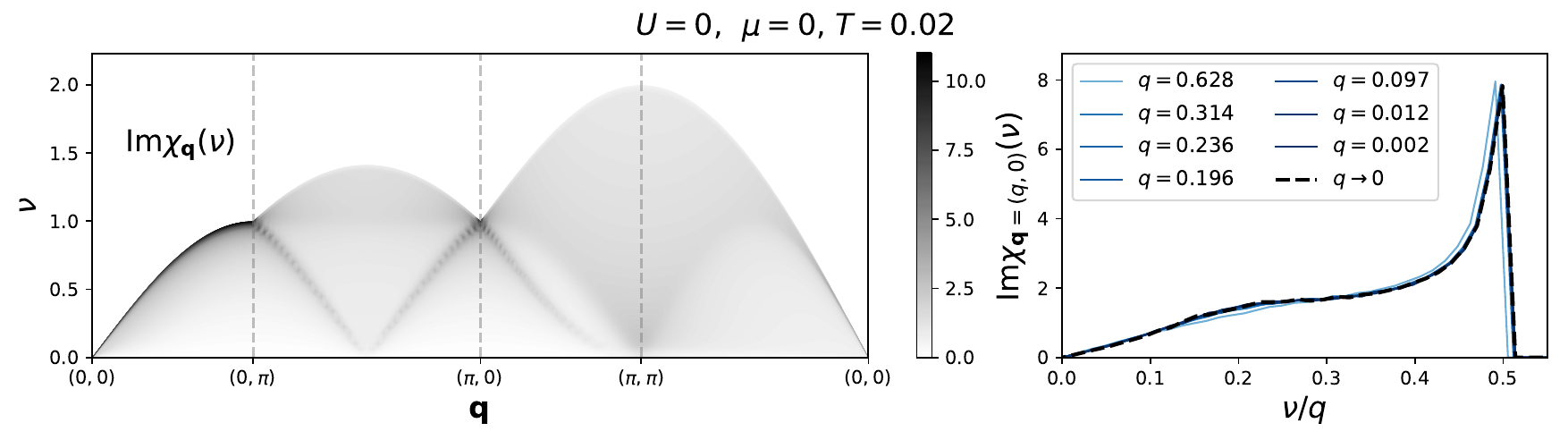} 
 \caption{ Non-interacting case, half-filling. Left: imaginary part of the charge-charge correlation function along a high-symmetry path in the BZ. Right: frequency dependence in the long-wavelength limit, for waves in the $x$-direction. The spectral weight drops off abruptly at $\nu/q=1/2$ - the apparent finite slope comes from the finite frequency resolution in our numerics. }
 \label{fig:sharp_plasmon_mode} 
\end{figure*}

\begin{figure*}[!ht]
 \includegraphics[width=\textwidth,trim=0cm 0cm 0cm 0cm]{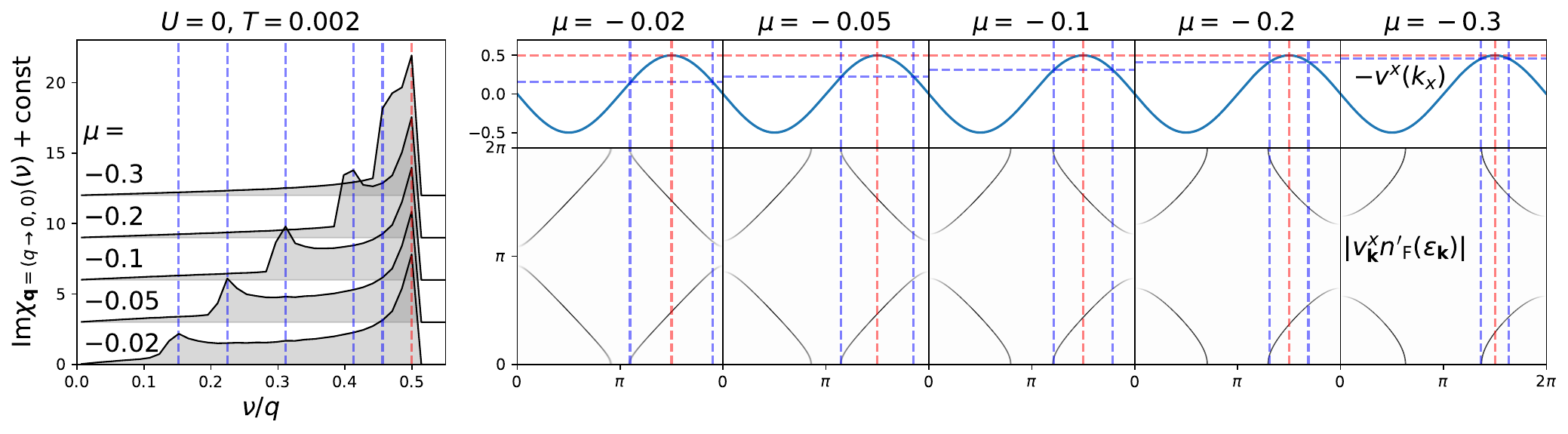}
 \caption{ Non-interacting case, various moderate dopings. Left: frequency dependence of the imaginary part of the charge-charge correlation function in the long-wavelength limit, for waves in the $x$-direction. Vertical blue and red dashed lines denote frequencies of two apparent peaks. Right: explanation for the appearance of two peaks. Top row: the electron velocity. Blue and red dashed lines denote the momenta where $v^x_\mathbf{k}=-\nu/q$, where $\nu$ is the frequency of the two peaks in the spectra on the left. Bottom row: the intensity plots (black-white scale) of the amplitude of the contribution to the charge-charge correlation function coming from different $\mathbf{k}$-vectors in the BZ. The blue and red dashed lines denote the contributions to the two peaks in the spectra on the left.
}
 \label{fig:second_peak_explanation} 
\end{figure*}

The imaginary-axis form Eq.~\ref{eq:chi_hyd_imag_axis} may still be useful in the $U\rightarrow 0$ limit.
The high-frequency asymptotics on the imaginary axis is determined by the entirety of the function on the real axis.
As the coupling constant is decreased, the weight of the function $\chi_{\mathbf{q}\rightarrow 0}$ on the real axis will be contained in an increasingly small range of low frequencies.
If wee assume that the hydrodynamic theory holds in some low frequency range, say $|\omega|<|\omega_\mathrm{max}|$, and that $\omega_\mathrm{max}$ saturates to a finite constant as $U\rightarrow 0$, then we can conclude that the imaginary-axis asymptotics of $\chi_{\mathbf{q}\rightarrow 0}$ will tend to Eq.\ref{eq:chi_hyd_imag_axis} as $U\rightarrow 0$. Clearly, non-universal features at high real frequencies will still be there, but they will not contribute significantly to the imaginary axis asymptotics.
Other scenarios are also possible, but in the following we work out the consequences of our expectation that the hydrodynamic law holds in a \emph{finite} range of frequency in the $U\rightarrow 0$ limit.
We start with Eq.~\ref{eq:chi_vs_Lambda_lattice}, which implies the long-wavelength asymptotics of the charge-charge correlation function of the form (see Appendix \ref{app:chi_vs_Lambda} for details)
\begin{equation}\label{eq:exact_chi_q0_asymptotics}
 \mathrm{Re}\chi_{\mathbf{q}\rightarrow 0}(i\nu\rightarrow i\infty) = -\frac{2t}{\nu^2}\sum_{\eta=\{x,y\}}\sum_\mathbf{k} q_\eta \sin k_\eta \mathbf{q}\cdot \nabla\langle  n_{\mathbf{k}}\rangle.
\end{equation}
This form is not necessarily isotropic. Nevertheless, one can take $\mathbf{q}=(q,0)$ and then, assuming a finite $\omega_\mathrm{max}$ and $U\rightarrow 0$, equate the RHS of Eq.~\ref{eq:exact_chi_q0_asymptotics} with the $i\nu\rightarrow i\infty$ limit of Eq.~\ref{eq:chi_hyd_imag_axis} to obtain
\begin{equation} \label{eq:DGamma_weak_coupling}
D \Gamma = -\frac{1}{\chi_c} \sum_\mathbf{k} v^x_\mathbf{k} \partial_{k_x} \langle  n_{\mathbf{k}}\rangle
\end{equation}
with $v^x_\mathbf{k}=2t\sin k_x$.
Under the current assumption of the weak coupling limit, we can write further
\begin{equation}
D \Gamma = -\frac{1}{\chi_c} \sum_\mathbf{k} (v^x_\mathbf{k})^2 (2 n'_\mathrm{F}(\varepsilon_{\mathbf{k}})).
\end{equation}
At high temperature $T=1/\beta\rightarrow \infty$, the first derivative of the Fermi distribution $n'_\mathrm{F}(\omega)\sim -\beta/4$, and $\chi_c = \frac{\partial \langle n \rangle}{\partial \mu}=-\int \mathrm{d}\varepsilon \rho(\varepsilon) 2 n'_\mathrm{F}(\varepsilon) \sim  \frac{2\beta}{4} \int \mathrm{d}\varepsilon \rho(\varepsilon) = \beta/2$. We also have $\frac{1}{(2\pi)^2} \int \mathrm{d}\mathbf{k} \sin k_x = \frac{1}{2}$. We conclude that in the weak coupling limit and high temperature, the effective hydrodynamic theory formulated by Eq.\ref{eq:continuity} and Eq.\ref{eq:constitutive} for the square-lattice Hubbard model (Eq.\ref{eq:hubbard_hamiltonian}), if valid in a finite range of real frequency, must satisfy
\begin{equation} \label{eq:DGamma_highT_limit}
\lim_{\substack{U\rightarrow 0 \\ T\rightarrow \infty}} D \Gamma = 2t^2.
\end{equation}
Thus, the equation of motion Eq.~\ref{eq:microscopic_constitutive} provides some microscopic support for the effective hydrodynamic theory.
As already mentioned, Eq.~\ref{eq:DGamma_highT_limit} indeed coincides with numerical results, and is roughly satisfied in a broad range of temperatures, even at strong coupling (see Section \ref{sec:weak_coupling}).
Finally, we note that the imaginary axis asymptotics (Eq.~\ref{eq:exact_chi_q0_asymptotics}) combined with $\sigma_\mathrm{dc}=\chi_c D$ (Eq.~\ref{eq:sigmaxx_hyd}) reveals that the hydrodynamic theory (Eq.~\ref{eq:chi_hyd_imag_axis}), taken to be valid at any frequency, is consistent with the Boltzmann expression for the dc conductivity
\begin{equation}\label{eq:boltzmann_sigmadc}
 \sigma_\mathrm{dc} = -\frac{1}{\Gamma} \sum_\mathbf{k} v^x_\mathbf{k} \partial_{k_x} \langle  n_{\mathbf{k}}\rangle.
\end{equation}

\section{Results}
\label{sec:results}

\subsection{Non-interacting limit}
\label{sec:noninteracting}

\begin{figure}[!ht]
 \includegraphics[width=3.2in,trim=0cm 0cm 0cm 0cm]{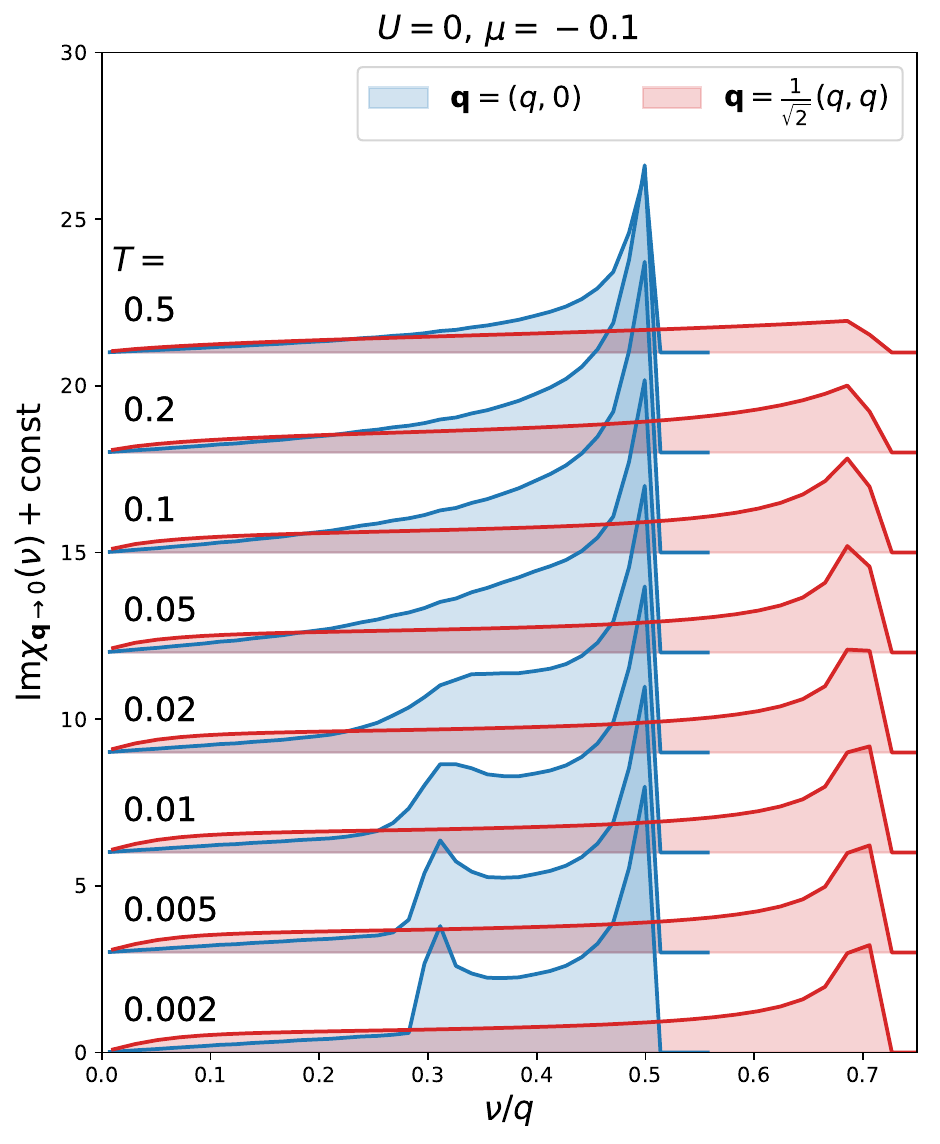}
 \caption{ Non-interacting case, moderate doping, various temperatures. Blue and red curves correspond to the long-wavelength limit of the imaginary part of the charge-charge correlation function for waves in the $x$ and $x=y$ directions. }
\label{fig:U0_anisotropy} 
\end{figure}

\begin{figure*}[!ht]
 \includegraphics[width=6.0in,trim=0cm 0cm 0cm 0cm]{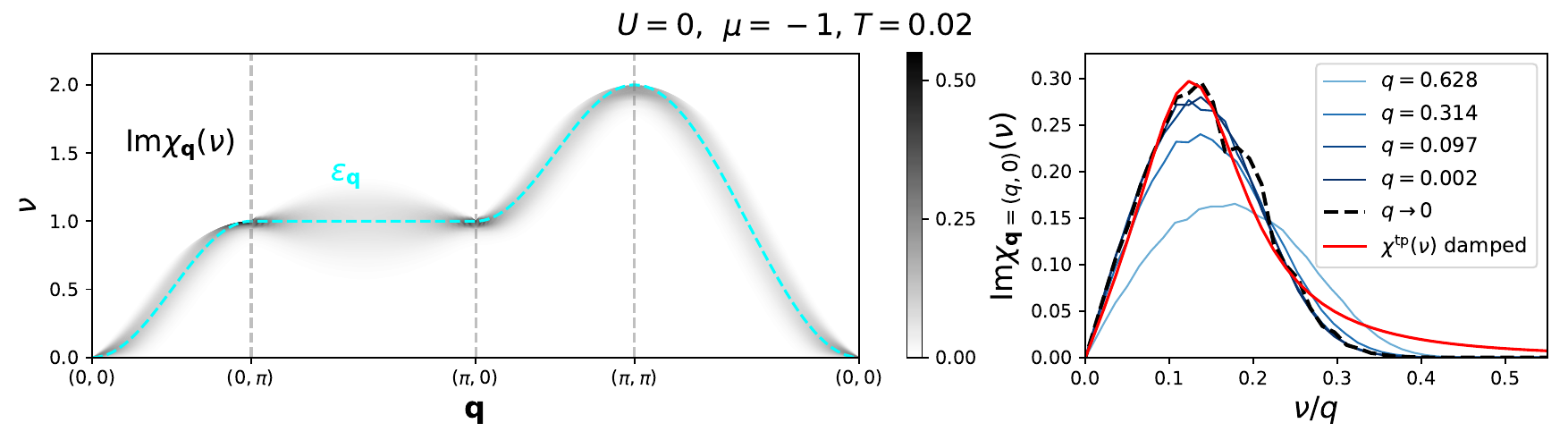}
 \caption{ Non-interacting case, nearly empty limit. Left and right: same as Fig.\ref{fig:sharp_plasmon_mode}. Cyan line on the left: electron dispersion. Red line on the right: fit to the frequency dependent part of the hydrodynamic theory $\chi^\mathrm{tp}$ as defined in Eq.~\ref{eq:chitp_z}; best fit corresponds to the damped oscillations (or ballistic) regime ($r$ purely real).
}
\label{fig:nearly_empty} 
\end{figure*}

We are interested in calculating two-particle correlation functions, in particular for the charge and current. In the non-interacting limit, these can be obtained numerically exactly, to a high accuracy, from the general (Kubo) bubble formula
\begin{equation}
 Q_\mathbf{q}[\varphi,\phi](\tau) = 
 2\sum_\mathbf{k} \varphi_{\mathbf{k},\mathbf{q}} G_{0,\mathbf{k}}(\tau) G_{0,\mathbf{k}+\mathbf{q}}(-\tau) \phi_{\mathbf{k}+\mathbf{q},-\mathbf{q}}
\end{equation}
The factor $2$ in front is due to summation over $\sigma$.
We denote $G_0$ the bare propagator, which is, at a finite temperature $T=\frac{1}{\beta}$, defined in the imaginary-time window $\tau\in[-\beta,\beta]$ as
\begin{equation}\label{eq:bare_greens_function}
 G_{0,\mathbf{k}}(\tau) = -\mathrm{sign}(\tau)e^{-\varepsilon_\mathbf{k}\tau}n_\mathrm{F}(-\mathrm{sign}(\tau)\varepsilon_\mathbf{k})
\end{equation}
with $n_\mathrm{F}(\omega)=\frac{1}{e^{\beta\omega}+1}$ the Fermi-Dirac distribution function.

The ``vertex factors'' $\varphi$ and $\phi$ correspond to the operators for which the correlation function is calculated (in general ${\cal O}_\mathbf{q}= \sum_{\sigma,\mathbf{k}} \varphi_{\mathbf{k},\mathbf{q}} c^\dagger_{\sigma,\mathbf{k}+\mathbf{q}} c_{\sigma,\mathbf{k}}$). 
We then simply have $\chi_\mathbf{q} = Q_\mathbf{q}[\varphi=1,\phi=1]$, and  $\Lambda^{\eta,\eta'}_\mathbf{q} = Q_\mathbf{q}[\varphi=v^\eta,\phi=v^{\eta'}]$, with 
$v^\eta_{\mathbf{k},\mathbf{q}} = it(e^{-i(k_\eta+q_\eta)}-e^{ik_\eta})$ .
%
In the entire complex frequency plane, one can then write
\begin{eqnarray}  \nonumber
 Q_\mathbf{q}[\varphi,\phi](z)  = 2\sum_\mathbf{k} \varphi_{\mathbf{k},\mathbf{q}}   \phi_{\mathbf{k}+\mathbf{q},-\mathbf{q}} 
  \frac{ n_\mathrm{F}(\varepsilon_{\mathbf{k}+\mathbf{q}}) - n_\mathrm{F}(\varepsilon_\mathbf{k}) 
 }{z-(\varepsilon_\mathbf{k}-\varepsilon_{\mathbf{k}+\mathbf{q}})}\\ \label{eq:Qqz_final}
\end{eqnarray}

We now consider the long-wavelength limit for the charge-charge correlation function $\chi$.
At small enough $\mathbf{q}$, one can write further:
$
 \varepsilon_{\mathbf{k}+\mathbf{q}} = \varepsilon_{\mathbf{k}} + \mathbf{q}\cdot \nabla \varepsilon_{\mathbf{k}}
$
and
$
 n_\mathrm{F}(\varepsilon_{\mathbf{k}+\mathbf{q}}) = n_\mathrm{F}(\varepsilon_{\mathbf{k}}) + (\mathbf{q}\cdot \nabla \varepsilon_{\mathbf{k}}) n_\mathrm{F}'(\varepsilon_{\mathbf{k}})
$.
These yield
\begin{eqnarray}
 \chi_{\mathbf{q}\rightarrow 0 }(z) = 
 2\sum_\mathbf{k}\frac{(\mathbf{q}\cdot \nabla \varepsilon_{\mathbf{k}}) n_\mathrm{F}'(\varepsilon_{\mathbf{k}})}{z+\mathbf{q}\cdot \nabla \varepsilon_{\mathbf{k}}}
\end{eqnarray}

We see that at small $q\equiv|\mathbf{q}|$, the frequency dependence no longer depends on $q$. In the denominator, $q$ multiplies the number which determines the position of a pole on the energy axis. Therefore, $q$ sets the energy scale, which means that with a proper rescaling of the $\nu$-axis, $\chi$ results for different small $\mathbf{q}$ along a given direction can be collapsed onto a single curve.
In the special case $\mathbf{q}=(q,0)$, the gradient of the dispersion will simply yield the velocity $v^{x}_{\mathbf{k}}\equiv v^{x}_{\mathbf{k},\mathbf{q}=0}=2t\sin k_x$, and one arrives at
\begin{equation}\label{eq:U0_chi_smallq}
 \lim_{q\rightarrow 0}\chi_{\mathbf{q}=(q,0) }(z)= 2\sum_\mathbf{k} \frac{v^x_\mathbf{k} n_\mathrm{F}'(\varepsilon_\mathbf{k}) }{z/q+v^x_\mathbf{k}}.
\end{equation} 

On the most general physical grounds, it is not expected that in the non-interacting limit an effective hydrodynamics governs the charge fluctuations at however long the wavelengths.
The diffusive motion of carriers at lengthscales $\lambda > 2\pi/q_D$ ultimately comes from a finite-lasting memory the electrons have of momentum; in the non-interacting case, the momentum eigenstates are infinitely long lived.
It is clear that no identification between Eq.\ref{eq:U0_chi_smallq} and Eq.\ref{eq:chi_hyd} is possible. In fact, the non-interacting case presents strongly non-universal, parameter dependent, and even anisotropic behavior that we illustrate in the following. 

We obtain the $\chi_\mathbf{q}(\nu)$ along a high-symmetry path in the Brillouin zone (BZ) using a $6000\times6000$-site lattice, and adaptive frequency grids to ensure sufficient frequency-resolution at all $\mathbf{q}$-vectors. We colorplot $\mathrm{Im}\chi_\mathbf{q}(\nu)$ and show the frequency dependent part at small $\mathbf{q}$ in Figures \ref{fig:sharp_plasmon_mode}-\ref{fig:nearly_empty}.

In Fig.~\ref{fig:sharp_plasmon_mode} we show results for the half-filled case $\mu=0$, $T=0.02$. In the long wavelength limit, we observe a sharp peak at the edge of the spectrum, at $\nu\sim q$. The peak is highly asymmetric, as the spectral weight drops off abruptly on the higher-freq. side. The single peak structure at $\nu\sim q$ is the expected linear zero-sound mode\cite{Hafermann2014}.

As the system is doped away from half filling, we start to observe a two peak structure at long-wavelengths (Fig.~\ref{fig:second_peak_explanation}). This can be understood by analyzing Eq.\ref{eq:U0_chi_smallq}. In Fig.~\ref{fig:second_peak_explanation} we illustrate how the contributions to $\mathrm{Im}\chi_{\mathbf{q}=(q,0)}(\nu)$ at a given energy $\nu$ comes from a line in the Brillouin zone (BZ) where $-v^{x}_{\mathbf{k}}=\nu/q$. The amplitude of a contribution at a given $\mathbf{k}$ is given by $v^{x}_{\mathbf{k}}n'_\mathrm{F}(\varepsilon_\mathbf{k})$, which roughly selects the Fermi surface. Therefore, one gets a peak at frequencies where $v^{x}_{\mathbf{k}}$ is maximal, but also where the Fermi surface is parallel to the $k_y$-axis. This calculation resembles a histrogram of a 1D function, and thus the spectrum resembles a typical density of states of a 1D tight-binding chain.

On Fig.~\ref{fig:U0_anisotropy} we illustrate the great level of anisotropy, by comparing the $q\rightarrow 0$ limit for $\mathbf{q}=(q,0)$ and $\mathbf{q}=\frac{1}{\sqrt{2}}(q,q)$. It is interesting that, as the temperature is increased, the anisotropy at low frequency becomes somewhat reduced.

Doping all the way to the near-empty limit, one observes a completely different behavior (see Fig.~\ref{fig:nearly_empty}). 
The charge fluctuation spectrum closely resembles the electron dispersion.
This indicates that in the single-particle limit, due to the irrelevance of the Fermi-Dirac statistics, the charge and the electron become the same.

\subsubsection{CDW amplitude evolution}

It is of interest to understand these $\chi_\mathbf{q}(\nu)$ results in the context of the quench setup studied in Ref.~\onlinecite{Brown2019} and already mentioned in Section~\ref{sec:hydrodynamic_model}.
Namely, we wish to investigate the time-evolution of the amplitude of a relaxing charge density wave (CDW). If the initial CDW is weak, we can work within the linear response theory, which can be solved numerically, by plugging the Eq.\ref{eq:Qqz_final} with $\varphi,\phi=1$ in Eq.\ref{eq:linear_response_theory}. The Fourier transform needed for this step is performed analogously to Eq.\ref{eq:chitp_t}. Then, to perform the integral in Eq.\ref{eq:linear_response_theory} analytically, it is necessary to regularize the integrand function, first. As is always done when working with retarded quantities, we take that the poles are located slightly below the real-axis. We obtain
\begin{equation} \label{eq:noninteracting_bubble_nqt}
 \langle n_\mathbf{q}(t)\rangle \sim \sum_\mathbf{k} 
 \frac{ n_\mathrm{F}(\varepsilon_{\mathbf{k}+\mathbf{q}}) - n_\mathrm{F}(\varepsilon_\mathbf{k})  }
 {\varepsilon_\mathbf{k}-\varepsilon_{\mathbf{k}+\mathbf{q}}}
  e^{-it(\varepsilon_\mathbf{k}-\varepsilon_{\mathbf{k}+\mathbf{q}})}.
\end{equation}

\begin{figure*}[!ht]
 \includegraphics[width=\textwidth,trim=0cm 0cm 0cm 0cm]{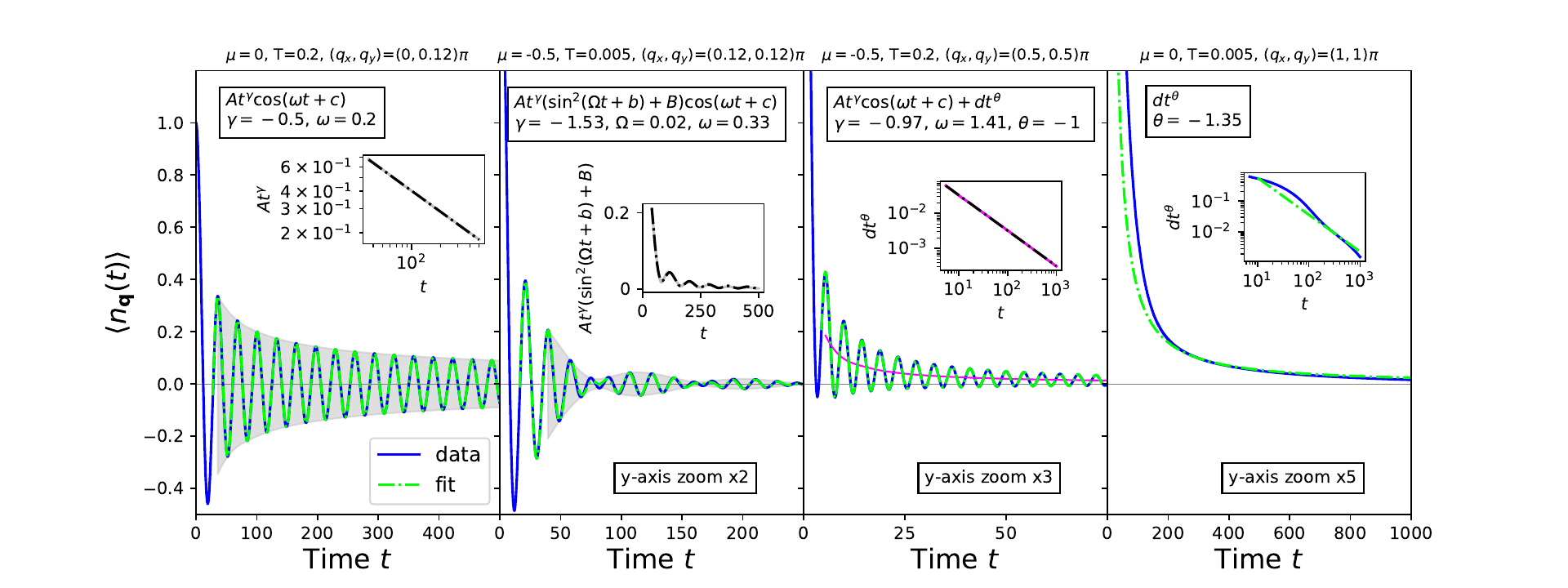} 
 \caption{ Non-interacting case, various dopings and temperatures. CDW amplitude vs. time at various wave-vectors, calculated within linear response theory, normalized to the initial amplitude of the CDW. Text boxes show the fitting function and its main parameters. The insets in the two plots on the left show the amplitude of damped oscillations vs. time, and the corresponding fit. The insets in the two plots on the right show the background, i.e. non-oscillatory components and the corresponding fits. Full lines are data, dot-dashed lines are fits.}
\label{fig:U0_At_examples} 
\end{figure*}

\begin{figure*}[!ht]
 \includegraphics[width=\textwidth,trim=0cm 0cm 0cm 0cm]{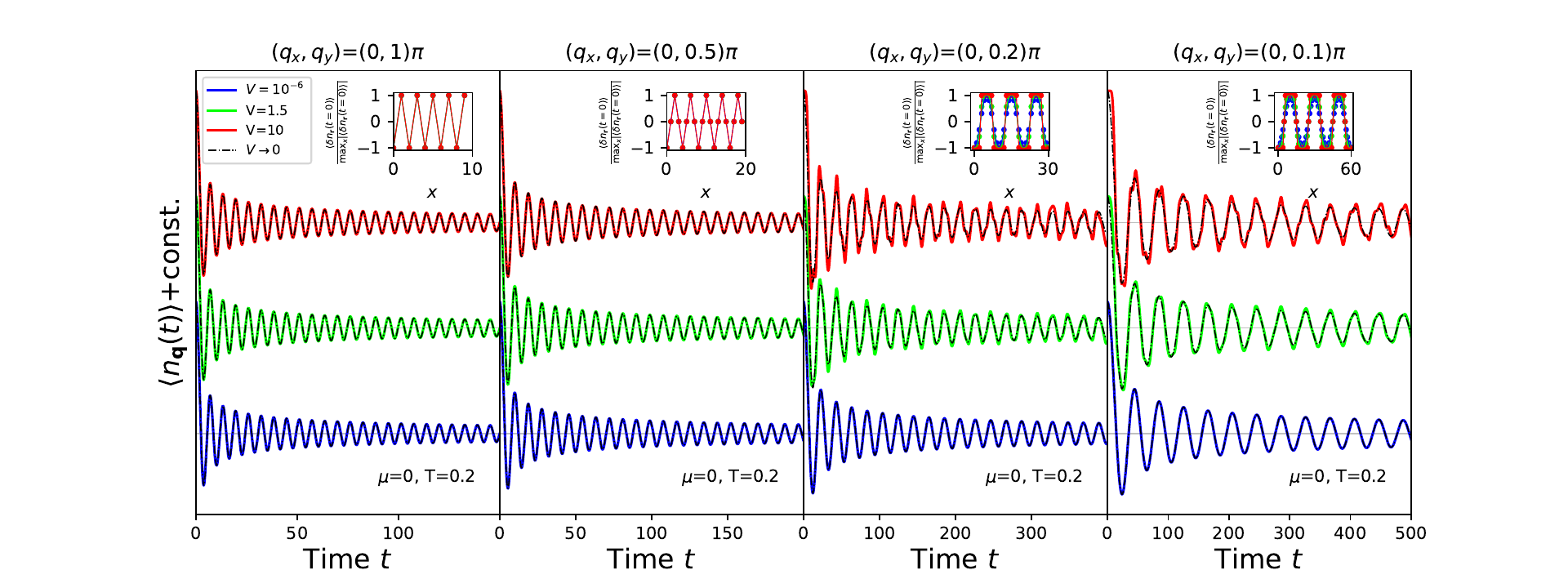}
 \caption{ Non-interacting case, half-filling, moderate temperature. CDW amplitude vs. time at various wave-vectors along the $x$-axis, calculated within linear response theory and from the full Kadanoff-Baym 3-piece contour formalism, assuming different amplitudes of the density-modulating field $V$ at $t<0$. CDW amplitude is normalized to the initial amplitude of the CDW. Insets: density profile of the initial CDW at different strengths of the field $V$. See text for details. }
\label{fig:U0_At_StrongV} 
\end{figure*}

\begin{figure}[!ht]
 \includegraphics[width=3.2in,trim=0cm 0cm 0cm 0cm]{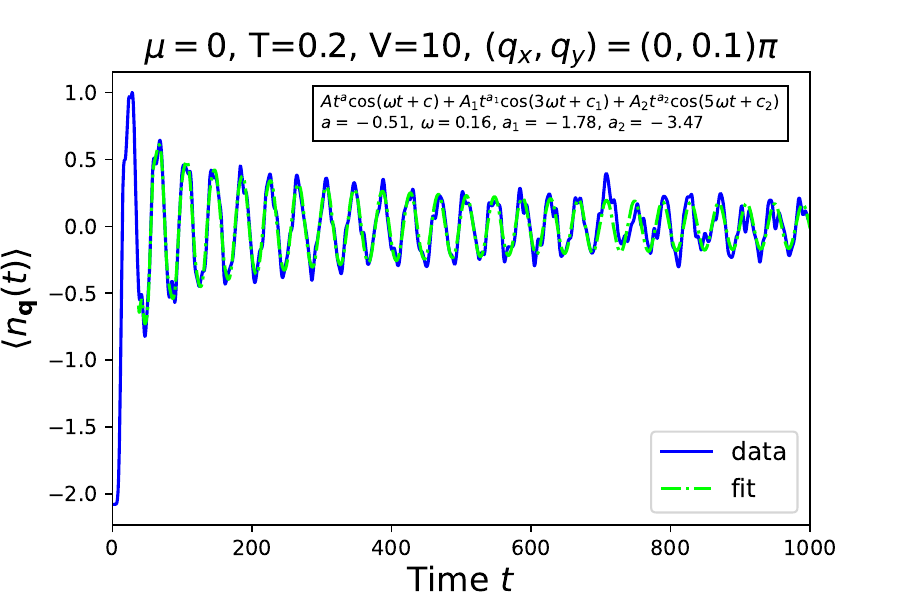}
 \caption{ Non-interacting case, half-filling, moderate temperature. Fit to the CDW amplitude vs. time, starting from a saturated CDW with a short wave-vector. }
\label{fig:U0_At_StrongV_fit} 
\end{figure}

We show several examples of this calculation in Fig.~\ref{fig:U0_At_examples}. We find numerous categories of solutions, and we illustrate some of them on the panels of Fig.~\ref{fig:U0_At_examples}, left to right:
\begin{itemize}
 \item power-law damped oscillations
 \item power-law damped oscillations with a breathing amplitude
 \item power-law damped oscillations with a decaying non-oscillatory component
 \item power-law decaying non-oscillatory behavior
\end{itemize}
The behavior at $\mathbf{q}=(\pi,\pi)$ is drastically different from the behavior at long wavelengths. At $(\pi,\pi)$ there is no clear peak in the spectrum, i.e. no characteristic frequency to produce oscillatory behavior. In particular, as $T\rightarrow 0$, the charge-charge correlation function (which is in the non-interacting limit equal to the spin-spin correlation function) approaches the form of a second-order pole $\sim 1/z^2$, which signals the instabiility towards order. One therefore finds only a non-oscillatory decay of the initial CDW amplitude, somewhat reminiscent of the diffusive regime of the hydrodynamic theory.

\subsubsection{Beyond linear response}

To cross-check these results and to be able to access the regime beyond the linear response (corresponding to initial density wave of a bigger amplitude) we perform the corresponding  Kadanoff-Baym 3-piece contour calculation\cite{AokiRMP2014}. 

The external field couples to the density wave at the wave-vector $\mathbf{q}=(q,0)$,
\begin{eqnarray}\nonumber
H[V] &=& H_0 - V\sum_\mathbf{\sigma,r} \cos (\mathbf{q}\cdot\mathbf{r}) n_{\sigma,\mathbf{r}} \\ 
     &=& H_0 - \frac{V}{2}\sum_\mathbf{\sigma,k}  \Big(c^\dagger_{\sigma,\mathbf{k}+\mathbf{q}}c_{\sigma,\mathbf{k}}+\mathrm{H.c.}\Big) \\
     &=& H_0 - \frac{V}{2} (n_\mathbf{q}+n_{-\mathbf{q}})
\end{eqnarray}
where $ V $ is the strength of the field.
We assume the field was turned on slowly at $t=-\infty$, and that by the time $t=0$, the system is already thermalized. Then, at $t=0$, the field is turned off abruptly. Therefore
\begin{equation}
 H(t < 0) = H[V],\;\;\;H(t > 0) = H_0
\end{equation}
In general, the expectation value of an operator $\mathcal{O}$ at time $t$ following the quench of the field $V$ is given by
\begin{equation}
\langle \mathcal{O}(t) \rangle
  = \frac{\mathrm{Tr}[e^{-\beta H[V]}e^{i H_0 t}\mathcal{O}e^{-i H_0 t}]}{\mathrm{Tr}[e^{-\beta H[V]}]}.
\label{expectation_value3}
\end{equation}
After the quench, the Hamiltonian has the following diagonal form
\begin{equation}
H_0=\sum_{\textbf{k}\sigma}\varepsilon_{\textbf{k}}c_{\sigma,\textbf{k}}^{\dagger}c_{\sigma,\textbf{k}},
\label{hamiltonian_t>0_2}
\end{equation}
whereas before the quench, the diagonal form is
\begin{equation}
H[V] = \sum_{\tilde{k}_x \mathcal{V} k_y} \sum_{\sigma}\varepsilon_{(\tilde{k}_x \mathcal{V} k_y)\sigma}c_{(\tilde{k} \mathcal{V} k_y)\sigma}^{\dagger}c_{(\tilde{k} \mathcal{V} k_y)\sigma}.
\end{equation}
Because of the symmetry breaking field, there is a reduction of the Brillouin zone, i.e. $\tilde{k}_x\in[0,2\pi/\lambda)$, where $\lambda=2\pi/q$ is the wave-length, or the number of sites in the unit cell; the additional quantum number arising due to the reduction of the BZ is $\mathcal{V}$. 

The time-evolution of density at a given point in space $\mathbf{r}=(x,0)$ (the translational symmetry is not broken along the $y$-axis: nothing changes if we take a more general $\mathbf{r}=(x,y)$) is given by
\begin{eqnarray} \nonumber
&& \langle n_\mathbf{r}(t) \rangle  \equiv \left\langle \sum_\sigma c_{\textbf{r}\sigma}^{\dagger} (t) c_{\textbf{r}\sigma}(t) \right\rangle  \\ \nonumber
&& = \frac{2}{N}  \sum_{\tilde{k_x},k_y\in\mathrm{RBZ}}\sum_{c,c'\in[0,\lambda)} \sum_{\mathcal{V}} 
e^{i x (c-c') q} e^{-i(\epsilon_{\textbf{k}}-\epsilon_{\textbf{k'}})t} \\
&& \;\;\;\times
\langle \tilde{k}_x \mathcal{V} k_y\vert k'_x k_y\rangle 
\langle k_x k_y \vert \tilde{k}_x \mathcal{V} k_y\rangle 
n_\mathrm{F}\big(\varepsilon_{\tilde{k}_x \mathcal{V} k_y}\big)
\label{eq:KadanoffBaym_nqt}
\end{eqnarray}
where we take $ k_x = \tilde{k}_x +c q$, $ k'_x = \tilde{k}_x +c' q$. The eigenstates of $H_0$ are denoted $|k_xk_y\rangle$, and the eigenstates of $H[V]$ are denoted $|\tilde{k}_x\mathcal{V}k_y\rangle$. The amplitude of the charge density wave $n_\mathbf{q}$ is given by the deviation of $n_\mathbf{r}$ from the lattice-averaged density, at the antinode of the wave, say $\mathbf{r}=(0,0)$.

We find perfect agreement between the results of Eq.\ref{eq:KadanoffBaym_nqt} with $V$ taken small and Eq.\ref{eq:noninteracting_bubble_nqt} which is in the strict $V\rightarrow 0$ limit. The full Kadanoff-Baym calculation is clearly more computationally expensive, but it allows us to set $V$ to stronger values and investigate the behavior starting from CDWs of finite amplitude. This is shown in Fig.~\ref{fig:U0_At_StrongV}.
In the two panels on the right,  we see that regular damped oscillations are replaced by a superposition of multiple waves as $V\rightarrow \infty$. This can be understood by looking at the density profile $n_\mathbf{r}$ at $t<0$ (shown in the insets of Fig.~\ref{fig:U0_At_StrongV}). One cannot place more than two electrons on a single site, which means that at strong values of $V$, the CDW is no longer harmonic; as $V\rightarrow \infty$ it becomes similar to the step function. This density profile corresponds to having multiple CDWs at the same time - at $q$, $3q$, $5q$, etc. All these CDWs will oscillate at different frequencies, but one also expects interactions between the waves. It is not easy to exaplain the detailed structure of $n_\mathbf{q}(t)$ beyond the linear response regime. However, in the long wavelength limit, the characteristic frequency of CDWs is proportional to $q$, and we are able to roughly fit the resulting $\langle n_\mathbf{q}(t)\rangle$ to a superposition of waves $\sim \sum_{l=1,3,5,...} t^{a_l}\cos(l\omega t + C_l)$. This is shown in Fig.~\ref{fig:U0_At_StrongV_fit}. 
However, the two panels on the left in Fig.~\ref{fig:U0_At_StrongV} show that at shortest wavelengths, one observes no change in behavior as $V$ is increased. This is because the density profile $n_\mathbf{r}(t<0)$ cannot change - there are no shorter waves to be excited by the increasing field. 

%

\subsection{Weak coupling theory}
\label{sec:weak_coupling}

\begin{figure}[ht!]
 \centering{ 
 \includegraphics[width=2.in,trim=0cm 0cm 0cm 0cm, clip]{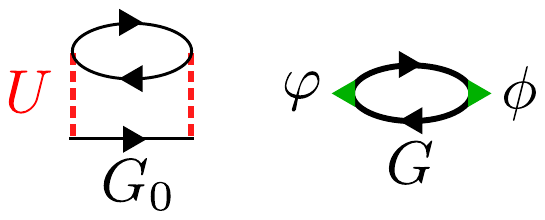}
 }
 \caption{ Illustration of our weak-coupling theory. Left: second-order self-energy diagram comprising the dynamical part of the self-energy, formulated in terms of the bare fermionic propagator (Eq.~\ref{eq:bare_greens_function}). Right: generalized (Kubo) bubble approximation for two-particle correlation functions, formulated in terms of the ``dressed'' Green's function (Eq.~\ref{eq:dressed_greens_function}).
}
\label{fig:weak_coupling_illustration} 
\end{figure}

\begin{figure*}[!ht]
 \centering{ 
 \includegraphics[width=6.4in,trim=0cm 0cm 0cm 0cm, clip]{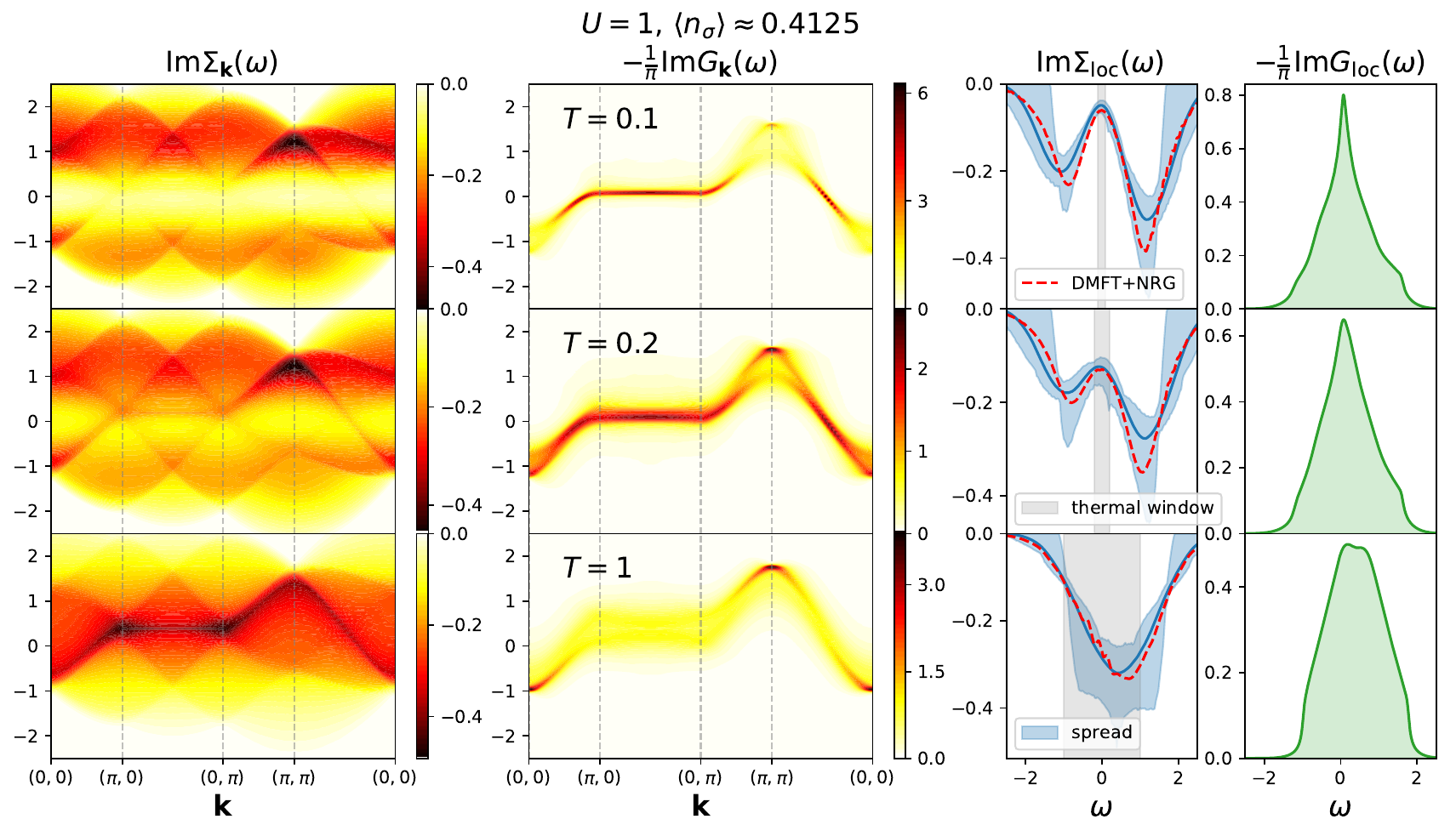}
 }
 \caption{ Moderate coupling, moderate doping, various temperatures. Self-energy at the level of second-order perturbation theory, and the corresponding Green's functions. Third column: blue line is the local self-energy; blue shading is the $\mathbf{k}$-spread, i.e. the range of values of $\mathrm{Im}\Sigma_\mathbf{k}$ at the given real-frequeny  $\omega$; red line DMFT(NRG) result for local self-energy; gray shading is the thermal window, denoting $\omega\in[-T,T]$. Rightmost column: the local density of states.
}
\label{fig:Sigma_examples} 
\end{figure*}

\subsubsection{Self-energy}

We start by calculating the self-energy up to the second-order in the coupling constant.
\begin{eqnarray}
\Sigma_\mathbf{k}(z) &=& U\langle n_{\bar{\sigma}} \rangle  + U^2 \tilde{\Sigma}_\mathbf{k}(z)\\ \nonumber
\tilde{\Sigma}_\mathbf{k}(z) &=& \sum_{\mathbf{k'},\mathbf{q}}  
  \frac{ 
    \sum_{s=\pm1} n_F(-s\varepsilon_{\mathbf{k}'}) n_F(s\varepsilon_{\mathbf{k}'+\mathbf{q}}) n_F(s\varepsilon_{\mathbf{k}-\mathbf{q}})  
  }
  { z-\varepsilon_{\mathbf{k}-\mathbf{q}}-\varepsilon_{\mathbf{k}'+\mathbf{q}}+\varepsilon_{\mathbf{k}'} } \\  \label{eq:Sigma_simple}   
\end{eqnarray}
The first term is the instantaneous Hartree shift, and the second term is the dynamic part, described by the second-order Feynman diagram illustrated in Fig.~\ref{fig:weak_coupling_illustration}.
The calculation of $\tilde{\Sigma}$ is expensive. In Appendix \ref{app:optmal_selfenergy} we describe a fast algorithm we used for this calculation, which allowed us to scan the phase diagram in considerable detail.

As we will see, for the practical calculations of conductivity in the limit $U\rightarrow 0$, the Hartree shift vanishes. At a finite $U$, we absorb the Hartree shift in the chemical potential, i.e. $\mu \rightarrow \mu-U\langle n_\sigma \rangle$. This means that at a finite $U$, we compute the second-order self-energy diagram with Hartree-shifted propagators. Therefore, the frequency dependence of the dynamical part of the self-energy does not change with increasing $U$, if $\tilde{\mu}=\mu-U\langle n_\sigma \rangle$ is kept fixed. In practice, we compute the occupancy a posteriori, and infer $\mu$ from $U\langle n_\sigma \rangle$ and $\tilde{\mu}$. The other possibility is to compute the self-energy diagram using bare propagators. The difference between the two approaches disappears when higher order diagrams are also computed (under the condition that both series converge), as well as in the $U\rightarrow 0$ limit.

We show examples of the self-energy results in Fig.~\ref{fig:Sigma_examples}.
It is interesting that at low temperature, the frequency dependence of the self-energy generally features two peaks, while at high temperature, it features a single peak. At the highest temperatures, the peak follows the shape of the electron dispersion.

\begin{figure*}[!ht]
 \centering{ 
 \includegraphics[width=6.0in, trim=0cm 0cm 1cm 0cm]{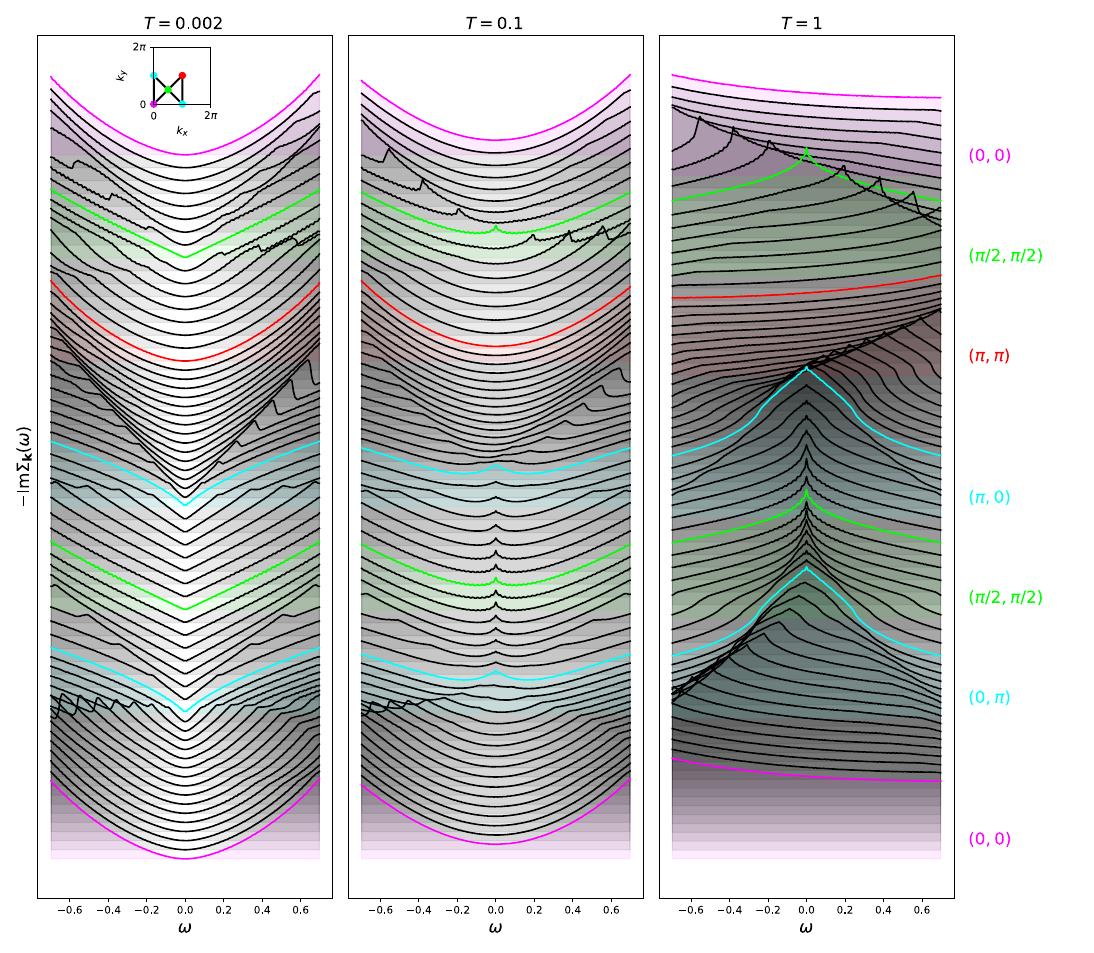}
 }
 \caption{ Half-filling, various temperatures. Second-order self-energy along a high-symmetry path in the BZ - zoom in on low frequencies. }
\label{fig:Sigma_examples_zoom} 
\end{figure*}

In Fig.~\ref{fig:Sigma_examples_zoom} we zoom in on the low-frequency part, along a high-symmetry path in the BZ.
We see that the scaling with $\omega$ around $\omega=0$ takes different forms depending on $\mathbf{k}$ and parameters of the model. 
The most interesting is the half-filling case, where we see that $\mathbf{k}=(\pi,0)$ and $\mathbf{k}=(\frac{\pi}{2},\frac{\pi}{2})$ are special points where in the $T\rightarrow 0$ limit one approaches $\mathrm{Im}\tilde{\Sigma}(\omega\rightarrow 0) \sim |\omega|^\alpha$ with $\alpha\approx \frac{4}{5}$ and $\alpha=1$, respectively. More precisely, the linear scaling $\alpha=1$ is observed along the path connecting $(0,\pi)$ and $(\pi,0)$, but is modified abruptly to $\alpha\approx \frac{4}{5}$ at those points. The linear scaling has been noted before\cite{Rohe2020}. However, the apparent $T\rightarrow 0$ limit of our second-order self-energy should only apply in the strict $U\rightarrow 0$ limit. At any finite coupling and low enough temperature, higher perturbation orders will play a role, and produce an insulating state\cite{Simkovic2020,Kim2020,Klett2020,Schaefer2021}.

We also note a large number of kinks in the frequency dependence of $\mathrm{Im}\tilde{\Sigma}$. The prominent peaks that appear at high temperature are not smooth: at the maximum no derivatives appear to be well defined.

\subsubsection{Green's function and compressibility}

Once we have the self-energy, we can plug it in the expression for the Green's function
\begin{equation}\label{eq:dressed_greens_function}
 G_\mathbf{k}(\omega) = \frac{1}{\omega-\varepsilon_\mathbf{k}-\Sigma_\mathbf{k}(\omega)}
\end{equation}
Examples of the Green's function are shown in Fig.\ref{fig:Sigma_examples}, as well as for the local density of states, $-\frac{1}{\pi}\sum_\mathbf{k}\mathrm{Im}G_\mathbf{k}(\omega)$. We observe that the sharp structures in the self-energy at intermediate temperature lead to a splitting of the peak in the single-particle spectrum at $\mathbf{k}\approx(\pi,\pi)$.

Ultimately, from the Green's function we get the average density $\langle n \rangle=-\frac{2}{\pi}\int \mathrm{d}\omega \sum_\mathbf{k} \mathrm{Im} G_\mathbf{k}(\omega) n_\mathrm{F}(\omega)$, and from it, the charge compressibility $\chi_\mathrm{c}=\frac{\partial \langle n \rangle}{\partial \mu}$, which will be needed to estimate the diffusion constant. At finite $U$, in practice, what enters the calculation is $\tilde{\mu}=\mu-U\langle n_\sigma\rangle$. It is then easiest to compute the quantity $\tilde{\chi}_\mathrm{c} = \frac{\partial \langle n \rangle}{\partial \tilde{\mu}}$. To get to the physical charge compressibility, one uses $\chi_\mathrm{c}=(\tilde{\chi}_\mathrm{c}^{-1}+U/2)^{-1}$.

\subsubsection{Bubble approximation for two-particle correlation functions}

It is of great interest to see how the long-wavelength behavior of the charge-charge (or equivalently the current-current) correlation functions changes due to weak interactions. The simplest approach is to just calculate the bubble approximation for $\chi$ or $\Lambda$ (illustrated in Fig.~\ref{fig:weak_coupling_illustration}). 
The bubble expression is simply the real-frequency formulation of Eq.\ref{eq:Qqz_final}, with the replacement $G_0\rightarrow G$. One obtains (under assumption that $\varphi_{\mathbf{k},\mathbf{q}}\phi_{\mathbf{k}+\mathbf{q},-\mathbf{q}}$ is purely real)
\begin{eqnarray} \label{eq:ImQqnu_bubble}
&& \mathrm{Im}Q_\mathbf{q}[\varphi,\phi](\nu)  \\ \nonumber
&&=\frac{2}{\pi} \sum_\mathbf{k} \varphi_{\mathbf{k},\mathbf{q}}\phi_{\mathbf{k}+\mathbf{q},-\mathbf{q}} \int \mathrm{d}\omega\\ \nonumber
&&\;\;\;\;\;\times  \mathrm{Im}G_{\mathbf{k}}(\omega)\mathrm{Im}G_{\mathbf{k}+\mathbf{q}}(\omega+\nu) [n_\mathrm{F}(\omega)-n_\mathrm{F}(\omega+\nu)]
\end{eqnarray}
where $2$ in front comes from the summation over spin, and $\frac{1}{\pi}=\frac{1}{\pi^2}\pi$ comes from the double Hilbert transform, and taking the delta-peak part of the integral (for detailed derivation in a more general case see Ref.~\onlinecite{VucicevicPRB2021}).
We have implemented this calculation and show results below.

However, the bubble approximation is not sufficient to properly address the question of whether the hydrodynamic form for $\chi$ (Eq.\ref{eq:chi_hyd}) or $\Lambda$ (Eq.\ref{eq:Lambda_hyd}) is valid at small $\mathbf{q}$. By construction, the bubble does \emph{not} satisfy the continuity equation. The reason for this is simple: the bubble expression is \emph{formally} an exact solution for a non-interacting system coupled to an external fermionic bath, the hybridization being the dynamical part of the self-energy. The bubble approximation for $\chi_\mathbf{q}(\nu)$ will therefore be manifestly wrong at $\mathbf{q}=0$, as one will get $\chi_{\mathbf{q}=0}(\nu\neq0)\neq 0$. Similarly, the bubble approximation for $\mathrm{Im}\Lambda_\mathbf{q}(\nu)$ will be manifestly wrong at $\mathbf{q}\neq0, \nu\rightarrow 0$, as it will scale as $\sim\nu$, and thus signal a finite conductivity. Clearly, if the system is open, a static wave of the electric field will scatter the incoming particles and maintain a current wave. The bubble approximations for $\chi$ and $\Lambda$ do not satisfy Eq.\ref{eq:chi_vs_Lambda_lattice}, and are not connected in a simple way. To restore physical properties, one needs to include vertex corrections, even at tiny couplings. 

On the other hand, the bubble approximation for $\mathrm{Im}\Lambda^{xx}_{\mathbf{q}=0}(\nu)$ will not be a priori unphysical, and can be considered a reasonable approximation for this object at low couplings. The parameters of the hydrodynamic model are encoded in $\mathrm{Im}\Lambda^{xx}_{\mathbf{q}=0}(\nu)$, or equivalently in $\sigma^{xx}_{\mathbf{q}=0}(\nu)$.
One can check whether $\mathrm{Im}\Lambda^{xx}_{\mathbf{q}=0}(\nu)$ satisfies Eq.\ref{eq:Lambdaq0_hyd}.
However, one has to keep in mind that different theories may reduce to the same form of $\Lambda^{xx}$ at $\mathbf{q}=0$. Even if $\mathrm{Im}\Lambda^{xx}_{\mathbf{q}=0}(\nu)$ satisfies Eq.\ref{eq:Lambdaq0_hyd} to a good degree, this cannot serve as proof that the hydrodynamic theory is valid. Nevertheless, assuming that the hydrodynamic theory is valid, one could use $\mathrm{Im}\Lambda^{xx}_{\mathbf{q}=0}(\nu)$ (or $\sigma^{xx}_{\mathbf{q}=0}(\nu)$) to extract the parameters for the hydrodynamic model, and investigate how they change with the microscopic parameters, which is what we will present in the following.

\subsubsection{Optical conductivity in the weak-coupling limit}

We distinguish here between the general weak coupling regime (say $U<1$) and the strict $U\rightarrow 0$ regime, i.e. the \emph{weak coupling limit}. To have a finite conductivity it is necessary to have scattering, so one cannot simply take $U=0$, but must rather consider an infinitesimal $U$. In the weak coupling limit, the bubble calculation for $\sigma^{xx}_{\mathbf{q}=0}(\nu)$ simplifies. In this way, one obtains the scaling of quantities in terms of $U$ in the $U\rightarrow 0$ limit.
In the following, we will distinguish between the simplified, \emph{weak coupling bubble} and the \emph{full bubble} calculations. The latter is computed for a finite $U$, using Eq.~\ref{eq:ImQqnu_bubble}.
The weak coupling bubble is inexpensive and we use it to cover the entire phase diagram. We also perform some full bubble calculations at $U=0.75$ and $U=1.0$ and show results below.

The weak coupling limit simplification of the bubble can be understood as follows.

If $U$ is sufficiently small, then the peaks in the spectral function start to resemble Lorentzians centered at $\omega=\varepsilon_\mathbf{k}$:
\begin{equation}
 G_\mathbf{k}(\omega\approx \varepsilon_\mathbf{k}; U\rightarrow 0) = \frac{1}{\omega-\varepsilon_\mathbf{k}-i\mathrm{Im}\Sigma_\mathbf{k}(\varepsilon_\mathbf{k})}
\end{equation}
The shifts coming from $\mathrm{Re}\Sigma_\mathbf{k}(\omega)$ can be neglected, and $\mathrm{Im}\Sigma_\mathbf{k}(\omega)$ can be considered to be constant at the scale of the width of the peak.
Away from $\omega\approx \varepsilon_\mathbf{k}$, $\mathrm{Im}G(\omega)$ can be considered zero.
Furthermore, the optical conductivity is expected to be non zero only at tiny frequencies, which also simplifies the Fermi-Dirac factor. We are ultimately able to employ the integral
\begin{eqnarray}
 \int \mathrm{d}x \mathrm{Im}\frac{1}{x\pm iy}\mathrm{Im}\frac{1}{x\pm\Delta\pm iy} 
 &=& \frac{\pi}{2y}\frac{1}{\left(\frac{\Delta}{2y}\right)^2+1}
\end{eqnarray}
which in the limit $\Delta = 0$ reduces to $\frac{\pi}{2y}$. In total we obtain
\begin{equation}\label{eq:sigma_weak_coupling}
 \sigma^{xx}_{\mathbf{q}=0}(\nu) = \sum_\mathbf{k} \frac{(v^x_\mathbf{k})^2 n'_F(\varepsilon_\mathbf{k})}{\mathrm{Im}\Sigma_\mathbf{k}(\varepsilon_\mathbf{k})}
 \left(\left(\frac{\nu}{2\mathrm{Im}\Sigma_\mathbf{k}(\varepsilon_\mathbf{k})}\right)^2+1\right)^{-1}.
\end{equation}
It is important to compare this expression to the Boltzmann expression for the dc conductivity Eq.~\ref{eq:boltzmann_sigmadc}. The two expressions do coincide, but only under assumption that $\mathrm{Im}\Sigma_\mathbf{k}(\varepsilon_\mathbf{k})$ does \emph{not} depend on $\mathbf{k}$, in which case one would have $\Gamma=-2\mathrm{Im}\Sigma_\mathbf{k}(\varepsilon_\mathbf{k})$. However, $\mathrm{Im}\Sigma_\mathbf{k}(\varepsilon_\mathbf{k})$ retains considerable $\mathbf{k}$-dependence even at infinite temperature as $\lim_{T\rightarrow\infty}\mathrm{Im}\tilde{\Sigma}_\mathbf{k}(\varepsilon_\mathbf{k})=-\frac{\pi}{4}\sum_{\mathbf{k}'\mathbf{q}}\delta(\varepsilon_\mathbf{k}-\varepsilon_{\mathbf{k}+\mathbf{q}}-\varepsilon_{\mathbf{k}'+\mathbf{q}}+\varepsilon_{\mathbf{k}'}$).
The expression Eq.\ref{eq:sigma_weak_coupling} presents a sum of Lorentzians of different heights and widths, and the end result might not fit well to the Lorenzian shape.
Our weak coupling theory does not a priori reduce to Eq.~\ref{eq:boltzmann_sigmadc} or the hydrodynamic Eq.~\ref{eq:sigmaxx_hyd}.
At infinite temperature, the $D$ and $\Gamma$ we might extract from our results are a priori separate objects - their product $D\Gamma$ will depend on the precise form of the self-energy. 

We now pull the $U^2$ factor out of the self-energy to obtain
\begin{equation}\label{eq:sigma_U2_pref}
 \tilde{\sigma}^{xx}_{\mathbf{q}=0}(\tilde{\nu}) \equiv \sum_\mathbf{k} \frac{(v^x_\mathbf{k})^2 n'_F(\varepsilon_\mathbf{k})}{\mathrm{Im}\tilde{\Sigma}_\mathbf{k}(\varepsilon_\mathbf{k})}
 \left(\left(\frac{\tilde{\nu}}{2\mathrm{Im}\tilde{\Sigma}_\mathbf{k}(\varepsilon_\mathbf{k})}\right)^2+1\right)^{-1}
\end{equation}
with the definitions
\begin{eqnarray} \label{eq:sigma_U2_law}
 \sigma(\nu=\tilde{\nu}U^2) = \frac{\tilde{\sigma}(\tilde{\nu})}{U^2} 
\end{eqnarray}
At low frequency, we can now equate the hydrodynamic form Eq.\ref{eq:sigmaxx_hyd} with the above equation, to reach the following
\begin{eqnarray} \label{eq:D_U2_pref}
 \tilde{D} = \frac{\tilde{\sigma}^{xx}_{\mathbf{q}=0}(\tilde{\nu}=0)}{\chi_c}&,\;\;\;\;& D = \frac{\tilde{D}}{U^2} \\ \label{eq:Gamma_U2_pref}
 \tilde{\Gamma} = \delta \left(1-\frac{\tilde{\sigma}^{xx}_{\mathbf{q}=0}(\tilde{\nu}=0)}{\tilde{\sigma}^{xx}_{\mathbf{q}=0}(\tilde{\nu}=\delta)}\right)^{-\frac{1}{2}}&,\;\;\;\;& \Gamma=\tilde{\Gamma} U^2
\end{eqnarray}
where we take $\delta$ small.

\begin{figure}[!ht]
 \centering{
 \includegraphics[width=3.2in,trim=0cm 0cm 0cm 0cm]{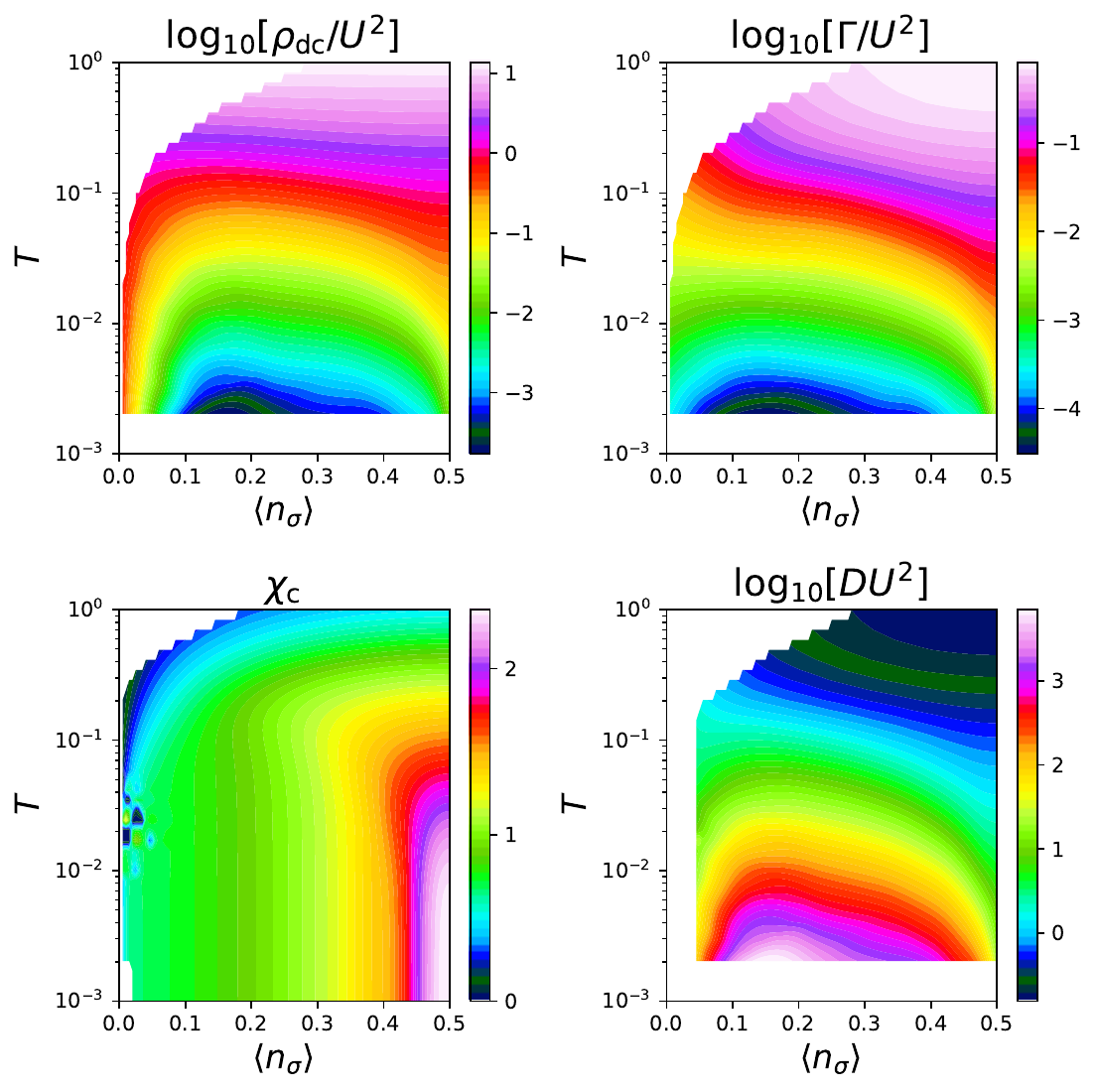}
 }
 \caption{ Summary of weak-coupling results: doping-temperature phase diagram. Color plots of dc resistivity, diffusion constant, momentum-relaxation rate in the $U\rightarrow 0$ limit, extracted from the weak coupling bubble calculation Eq.~\ref{eq:sigma_U2_pref}, and the corresponding non-interacting compressibility. }
\label{fig:weak_coupling_summary}
\end{figure}

\begin{figure}[!ht]
\begin{centering}
 \includegraphics[width=3.2in,trim=0cm 0cm 0cm 0cm]{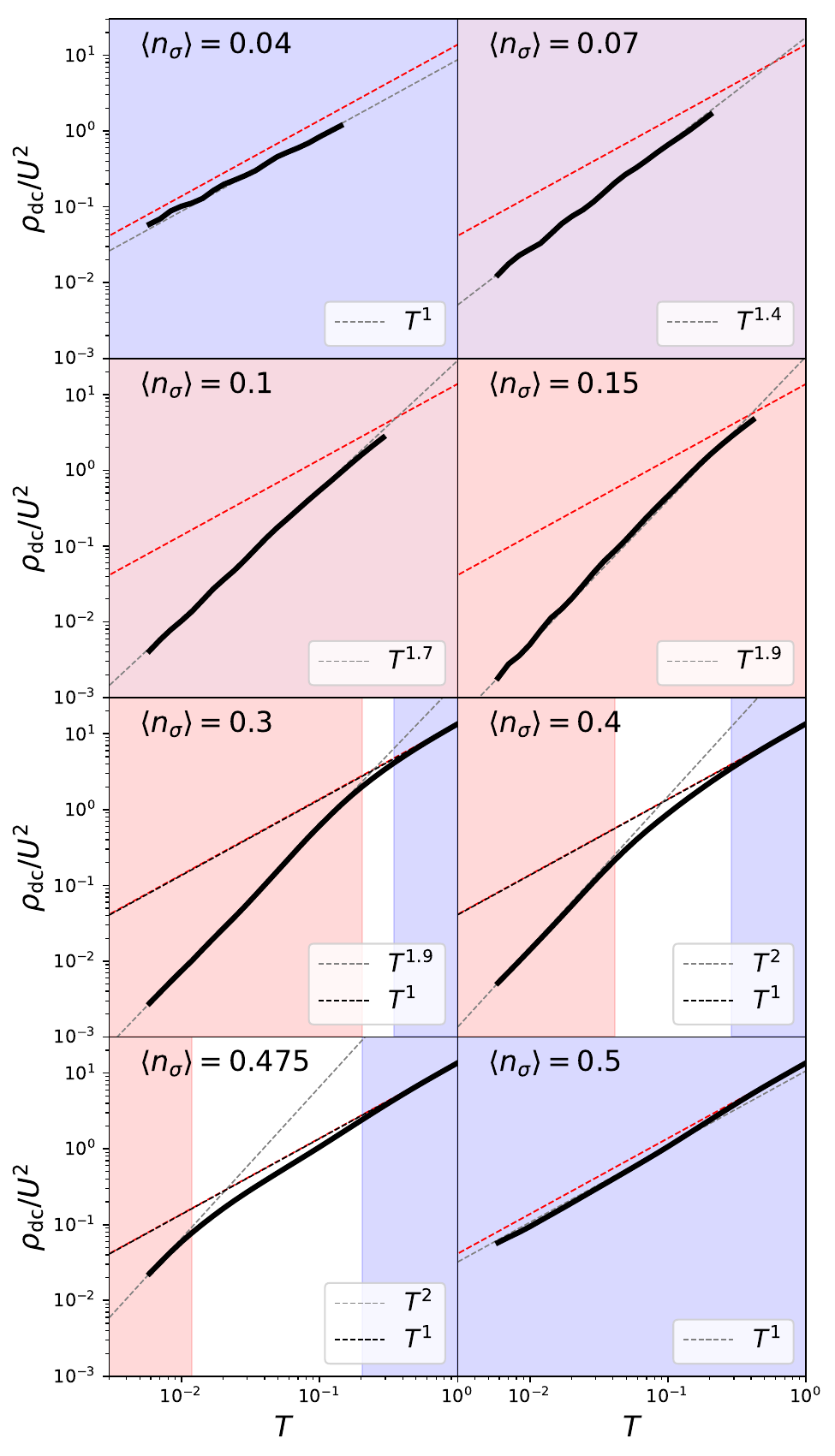}
 \includegraphics[width=3.2in,trim=0cm 0cm 0cm 0cm]{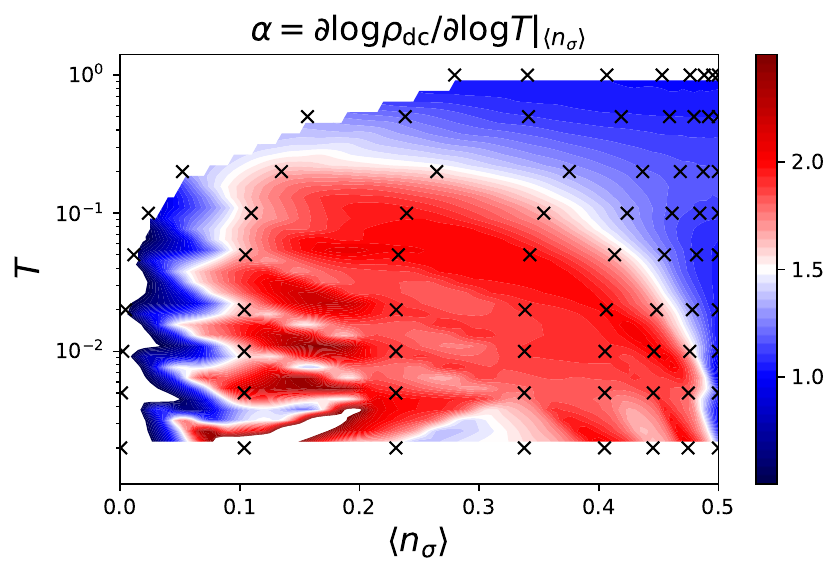}
\end{centering}
 \caption{ Weak coupling bubble dc resistivity results. Upper panels: temperature-dependence at different dopings. Lower panel: effective exponent of the temperature dependence, color-plotted in the doping-temperature plane; black crosses are actual data points - the rest is obtained by interpolation.
 }
\label{fig:weak_coupling_rhodc}
\end{figure}

This result gives us the estimate of how $\Gamma$ and $D$ behave as functions of $U^2$, at low coupling. 
Additionally, one can conclude that the diffusive regime extends to shorter wavelengths as coupling is increased, i.e.
\begin{equation} \label{eq:qD_vs_U2}
 q_D = \sqrt{\frac{\Gamma}{4D}} = U^2 \sqrt{\frac{\tilde{\Gamma}}{4\tilde{D}}} \equiv U^2 \tilde{q}_D
\end{equation}
The coefficients $\tilde{D}$ and $\tilde{\Gamma}$ depend on the microscopic parameters, and we extract them from $\tilde{\sigma}^{xx}_{\mathbf{q}=0}(\tilde{\nu})$, calculated by Eq.\ref{eq:sigma_U2_pref}.

Even though $\tilde{\sigma}(\tilde{\nu})$ might not have the shape of a Lorenzian, the property $\tilde{\sigma}(\tilde{\nu}\rightarrow 0) \sim \tilde{\sigma}_\mathrm{dc}(1-\tilde{\nu}^2)$ is guaranteed. Therefore, the form Eq.\ref{eq:sigmaxx_hyd} is bound to hold at least at the lowest frequencies, and one can certainly extract the effective $\tilde{\Gamma}$ via Eq.\ref{eq:Gamma_U2_pref}. It is interesting to see in what range of frequencies will the hydrodynamic Eq.\ref{eq:sigmaxx_hyd} be valid.


\subsubsection{Results for dc resistivity}

We first focus on the $\tilde{\rho}_\mathrm{dc}=1/\tilde{\sigma}_\mathrm{dc}$ results. The colorplot of $\tilde{\rho}_\mathrm{dc}$ as a function of doping and temperature is given in Fig.~\ref{fig:weak_coupling_summary}. We show the $T$-dependence at different dopings in the upper part of Fig.~\ref{fig:weak_coupling_rhodc}. We see the following trends.
At half-filling one observes two separate $\tilde{\rho}_\mathrm{dc}\sim T$ regimes, one at low, the other at high temperature.
The high-temperature limit of Eq.~\ref{eq:sigma_U2_pref} can be easily computed for the half-filled case based on the high-temperature asymptotic form of $\tilde\Sigma_\mathbf{k}(\varepsilon_\mathbf{k})$. One obtains $\tilde{\rho}_\mathrm{dc}(T) \approx 13.8 T$. This appears to be the high-temperature asymptotic behavior at least at moderate dopings, as well.
As one dopes away from half-filling, a $\tilde{\rho}_\mathrm{dc}\sim T^2$ regime emerges at ever higher temperatures, while the $\tilde{\rho}_\mathrm{dc}\sim T$ is pushed to higher $T$. Starting from around $\langle n_\sigma \rangle =0.3$, the low-$T$ regime transforms into $\tilde{\rho}_\mathrm{dc}\sim T^{1.9}$. At $\langle n_\sigma \rangle =0.15$ we no longer observe  $\tilde{\rho}_\mathrm{dc}\sim T$ in the accessible range of temperature, but further doping continuously reduces the exponent in the FL-like regime. At very low fillings, we see $\tilde{\rho}_\mathrm{dc}\sim T$ in the accessible range of $T$. The effective exponent $\alpha$ of the $T$-dependence of resistivity can be obtained as $\alpha=\left.\frac{\partial \log \rho_\mathrm{dc}(T)}{\partial \log T}\right|_{\langle n \rangle}$\cite{VucicevicPRB2013} and is color-coded in the bottom part of Fig.~\ref{fig:weak_coupling_rhodc}.

\begin{figure}[!ht]
\begin{centering}
 \includegraphics[width=3.2in,trim=0cm 0cm 0cm 0cm]{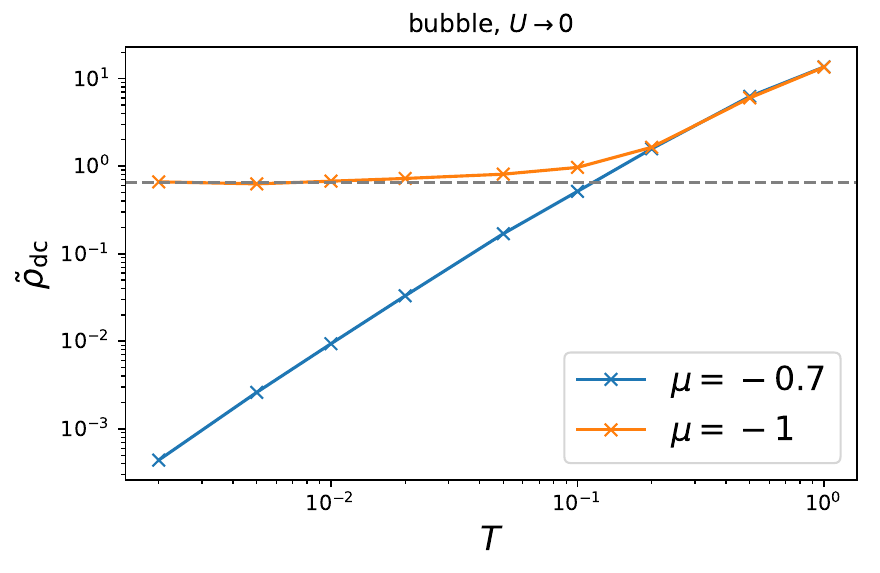}
\end{centering}
 \caption{  Weak coupling bubble ($U\rightarrow 0$ limit) dc resistivity results at a fixed chemical potential. Horizontal line denotes 0.65, which is the value $\rho_\mathrm{dc}$ apparently converges to at $\mu=-1$ and $T\rightarrow 0$.
 }
\label{fig:band_edge_rhodc}
\end{figure}

\begin{figure*}[t]
 \centering{
 \includegraphics[width=6.4in,page=2, trim=0cm 0cm 0cm 0cm]{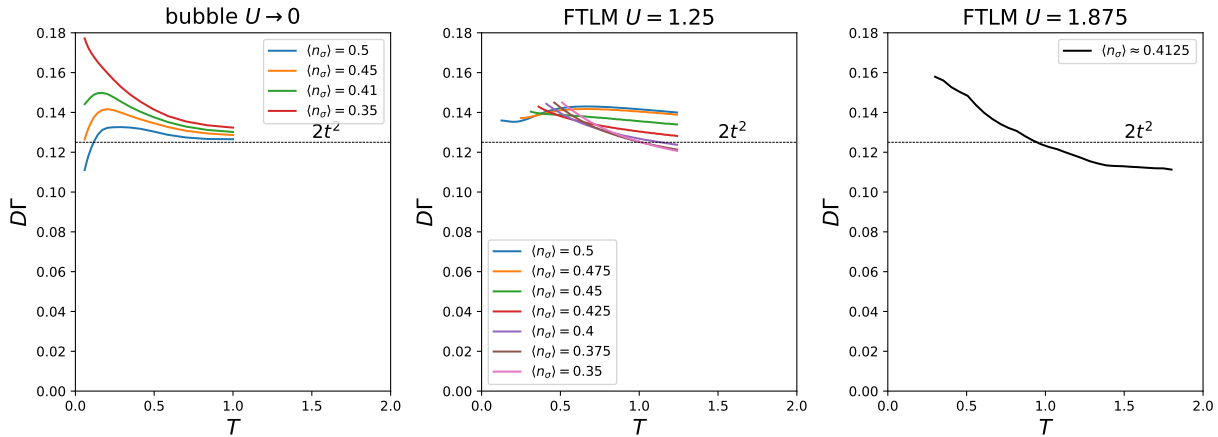}
 }
 \caption{ Weak coupling bubble  ($U\rightarrow 0$ limit) in comparison with moderate and strong coupling. At weak coupling, $D$ and $\Gamma$ are computed using Eq.~\ref{eq:Gamma_U2_pref} and Eq.~\ref{eq:D_U2_pref} at various dopings. At moderate and strong coupling, FTLM result is the best available result. At strong coupling, data is reconstructed from Ref.\cite{Brown2019}, but only for a single doping.
 }
\label{fig:weak_coupling_DGamma}
\end{figure*}

It is interesting to inspect the case of fixed $\mu=-1$: this means that strictly $k_F=0$ and all occupancy comes from thermal excitations. There we observe roughly $\tilde{\rho}_\mathrm{dc}\rightarrow 0.65$ as $T\rightarrow 0$ (see Fig.~\ref{fig:band_edge_rhodc}). This can be understood as follows: at low temperature, the contribution will come from an increasingly small vicinity of $\mathbf{k}=0$. We observe that $\mathrm{Im}\Sigma_{\mathbf{k}=0}(\omega=0) \sim T$. On the other hand, the velocity of electrons will decrease as temperature is lower. Ultimately, the amplitude of contributions will reduce to the integral $\int d\mathbf{k} k_x^2 e^{-\beta k^2} \sim \int_0^\infty dk\; k^3 e^{-\beta k^2} \sim T$. This means that the increased coherence of the electrons will be canceled exactly by their decreasing velocity, and the resistivity will converge to a constant as $T\rightarrow 0$. At $\mu$ slightly above $-1$ one expects the resistivity to go to 0, whereas for $\mu$ slightly below, one expects it to go to infinity.

\begin{figure}[h]
 \centering{
 \includegraphics[width=2.8in,trim=0cm 0cm 0cm 0cm]{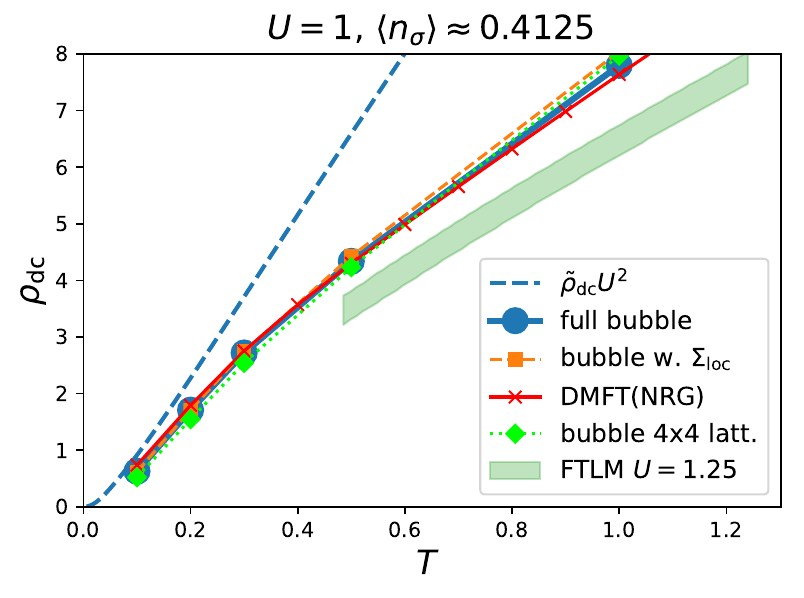}
 }
 \caption{ Moderate coupling $U=1$, moderate doping, results for dc resistivity. Dashed blue: weak coupling bubble. Blue line and dots: full bubble calculation. Orange dashed and squares: full bubble calculated with only the local component of self-energy. Red with crosses: full bubble with the local DMFT(NRG) self-energy result. Lime dotted with diamonds: full bubble computed on a $4\times 4$ lattice with the full $\mathbf{k}$-dependent second-order self-energy. Green stripe: FTLM $4\times 4$ result at a larger coupling $U=1.25$ (including the vertex corrections).
 }
\label{fig:rhodc_Boltzman_vs_bubble_vs_DMFT}
\end{figure}

\subsubsection{Results for hydrodynamic parameters}

The results for $\tilde{D}$ and $\tilde{\Gamma}$ are summarized on Fig.~\ref{fig:weak_coupling_summary}. It is apparent that roughly $\tilde{D}\sim 1/\tilde{\Gamma}$.
This can be understood intuitively - the more coherent the quasiparticles, the bigger the dc conductivity and the narrower the Drude peak. 
However, the inverse proportionality coefficient, i.e. $D\Gamma$ value is a priori unclear. 
We plot $D\Gamma$ in Fig.\ref{fig:weak_coupling_DGamma} and find that results approach $2t^2$ at high temperature. 
This is in agreement with the loose expectation based on Eq.~\ref{eq:microscopic_constitutive}, but also in agreement with the hydrodynamic theory (Eq.\ref{eq:DGamma_highT_limit}).
We also plot the corresponding result from FTLM computed at a moderate and a strong value of coupling (strong coupling data was reconstructed from Ref.~\onlinecite{Brown2019}), and find a similar result.
It is striking that $D\Gamma$ is within $\approx 20\%$ of $2t^2$ in a large range of temperature and even at strong coupling.

\begin{figure*}[t]
 \centering{
 \includegraphics[width=5.4in,trim=0cm 0cm 0cm 0cm]{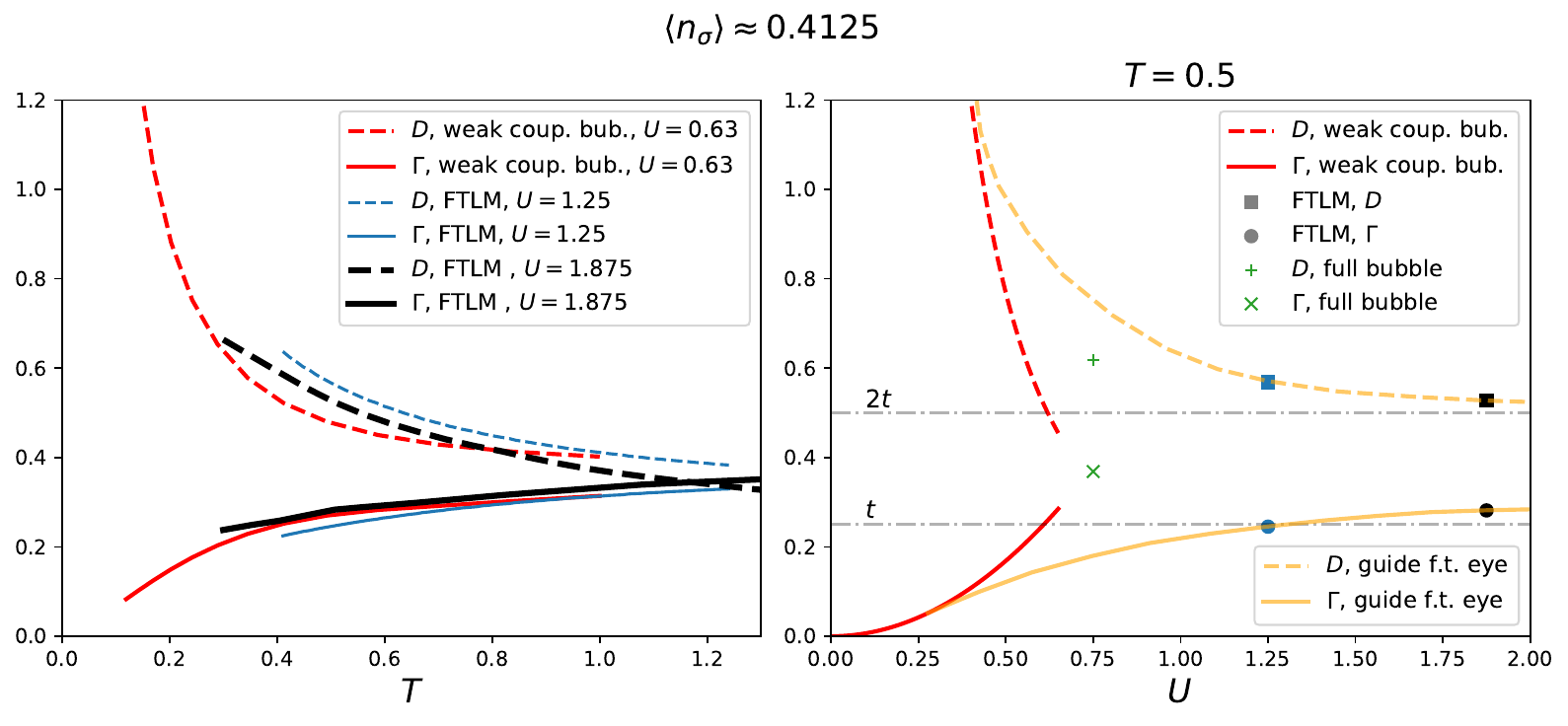}
 }
 \caption{ Left panel: Temperature dependence of $D$ and $\Gamma$ in the weak, moderate and strong coupling regimes. Right panel: Coupling dependence of $\Gamma$ and $D$, as obtained from the weak coupling bubble, full bubble and FTLM. The orange lines are guides for the eye, a possible scenario connecting results in the weak and moderate-to-strong coupling regimes.
}
\label{fig:weak_coupling_D_and_Gamma}
\end{figure*}

In Fig.~\ref{fig:rhodc_Boltzman_vs_bubble_vs_DMFT} we cross-check our weak-coupling $\rho_\mathrm{dc}$ result based on Eq.\ref{eq:sigma_weak_coupling} with the corresponding full bubble calculation (Eq.\ref{eq:ImQqnu_bubble}) at $U=1$.
As expected, the agreement is better at lower temperature and lower coupling (for the latter, the data is not shown), i.e. in cases where scattering rates $\mathrm{Im}\tilde{\Sigma}_\mathbf{k}(\varepsilon_\mathbf{k})$ are smaller.
We also compare our full bubble result to DMFT\cite{Georges1996} calculation at $U=1$ (implemented with the numerical renormalization group, NRG, real-frequency impurity solver\cite{wilson1975,krishna1980a,bulla2008,resolution,VucicevicPRL2019,VranicPRB2020,VucicevicPRB2021}) and surprisingly, find excellent agreeement.
Our second-order self-energy $\tilde\Sigma$ is clearly non-local (and remains non-local up to infinite temperature; see Fig.~\ref{fig:Sigma_examples}), yet the non-local part does not seem to play a big role in the value of dc resistivity. We check this explicitly by computing the bubble with only the local part of our self-energy - we find very similar result.
Moreover, the agreement with DMFT suggests that the local part of our second-order self-energy agrees well with DMFT. 
We confirm this in Fig.~\ref{fig:Sigma_examples}, especially in the thermal window $\omega \in [-T,T]$, which is the range of frequencies relevant for the conductivity calculation.
We also compute the full bubble on a small $4\times 4$ lattice. In the previous work of some of us\cite{VucicevicPRL2019}, it was shown that finite size effects subside at high temperature, and that the $4\times 4$-lattice FTLM calculation was correct at $T\gtrsim0.3$. However, this was at the value of coupling $U=2.5$ - we now see similar lack of finite-size effects even at $U=1$, which is somewhat unexpected (at a lower $U$ the relevant correlation lengths should be greater, and the systematic errors due to finite system size more pronounced). Comparing our full bubble result for $\rho_\mathrm{dc}$ at $U=1$ with the FTLM result at $U=1.25$ we can conclude that the vertex corrections are still sizeable, and affect the result in a similar fashion as at $U=2.5$, i.e. the vertex corrections present a roughly contant shift towards lower resistivity.

\begin{figure}[t]
 \centering{
 \includegraphics[width=3.2in,trim=0cm 0cm 0cm 0cm]{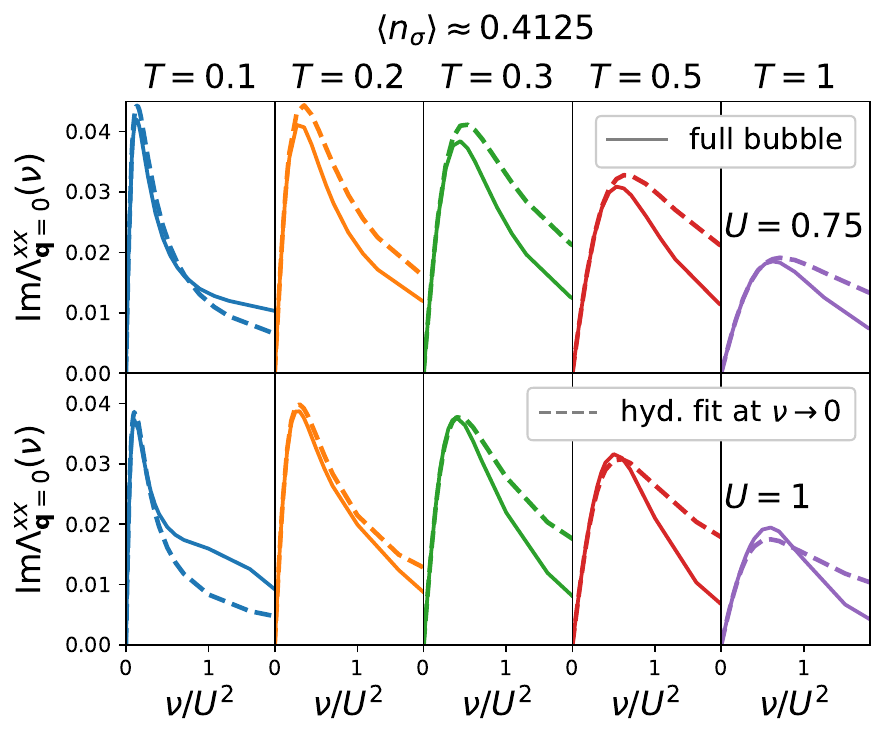}
 }
 \caption{ Moderate coupling, moderate doping. Full lines: full bubble result (Eq.~\ref{eq:ImQqnu_bubble}) for the uniform ($\mathbf{q}=0$) longitudinal current-current correlation function. Dashed lines: hydrodynamic form Eq.~\ref{eq:Lambdaq0_hyd} fitted at $\nu\rightarrow 0$.}
\label{fig:weak_coupling_Lambda}
\end{figure}

\begin{figure*}[!ht]
 \centering{ 
 \includegraphics[width=6.0in,trim=0cm 0cm 0cm 0cm]{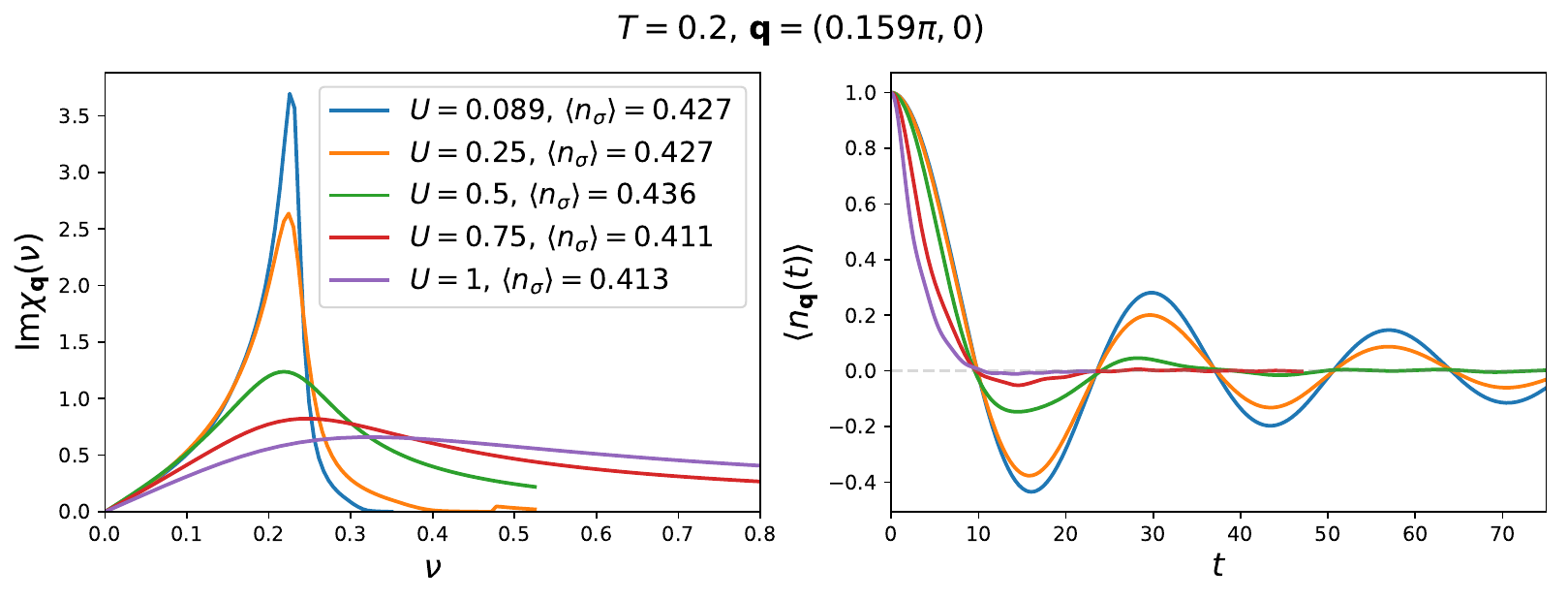}
 }
 \caption{ Weak to moderate coupling, moderate doping, moderate temperature. Left: full bubble calculation (Eq.~\ref{eq:ImQqnu_bubble}) for the charge-charge correlation function at a fixed wave vector at different values of coupling. Right: corresponding CDW amplitude vs. time curves (Eq.\ref{eq:linear_response_theory}). Prediction based on Eq.~\ref{eq:qD_vs_U2} for this doping and temperature is that $q_D(U=1)\approx 0.15\pi$, in agreement with the data in the plot on the right. }
\label{fig:bubble_chiq_vs_U} 
\end{figure*}

We compare $D$ and $\Gamma$ obtained from the weak coupling bubble at $U=0.63$ (Eqs.\ref{eq:D_U2_pref} and \ref{eq:Gamma_U2_pref}) to the FTLM result at $U=1.875$ and $U=1.25$ and find surprising similarity (see Fig.~\ref{fig:weak_coupling_D_and_Gamma}).
As was already apparent from Refs.\onlinecite{Brown2019} and \onlinecite{VucicevicPRL2019}, the bubble approximation tends to overestimate $\rho_\mathrm{dc}$ (i.e. underestimate $D$) and overestimate $\Gamma$. This explains the apparent agreement between the weak coupling bubble approximation and the numerically exact result at strong coupling.
However, up to a prefactor, even our weak coupling bubble results for $\tilde{D}(T)$ and $\tilde{\Gamma}(T)$ display a shape very similar to the FTLM result at $U=1.875$.
The behavior of the hydrodynamic parameters does not seem to change drastically going from weak to rather strong coupling. At a fixed temperatre $T=0.5$, the weak-coupling $\Gamma\sim U^2$ and $D\sim 1/U^2$ trends slow down at stronger coupling, so that the difference in $\Gamma$ and $D$ between $U=1.25$ and $U=1.875$ results is rather small. The full bubble computed at $U=0.75$ improves the result of the weak coupling bubble. It appears that in the strong coupling limit, $D\rightarrow 2t$, and roughly $\Gamma \rightarrow t$, which is consistent with $\lim_{U\rightarrow\infty} D\Gamma \approx 2t^2$.

\subsubsection{Deviations from the Lorentzian Drude peak}

As for the frequency-range of the validity of Eq.\ref{eq:sigmaxx_hyd}, i.e. the hydrodynamic form for the current-current correlation function Eq.\ref{eq:Lambdaq0_hyd}: it strongly depends on the microscopic parameters. At high temperature the agreement is excellent up to the peak of $\mathrm{Im}\Lambda(\nu)$, but, as expected, the high-frequency tail has a different scaling. This is shown in Fig.~\ref{fig:weak_coupling_Lambda} where we compare a fitted Eq.\ref{eq:Lambdaq0_hyd} with the result of the full bubble calculation Eq.\ref{eq:ImQqnu_bubble}. 

\subsubsection{Critical wavelength for diffusive behavior}

Finally, we go back to the simple prediction Eq.~\ref{eq:qD_vs_U2} that the characteristic wavelength for diffusive behavior will become shorter with increasing coupling. We check this directly in our $\chi_\mathbf{q}(\nu)$ results. As already mentioned, the bubble approximation is unsuitable for the investigation of $\chi$ at very long-wavelengths, but one might still want to inspect the results at somewhat bigger $\mathbf{q}$. In Fig.~\ref{fig:bubble_chiq_vs_U} we show the $\mathrm{Im}\chi_\mathbf{q}(\nu)$ results at a fixed $\mathbf{q}=(0.159,0)\pi$, and vary the coupling. The corresponding $\langle n_\mathbf{q}(t)\rangle$ results calculated via Eq.~\ref{eq:linear_response_theory} are presented on the panel on the right. For the occupancy $\langle n_\sigma \rangle \approx 0.4125$ and $T=0.2$, our weak coupling bubble calculation yields $\tilde{q}_D=\sqrt{\frac{\tilde\Gamma}{4\tilde{D}}}\approx0.15\pi$, which means that the behavior should become diffusive at wave vector $\mathbf{q}\approx(0.15\pi,0)$ at around $U=1$. This is in excellent agreement with the result we obtain directly from the full bubble approximation for $\chi$, as evidenced by Fig.~\ref{fig:bubble_chiq_vs_U}, panel on the right. Here we have plugged the full bubble result for $\chi_\mathbf{q}(\nu)$ in the linear response theory expression for the CDW amplitude $n_\mathbf{q}(t)$, Eq.~\ref{eq:linear_response_theory}

\subsection{Strong coupling}
\label{sec:strong_coupling}

We focus now on quantum Monte Carlo (QMC) results for the charge-charge and current-current correlation functions. We make use of the continuous-time interaction-expansion QMC, CTINT\cite{Rubtsov2005}. This method is numerically exact for a given lattice size. We calculate the inter-site $\chi_{ij}(i\nu)$ on the Matsubara frequency axis for a cyclic lattice of size $L\times L$ and then perform a periodization procedure, where we promote the intersite components to the real-space components of an infinite lattice, and thus obtain $\chi_{\mathbf{r}}(i\nu)$ of a finite range (components at $\mathbf{r}: r_\eta > L/2$ are considered 0). We then Fourier transform to obtain $\chi_\mathbf{q}(i\nu)$ with arbitrary resolution in the BZ. However, very short $\mathbf{q}$-vectors corresponding to wavelengths much greater than $L$ remain inaccessible. Nevertheless, we are able to obtain solid results for wavelengths up to 20 lattice spacings and that way cover the range of wavelengths studied in the cold-atom experiment by Brown et al.~\onlinecite{Brown2019}. We perform the exact same procedure for $\Lambda^{xx}$ as well.

\begin{figure*}[ht!]
 \centering{
 \includegraphics[width=6.0in,trim=0cm 0cm 0cm 0cm]{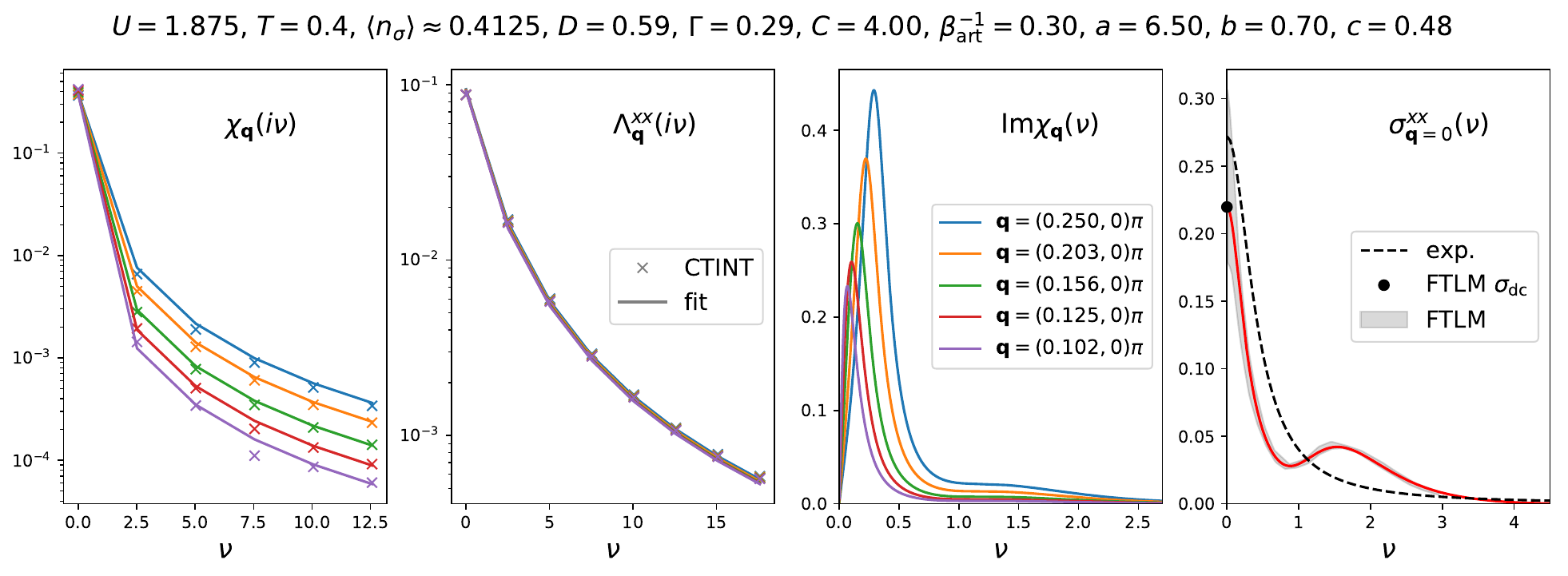}
 \includegraphics[width=6.0in,trim=0cm 0cm 0cm 0cm]{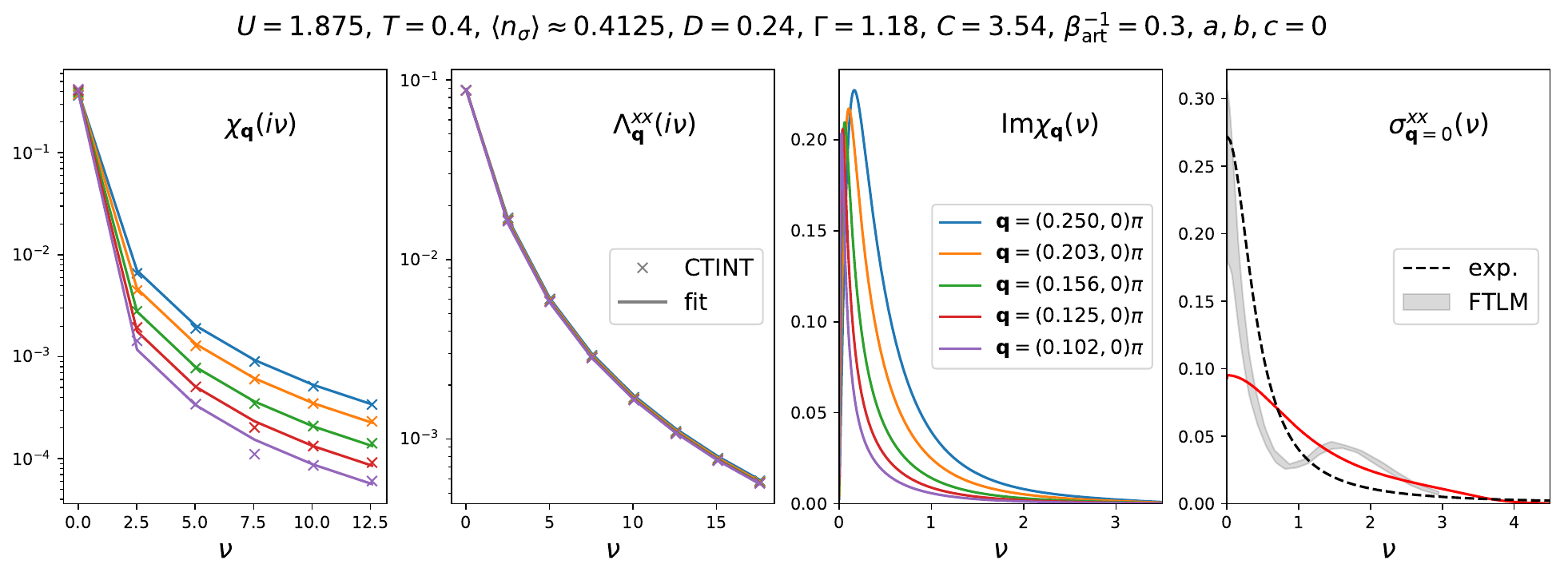}
 }
 \caption{ Strong coupling, moderate doping and temperature. Upper panels: modified hydrodynamic law with parameters hand-picked to reproduce FTLM result for optical conductivity (rightmost panel, gray stripe and black dot); two panels on the left: comparison of the modified hydrodynamic law the with corresponding CTINT results for the Matsubara axis charge-charge and current-current correlation functions; third panel: corresponding real-frequency charge-charge correlation function. Lower panels: unbiased fit of the modified hydrodynamic law without the high-frequency peak to the CTINT charge-charge and current-current correlation functions, simultaneously. The result has a strong bias for the values of $D$ and $\Gamma$.
 }
\label{fig:modified_hyd_vs_ctint_and_ftlm}
\end{figure*}

We perform a finite-size scaling analysis and observe that no obvious trends with $L$ are apparent in the results already between $L=4$ and $L=10$ (data not shown). This is consistent with the estimates from the recent Ref.~\onlinecite{Simkovic2021} where it was shown in a thermodynamic limit DiagMC calculation that the charge-correlations are short-ranged, with values becoming very small already at distances of about 5 lattice spacings. In our calculations, we consider the statistical errors to be the leading uncertainty, and use the $L=10$ results to perform analyses.

As already mentioned, it would make no sense to compare the hydrodynamic law to imaginary-axis data, because the hydrodynamic law predicts an unphysical asymptotic behavior of the current-current correlation function. To be able to compare the hydrodynamic theory with our Matsubara-frequency results, we propose a modified hydrodynamic form
\begin{eqnarray}\nonumber
 \mathrm{Im}\chi^{\mathrm{mh}}_\mathbf{q}(\nu) = \mathrm{Im}\chi^{\mathrm{hyd}}_\mathbf{q}(\nu) \big(1+L(\nu; a, b,c)\big) n_\mathrm{F}(\nu-C; \beta_\mathrm{art}) \\
 \label{eq:modified_hyd}
\end{eqnarray}
where $\chi^\mathrm{hyd}$ denotes the form in Eq.~\ref{eq:chi_hyd}, $L(\nu; a, b,c) = a \exp(-(\log\nu - b)^2/c^2)$ is the log-normal distribution, and $n_\mathrm{F}(\omega; \beta)=\frac{1}{e^{\beta\omega}+1}$ is the Fermi-Dirac distribution function. 
The $1+L$ part here is necessary to introduce high-frequency excitations to and from the upper Hubbard band, which are expected in the doped Mott insulator regime.
The Fermi-Dirac distribution function facilitates the exponential cutoff at high frequency.
This modified hydrodynamic form thus has 5 additional parameters: $a,b,c$ are the amplitude, position and width of the Hubbard peak, respectively, and $C$ and $\beta_\mathrm{art}$ are the cutoff frequency and the artificial temperature determining the rate at which the spectral weight is exponentially surpressed at cutoff. We make sure that $\beta_\mathrm{art}$ is small enough so that $n_F(-C,\beta_\mathrm{art})\approx 1$. Our choice of $\mathrm{Im}\chi^{\mathrm{mh}}_\mathbf{q}(\nu)$ ensures $\sim1/\nu^2$ Matsubara axis asymptotic behavior for $\Lambda^{xx}$. The prefactor of the asymptotic behavior will in general depend on the parameters $a,b,c,C,\beta_\mathrm{art}$.

We now check whether the modified hydrodynamic form Eq.~\ref{eq:modified_hyd} is consistent with the available CTINT and FTLM data.
We first hand-pick the parameters so that the FTLM result for the optical conductivity is reproduced. It is noteworthy that we can get a very good fit to FTLM data, and that the shape of the high-frequency peak is roughly a log-normal distribution. We then compare the resulting $\chi^{\mathrm{mh}}_\mathbf{q}(i\nu)$ and the corresponding $\Lambda^{xx,\mathrm{mh}}$ (obtained via Eq.~\ref{eq:chi_vs_Lambda_continuum}) to CTINT data. We find solid agreement, as shown in the upper part of Fig.~\ref{fig:modified_hyd_vs_ctint_and_ftlm}. 

In the lower part of Fig.~\ref{fig:modified_hyd_vs_ctint_and_ftlm} we illustrate how fitting the Matsubara data to a hydrodynamic law without the high-frequency peak will yield wrong results for $D$ and $\Gamma$, even if a proper high-frequency cutoff is used. We do a fully unbiased fit of $\chi^{\mathrm{mh}}_\mathbf{q}(i\nu)$ (with $a,b,c=0$ and $\beta_\mathrm{art}$ fixed to $0.3$, $D$,$\Gamma$ $\chi_c$,$C$ free), to reproduce at the same time $\chi$ and $\Lambda$ CTINT-results at 5 small $\mathbf{q}$-vectors. We get an excellent fit, but we get completely wrong values for $D$ and $\Gamma$. The optical conductivity contains two peaks, and fitting with only a single peak will compensate by making this one peak wider and shorter, thus underestimating $D$ and overestimating $\Gamma$. 
The Fig.~\ref{fig:modified_hyd_vs_ctint_and_ftlm} nicely illustrates the difficulty of analytical continuation - the fit function on the imaginary axis is almost indistinguishable between the top and bottom rows, yet corresponds to drastically different optical conductivity.

\begin{figure}[ht!]
 \centering{ 
 \includegraphics[width=3.2in,trim=0cm 0cm 0cm 0cm]{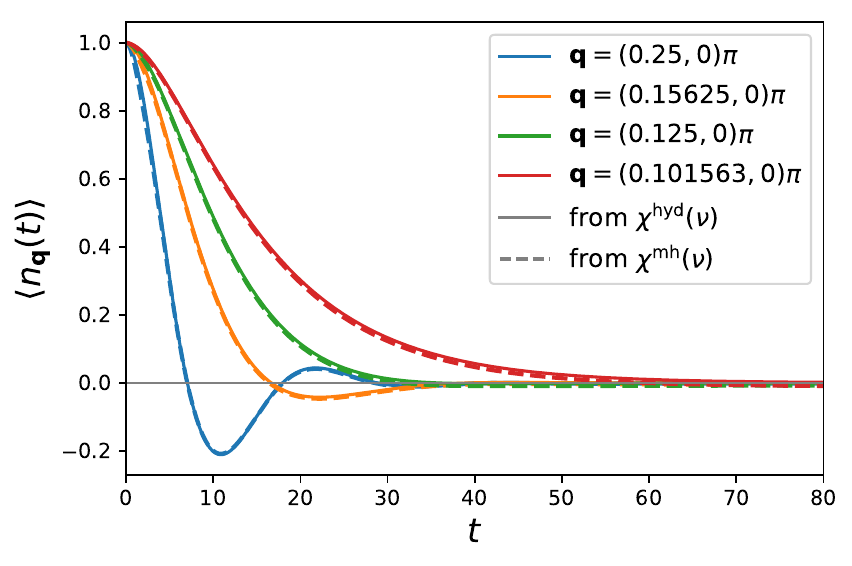} 
 }
 \caption{
 CDW amplitude vs. time curves, corrsponding to the modified hydrodynamic law from the upper panels of Fig.~\ref{fig:modified_hyd_vs_ctint_and_ftlm}, compared to the original hydrodynamic law Eq.~\ref{eq:chi_hyd} with the same $D$ and $\Gamma$.
}
\label{fig:nqt_hyd_vs_mh} 
\end{figure}

In Fig.~\ref{fig:nqt_hyd_vs_mh} we show the $\langle n_\mathbf{q}(t) \rangle$ curves, corresponding to $\chi^\mathrm{mh}$ parameters from the upper part of Fig.~\ref{fig:modified_hyd_vs_ctint_and_ftlm}. Comparing to the corresponding pure hydrodynamic law $\chi^\mathrm{hyd}$ (Eq.\ref{fig:nqt_hyd_vs_mh}), we find no visible difference - the inability of the hydrodynamic law to describe high-frequency features of $\chi$ are unlikely to have affected the fitting procedure in Ref.~\onlinecite{Brown2019}.

\section{Discussion and prospects} 
\label{sec:discussion}

Our work builds on the milestone study of charge fluctuations in the Hubbard model by Hafermann et al.\cite{Hafermann2014} 
In that work, the non-interacting charge-charge correlation function was calculated, but only at half-filling, and with a relatively low resolution - the striking two-linear-modes feature that we observe at finite doping was, therefore, overlooked.
More importantly, the Matsubara data were fitted to a law which only allows for a linear mode at long wavelengths, which may not be appropriate.
In light of more recent experimental evidence, and on general physical grounds, the emergence of diffusive behavior and a quadratic mode around $\mathbf{q}=0$ is expected.
Furthermore, in the work by Hafermann et al., vertex corrections had unlimited range, but were calculated as a diagrammatic extension of DMFT\cite{RohringerRMP2018}, which introduces systematic errors.
By using CTINT at $10\times10$ lattice size, we capture complete vertex corrections up to a medium range, yet the finite size effects in our theory are unlikely to have introduced significant systematic error.
Finally, Hafermann et al. have derived numerous useful identities relevant for charge-charge and current-current correlation function. However, the fully general Eq.~\ref{eq:chi_vs_Lambda_lattice} may have been overlooked so far.

Our weak coupling calculation is complementary to the semi-classical Boltzmann equation approach of Kiely and Mueller~\onlinecite{Kiely2021}. 
It is not clear that either of the two approaches yield exact results, even in the $U\rightarrow 0$ limit, thus it is important to cross check the results and look for robust, shared features.
Indeed, our results at low dopings are in excellent qualitative agreement with the Boltzmann equation - we observe linear resistivity at half-filling and an emerging $T^2$ at low temperature as one dopes away from half-filling. 
However, there is a significant quantitative difference in the values of $\tilde{\rho}_\mathrm{dc}$. 
The Boltzmann equation predicts the high-temperature asymptotic behavior $\tilde{\rho}_\mathrm{dc}=0.076 (T/t)(U/t)^2 = 4.864 (T/4t)(U/4t)^2$ while our Kubo bubble theory yields $\tilde{\rho}_\mathrm{dc}=13.8 (T/4t)$. The difference is half an order of magnitude. Comparing to numerically exact FTLM result at $U=1.25$, it is clear that our extrapolated $U\rightarrow 0$ theory strongly overestimates $\rho_\mathrm{dc}$ at high-temperature (see Fig.~\ref{fig:rhodc_Boltzman_vs_bubble_vs_DMFT}).
However, the coefficient for the linear high-$T$ asymptotics is overestimated by a factor of 2-3, at most.
Furthermore, we can compute the bubble result at a finite coupling to obtain much better results, the relative error unlikely being more than $30-40\%$ in the relevant range of temperature. 
There is also a striking qualitative difference in the $\rho_\mathrm{dc}(T)$ at large doping.
At $\langle n_\sigma \rangle=0.1$ Kiely and Mueller observe an exponential drop of resistivity at low temperature, in sharp contrast to our observations. 
Their finding was argued to be due to frustration of umklapp scattering. It is possible that, in our approach, vertex corrections are needed to observe this phenomenon. Further work is necessary to fully resolve the origin of this discrepancy.
Finally, our approach allows us to compute the full optical conductivity, and estimate $D$ and $\Gamma$ as separate objects, which, to the best of our understanding, could not have been done in their work. To our understanding, $\Gamma$ was extracted from $D$, \emph{assuming} the validity of the Boltzmann expression for conductivity (Eq.~\ref{eq:boltzmann_sigmadc}).
The analysis based on the asymptotic behavior of $\chi_{\mathbf{q}\rightarrow 0}(i\nu)$ that shows that the hydrodynamic theory (Eq.~\ref{eq:continuity}, Eq.~\ref{eq:constitutive}) is consistent with Eq.~\ref{eq:boltzmann_sigmadc} may have been previously overlooked.


Our analysis of the equation of motion for the current (Eq.~\ref{eq:microscopic_constitutive}), as well as our numerical results displaying $D\Gamma=2t^2$ at weak coupling and high-temperature (Fig.~\ref{fig:weak_coupling_DGamma}) provide some microscopic evidence for the validity of the hydrodynamic theory proposed in Ref.~\onlinecite{Brown2019}. 
Also, our Fig.~\ref{fig:bubble_chiq_vs_U} provides some support for $q_D=\sqrt{\Gamma/4D}$ which is a specific property of the hydrodynamic theory (Eq.~\ref{eq:continuity}, Eq.~\ref{eq:constitutive}).
However, the definite answer to the questions raised in this paper will have to come from more sophisticated methods.
It is essential to formulate the theory in real frequency and at the same time treat the thermodynamic limit and the vertex corrections. 
The recently developed real-frequency diagrammatic Monte Carlo (RFDiagMC)\cite{Taheridehkordi2019,VucicevicPRB2020,Taheridehkordi2020,Taheridehkordi2020b,VucicevicPRR2021} is a clear candidate, and our Kubo bubble and second-order self-energy theory is the first step in this approach.
Very recently, RFDiagMC was used to calculate the charge-charge correlation function in a slightly different model, at very weak coupling\cite{McNiven2022}. 
Pushing RFDiagMC to stronger coupling\cite{Wu2017,Simkovic2019,Rossi2020,VucicevicPRR2021,Kim2021} and higher resolution necessary to investigate the hydrodynamic behavior in the regime relevant for cold-atom experiments is a difficult task that we leave for future work. 

Finally, it is important to view our $T$-linear-resistivity result in the $U\rightarrow 0$ limit in light of the very recent work by Xu et al. \onlinecite{Xu2022}. In this work, a quantum critical line is observed to pass through $U=0$, $\langle n_\sigma \rangle = 0.5$, separating two distinct ordered phases in the ground-state $(\delta,U)$-phase-diagram of the Hubbard model. This is in line with our observation that the charge-charge and spin-spin susceptibilities diverge at $\mathbf{q}=(\pi,\pi)$ as $T\rightarrow 0$ at $U=0$, $\mu=0$ (See Appendix~\ref{app:chic}). The hypothesis considered in many works\cite{Cha2020} is that linear resistivity is expected above quantum critical points. The linear resistivity that we observe may, indeed, be intimately linked to instability towards order, i.e. a degeneracy of the ground state at $U\rightarrow 0$ at half-filling. The van Hove singularity at the Fermi level perhaps does not play the essential role here, in contrast to the conclusions in Ref.\onlinecite{Herman2019}. Whether resistivity remains linear all the way down to zero temperature when Fermi level is at a van Hove singularity in the density of states, regardless of any ordering instabilities, is currently unclear.

\section{Conclusions}
\label{sec:conclusions}
We have studied charge fluctuations and transport in the Hubbard model.
In the non-interacting limit, charge fluctuations are anisotropic, and can have multiple linear modes at long wavelengths.
Near the empty limit, the charge spectral function resembles that of the electron. 
At weak coupling, the self-energy presents several peculiar behaviors, including an abundance of kinks in the frequency-dependence. 
At low temperature we generally find two peaks in $\mathrm{Im}\Sigma_{\mathbf{k}}(\omega)$ at any $\mathbf{k}$.
At half-filling and $T\rightarrow 0$, we find $\mathrm{Im}\Sigma_{\mathbf{k}=(0,\pi)}(\omega)\sim|\omega|^{4/5}$ and, along the Fermi surface, $\mathrm{Im}\Sigma_{\mathbf{k}=(0,\pi)\leftrightarrow(\pi,0)}(\omega)\sim|\omega|$, the latter being in agreement with previous work\cite{Rohe2020}. 
As temperature is raised, a sharp peak in $\mathrm{Im}\Sigma_{\mathbf{k}}(\omega)$ rises at low frequency, splitting the quasiparticle peak in $\mathrm{Im}G_\mathbf{k}(\omega)$ at around $\mathbf{k}=(\pi,\pi)$.
At high temperature, we find that self-energy has a single peak as a function of frequency at around $\omega=\varepsilon_\mathbf{k}$, and it is not smooth.
We observe that the dc resistivity is linear at half-filling and at high temperature, in agreement with recent findings\cite{Kiely2021}.
Surprisingly, non-local self-energy components are found to have little effect on dc resistivity.
Precisely at the band insulator transition, our bubble appoximation predicts a finite resistivity at $T\rightarrow 0$, coming as a consequence of perfect cancellation of the reducing velocity and scattering rate, both scaling as $T$. 
We observe that the hydrodynamic parameters (diffusion constant and momentum relaxation rate) are roughly inversely proportional, in the bubble approximation at weak coupling, as well as in the numerically exact FTLM result at strong coupling. 
Their product appears to be $D\Gamma \approx 2t^2$, which coincides with one term in the microscopic equation of motion for the current, indicating that other terms might play less of a role.
This supports the hydrodynamic theory, for which we show that it must satisfy $D\Gamma=2t^2$ at weak coupling and high-temperature.
Finally, we propose a modified hydrodynamic law that has correct behavior at high-frequency, and find that it is consistent with both the numerically exact FTLM and the numerically exact CTINT.
Our results provide some evidence that the hydrodynamic theory is correct, but further work with better methods is needed to fully resolve this issue.
%

\begin{acknowledgments}
We acknowledge useful discussions with (in no particular order): Mihailo \v Cubrovi\' c,  Waseem Bakr, Rok \v Zitko, Jernej Mravlje, Jure Kokalj and Antoine Georges.
We acknowledge contributions from Pavle Stipsi\' c in the early stages of the work.
The FTLM $U=1.25$ $\rho_\mathrm{dc}(T)$ data was provided by Martin Ulaga and Jure Kokalj.
The DMFT data was provided by Jernej Mravlje, using the NRG code by Rok \v Zitko.
Rok \v Zitko communicated to us the Eq.\ref{eq:EOM_in_z} and its derivation, and contributed to the derivation of Eq.\ref{eq:chi_vs_Lambda_lattice}.
The CTINT method was implemented using the TRIQS library\cite{Parcollet2015}.
Computations were performed on the PARADOX supercomputing facility
(Scientific Computing Laboratory, Center for the Study of Complex
Systems, Institute of Physics Belgrade). J.~V. and S.~P. acknowledge funding
provided by the Institute of Physics Belgrade, through the grant by
the Ministry of Education, Science, and Technological Development of
the Republic of Serbia, as well as by the Science Fund of the Republic
of Serbia, under the Key2SM project (PROMIS program, Grant No. 6066160).
J.~V. acknowledges funding by the European Research Council, grant ERC-2022-StG: 101076100.
S.~P. acknowledges funding by the Deutsche Gesellschaft für Internationale Zusammenarbeit (GIZ)
on behalf of the German Federal Ministry for Economic Cooperation and Development (BMZ), within the Support for Returning Experts program.
This work was granted access to the HPC resources of TGCC and IDRIS under the allocation A0110510609 attributed by GENCI (Grand Equipement National de Calcul Intensif).
\end{acknowledgments}

\appendix

\section{Continuity equation on the lattice}
\label{app:continuity}

We start by noting the general expression for the time-derivative of a bosonic operator $\cal O$ in the Heisenberg picture
\begin{eqnarray} \nonumber
 \partial_t {\cal O}(t)  &=& \partial_t e^{itH} {\cal O} e^{-itH} \\
 &=&  i [H, {\cal O}  ] .
\end{eqnarray}

We will also need the general expression for commutators of the following general form, with $a^\dagger,b,c^\dagger,d$ fermionic creation/annihilation operators
\begin{eqnarray}
 [a^\dagger b, c^\dagger d] &=& 
    \delta_{bc} a^\dagger d - \delta_{ad} c^\dagger b.
\end{eqnarray}
This can be proven simply by using $ [AB,C] = A[B,C]+[A,C]B $,
$[A,BC] = B[A,C]+[A,B]C$ and therefore 
$ [AB,CD] = A(C[B,D]+[B,C]D) + (C[A,D]+[A,C]D)B $.

Using these we can then show
\begin{eqnarray} \nonumber
 &&\partial_t n_\mathbf{r}(t) \\ \nonumber
 && =  i [H_\mathrm{kin}, n_\mathbf{r} ] \\ \nonumber
 && =  -it \left[
                \sum_{\sigma',\mathbf{r}',s\in\{1,-1\},\eta\in\{x,y\}} c^\dagger_{\sigma',\mathbf{r}'} c_{\sigma',\mathbf{r}'+s\mathbf{e}_\eta} ,\;
                \sum_\sigma c^\dagger_{\sigma,\mathbf{r}} c_{\sigma,\mathbf{r}} 
         \right] \\ \nonumber
 && =  -it \sum_{\sigma,s\in\{1,-1\},\eta\in\{x,y\}}  \\ \nonumber
 && \;\;\;\times \Big(  
         \left[
                c^\dagger_{\sigma,\mathbf{r}} c_{\sigma,\mathbf{r}+s\mathbf{e}_\eta} ,\;
                c^\dagger_{\sigma,\mathbf{r}} c_{\sigma,\mathbf{r}} 
         \right]    
         +
         \left[
                c^\dagger_{\sigma,\mathbf{r}-s\mathbf{e}_\eta} c_{\sigma,\mathbf{r}} ,\;
                c^\dagger_{\sigma,\mathbf{r}} c_{\sigma,\mathbf{r}} 
         \right]
 \Big)  \\ \nonumber 
 && =  -it \sum_{\sigma,s\in\{1,-1\},\eta\in\{x,y\}}  
 \left(  
         - c^\dagger_{\sigma,\mathbf{r}}c_{\sigma,\mathbf{r}+s\mathbf{e}_\eta}
         + c^\dagger_{\sigma,\mathbf{r}-s\mathbf{e}_\eta} c_{\sigma,\mathbf{r}}
 \right) \\ 
 && =  -\sum_{\eta\in\{x,y\}} (j^\eta_\mathbf{r} -  j^\eta_{\mathbf{r}-\mathbf{e}_\eta})
\end{eqnarray}
with the definition $j^\eta_\mathbf{r} = it\sum_\sigma( c^\dagger_{\sigma,\mathbf{r}+\mathbf{e}_\eta} c_{\sigma,\mathbf{r}} - c^\dagger_{\sigma,\mathbf{r}}c_{\sigma,\mathbf{r}+\mathbf{e}_\eta})$. The expression $\sum_{\eta\in\{x,y\}} (j^\eta_\mathbf{r} -  j^\eta_{\mathbf{r}-\mathbf{e}_\eta})$ is the lattice-version of the divergence of current.

We can express the continuity equation in momentum space by Fourier transform of both sides
\begin{eqnarray}
\partial_t n_\mathbf{q} &=&  i[H,n_\mathbf{q}] = -\sum_{\eta\in\{x,y\}} (1-e^{iq_\eta})j^\eta_\mathbf{q}
\end{eqnarray}
where $n_\mathbf{q}=\sum_\mathbf{r} e^{i\mathbf{q}\cdot\mathbf{r}} n_\mathbf{r} = \sum_\mathbf{k} c^\dagger_{\mathbf{k}+\mathbf{q}} c_{\mathbf{k}}$. Notice that we distinguish between $n_\mathbf{k}=c^\dagger_\mathbf{k} c_\mathbf{k}$ and $n_\mathbf{q}$ solely by the choice of the symbol in the subscript. 

\section{Connection between the charge-charge and current-current correlation functions}
\label{app:chi_vs_Lambda}

Here we make use of the general equation of motion
\begin{eqnarray} \label{eq:EOM_in_z}
 &&z^2\langle\langle A; B \rangle \rangle_z \\ \nonumber
 &&= -z \langle [A,B] \rangle - \langle [[A,H],B] \rangle + \langle \langle [A,H];[H,B] \rangle\rangle_z
\end{eqnarray} 
where we denote with $\langle\langle A; B \rangle \rangle_z$ the correlator of operators $A$ and $B$ as a function of complex frequency $z$.
The full derivation of Eq.\ref{eq:EOM_in_z} is given in Appendix~\ref{app:eom}.

If we replace $A=n_\mathbf{q}$ and $B=n_{-\mathbf{q}}$, and using $[A,H]=-[H,A]$ and $1/i=-i$ we get:
\begin{eqnarray}
 &&z^2\langle\langle n_\mathbf{q}; n_{-\mathbf{q}} \rangle \rangle_z  \\ \nonumber
 &&=-z \langle [n_\mathbf{q},n_{-\mathbf{q}}] \rangle + i\sum_{\eta\in\{x,y\}} (1-e^{iq_\eta}) \langle [j^\eta_\mathbf{q}, n_{-\mathbf{q}}] \rangle \\ \nonumber
 &&\;\;\; + \sum_{\eta,\eta'\in\{x,y\}} (1-e^{iq_\eta}-e^{-iq_{\eta'}}+e^{i(q_\eta-q_{\eta'})}) \langle \langle j^\eta_\mathbf{q}; j^{\eta'}_{-\mathbf{q}} \rangle\rangle_z
\end{eqnarray}
\vfill\null
\pagebreak
Let's work out the two commutators:
\begin{eqnarray} \nonumber
 &&[n_\mathbf{q},n_{-\mathbf{q}}]\\ \nonumber
 &&=\sum_\sigma[\sum_\mathbf{k} c^\dagger_{\sigma,\mathbf{k}+\mathbf{q}} c_{\sigma,\mathbf{k}}, \sum_{\mathbf{k}'} c^\dagger_{\sigma,\mathbf{k}'} c_{\sigma,\mathbf{k}'+\mathbf{q} } ]\\ \nonumber
 &&=\sum_\sigma\sum_\mathbf{k}[ c^\dagger_{\sigma,\mathbf{k}+\mathbf{q}} c_{\sigma,\mathbf{k}}, c^\dagger_{\sigma,\mathbf{k}} c_{\sigma,\mathbf{k}+\mathbf{q} } ] \\ \nonumber
 &&=\sum_\sigma\sum_\mathbf{k} (n_{\sigma,\mathbf{k}+\mathbf{q}} - n_{\sigma,\mathbf{k}}) \\ \nonumber
 &&=\sum_\sigma (n_{\sigma,\mathbf{r}=0} - n_{\sigma,\mathbf{r}=0})\\ 
 &&=0 .
\end{eqnarray}
Therefore, the first term drops out. 

For the second term one gets
\begin{eqnarray} \nonumber
 &&[j^\eta_\mathbf{q}, n_{-\mathbf{q}}] \\ \nonumber
 &&=[\sum_\mathbf{k} v^\eta_{\mathbf{k},\mathbf{q}} c^\dagger_{\sigma,\mathbf{k}+\mathbf{q}} c_{\sigma,\mathbf{k}}, \sum_{\mathbf{k}'} c^\dagger_{\sigma,\mathbf{k}'} c_{\sigma,\mathbf{k}'+\mathbf{q} } ] \\ 
 &&=\sum_\sigma\sum_\mathbf{k} v^\eta_{\mathbf{k},\mathbf{q}} (n_{\sigma,\mathbf{k}+\mathbf{q}} - n_{\sigma,\mathbf{k}})
\end{eqnarray}
where $v^\eta_{\mathbf{k}\mathbf{q}} = it(e^{-i(k_\eta+q_\eta)}-e^{ik_\eta})$.
The overall prefactor is purely real for each $\mathbf{k}$:
\begin{eqnarray} \nonumber
 && i^2t(1-e^{iq_\eta})(e^{-i({k_\eta+q_\eta})}-e^{ik_\eta}) \\ \nonumber
 &&= -t(e^{-i(k_\eta+q_\eta)}-e^{ik_\eta}-e^{-ik_\eta}+e^{i(k_\eta+q_\eta)} ) \\ \nonumber
 &&= -2t(\cos(k_\eta+q_\eta)-\cos k_\eta )\\
 &&\equiv \Phi^\eta_{\mathbf{k},\mathbf{q}}
\end{eqnarray}
This leads us to the fully general expression Eq.~\ref{eq:chi_vs_Lambda_lattice}, and here we write it separately for the real and imaginary parts.
\begin{eqnarray} \label{eq:rechi_vs_reLambda_lattice}
 &&z^2 \mathrm{Re} \chi_\mathbf{q}(z) \\ \nonumber
 &&= \sum_{\eta=\{x,y\}}\sum_\mathbf{k}\Phi^\eta_{\mathbf{k},\mathbf{q}} (\langle n_{\mathbf{k}+\mathbf{q}}\rangle  - \langle n_{\mathbf{k}}\rangle) \\ \nonumber
 &&\;\;\; + \mathrm{Re} \sum_{\eta,\eta'\in\{x,y\}}\Big(1-e^{iq_\eta}-e^{-iq_{\eta'}}+e^{i(q_\eta-q_{\eta'})}\Big) \Lambda^{\eta,\eta'}_\mathbf{q}(z) \\ 
 \label{eq:imchi_vs_imLambda_lattice}
 &&z^2 \mathrm{Im} \chi_\mathbf{q}(z) \\ \nonumber
 &&= \mathrm{Im} \sum_{\eta,\eta'\in\{x,y\}} \Big(1-e^{iq_\eta}-e^{-iq_{\eta'}}+e^{i(q_\eta-q_{\eta'})}\Big) \Lambda^{\eta,\eta'}_\mathbf{q}(z) 
\end{eqnarray}
Notice that the constant shift in Eq.\ref{eq:rechi_vs_reLambda_lattice} is crucial to allow that both $\chi$ and $\Lambda$ scale as $1/\nu^2$ at high Matsubara frequency, which is expected on grounds of symmetry of these correlators in imaginary time. At large $\nu$ we get for the real part $\nu^2\frac{\mathrm{const}}{\nu^2} = \mathrm{const} + 1/\nu^2$ which reduces to $\mathrm{const}=\mathrm{const}$ as $\nu$ goes to infinity. This expression also reveals the high-frequency scaling which must hold in general
\begin{equation}
 \mathrm{Re}\chi_\mathbf{q}(i\nu\rightarrow i\infty) = -\frac{1}{\nu^2}\sum_{\eta=\{x,y\}}\sum_\mathbf{k}\Phi^\eta_{\mathbf{k},\mathbf{q}} (\langle n_{\mathbf{k}+\mathbf{q}}\rangle  - \langle n_{\mathbf{k}}\rangle)
\end{equation}

\begin{widetext}
\section{The constitutive equation}
\label{app:constitutive}

In the following we derive the time-derivative of current operator in Heisenberg picture. The derivation boils down to working out the following commutators
\begin{eqnarray}
 \partial_t j^\eta_\mathbf{r} 
 =  i [H, j^\eta_\mathbf{r} ] 
 =  i ( [ H_\mathrm{kin}, j^\eta_\mathbf{r} ] + [ H_\mathrm{int}, j^\eta_\mathbf{r} ] + [ H_\mathrm{chem}, j^\eta_\mathbf{r} ] ).
\end{eqnarray}
The commutator with kinetic energy reads
\begin{eqnarray} \nonumber
 && i [ H_\mathrm{kin}, j^\eta_\mathbf{r} ] \\ \nonumber
 &&=  i  \left[ -t\sum_{\mathbf{r}',\sigma',s\in\{1,-1\},\eta'\in\{x,y\}} c^\dagger_{\sigma',\mathbf{r}'} c_{\sigma',\mathbf{r}'+s\mathbf{e}_{\eta'}}, \;\;\;it \sum_\sigma(c^\dagger_{\sigma,\mathbf{r}+\mathbf{e}_\eta} c_{\sigma,\mathbf{r}} - c^\dagger_{\sigma,\mathbf{r}}c_{\sigma,\mathbf{r}+\mathbf{e}_\eta}  ) \right] \\ \nonumber
&&=  -t^2  \sum_{\sigma} \Bigg\{
      [ c^\dagger_{\sigma,\mathbf{r}+\mathbf{e}_\eta} c_{\sigma,\mathbf{r}} , c^\dagger_{\sigma,\mathbf{r}} c_{\sigma,\mathbf{r}+\mathbf{e}_\eta} ]
      -[ c^\dagger_{\sigma,\mathbf{r}} c_{\sigma,\mathbf{r}+\mathbf{e}_\eta} , c^\dagger_{\sigma,\mathbf{r}+\mathbf{e}_\eta} c_{\sigma,\mathbf{r}} ]  \\ \nonumber      
&&\;\;\;\;\;\;\;\;\;\;\;\;\;\;\;\;\;+\sum_{\mathbf{u}\in\{-\mathbf{e}_{\eta},\mathbf{e}_{\bar\eta},-\mathbf{e}_{\bar\eta}\}}
    \Bigg(
        [ c^\dagger_{\sigma,\mathbf{r}+\mathbf{u}} c_{\sigma,\mathbf{r}} , c^\dagger_{\sigma,\mathbf{r}} c_{\sigma,\mathbf{r}+\mathbf{e}_\eta} ]
    +   [ c^\dagger_{\sigma,\mathbf{r}+\mathbf{e}_\eta} c_{\sigma,\mathbf{r}+\mathbf{e}_\eta-\mathbf{u}} , c^\dagger_{\sigma,\mathbf{r}} c_{\sigma,\mathbf{r}+\mathbf{e}_\eta} ]
    \\ \nonumber
&&\;\;\;\;\;\;\;\;\;\;\;\;\;\;\;\;\;\;\;\;\;\;\;\;\;\;\;\;\;\;\;\;\;\;\;\;\;
    -   [ c^\dagger_{\sigma,\mathbf{r}} c_{\sigma,\mathbf{r}+\mathbf{u}}, c^\dagger_{\sigma,\mathbf{r}+\mathbf{e}_\eta} c_{\sigma,\mathbf{r}}]
    -   [ c^\dagger_{\sigma,\mathbf{r}+\mathbf{e}_\eta-\mathbf{u}} c_{\sigma,\mathbf{r}+\mathbf{e}_\eta}, c^\dagger_{\sigma,\mathbf{r}+\mathbf{e}_\eta} c_{\sigma,\mathbf{r}}]
    \Bigg)\Bigg\}  \\
&&=  -t^2  \sum_{\sigma} 
    \Bigg \{
         2 c^\dagger_{\sigma,\mathbf{r}+\mathbf{e}_\eta} c_{\sigma,\mathbf{r}+\mathbf{e}_\eta}
      -  2 c^\dagger_{\sigma,\mathbf{r}} c_{\sigma,\mathbf{r}} 
    +\sum_{\mathbf{u}\in\{-\mathbf{e}_{\eta},\mathbf{e}_{\bar\eta},-\mathbf{e}_{\bar\eta}\}}  \Bigg(     
         c^\dagger_{\sigma,\mathbf{r}+\mathbf{u}} c_{\sigma,\mathbf{r}+\mathbf{e}_\eta} 
    -     c^\dagger_{\sigma,\mathbf{r}}  c_{\sigma,\mathbf{r}+\mathbf{e}_\eta-\mathbf{u}}     
    + \mathrm{H.c} \Bigg) \Bigg\}.      
\end{eqnarray}
The first two terms comprise the lattice version of the gradient of charge in the direction of current. The other terms are longer range hoppings.

The commutator of the current with the total number of particles has to be zero, and we leave out the explicit derivation of $[ H_\mathrm{chem}, j^\eta_\mathbf{r} ]$.
The commutator with the local Hubbard interaction, on the other hand, is non-trivial:
\begin{eqnarray} \nonumber
 && i [ H_\mathrm{int}, j^\eta_\mathbf{r} ] \\ \nonumber
 &&= i \Bigg[ U\sum_{\mathbf{r}'} c^\dagger_{\up,\mathbf{r}'}c_{\up,\mathbf{r}'}c^\dagger_{\dn,\mathbf{r}'}c_{\dn,\mathbf{r}'}, 
 \;\;\;\;-it \sum_\sigma(c^\dagger_{\sigma,\mathbf{r}}c_{\sigma,\mathbf{r}+\mathbf{e}_\eta} - c^\dagger_{\sigma,\mathbf{r}+\mathbf{e}_\eta} c_{\sigma,\mathbf{r}} ) \Bigg]  \\ \nonumber
 &&= t U\sum_\sigma \Bigg\{ 
     n_{\bar\sigma,\mathbf{r}}\Bigg( 
             [ c^\dagger_{\sigma,\mathbf{r}}c_{\sigma,\mathbf{r}}, c^\dagger_{\sigma,\mathbf{r}}c_{\sigma,\mathbf{r}+\mathbf{e}_\eta}]             
             - [ c^\dagger_{\sigma,\mathbf{r}}c_{\sigma,\mathbf{r}}, c^\dagger_{\sigma,\mathbf{r}+\mathbf{e}_\eta} c_{\sigma,\mathbf{r}}  ]
     \Bigg) \\ \nonumber
 &&\;\;\;\;\;\;\;\;\;\;\;\;\;\;\;\;     +
     n_{\bar\sigma,\mathbf{r}+\mathbf{e}_\eta}\Bigg(
             [ c^\dagger_{\sigma,\mathbf{r}+\mathbf{e}_\eta}c_{\sigma,\mathbf{r}+\mathbf{e}_\eta}, c^\dagger_{\sigma,\mathbf{r}}c_{\sigma,\mathbf{r}+\mathbf{e}_\eta}]
             - [ c^\dagger_{\sigma,\mathbf{r}+\mathbf{e}_\eta}c_{\sigma,\mathbf{r}+\mathbf{e}_\eta}, c^\dagger_{\sigma,\mathbf{r}+\mathbf{e}_\eta} c_{\sigma,\mathbf{r}}  ]             
     \Bigg)
     \Bigg\} \\ \nonumber
 &&= t U\sum_\sigma \Bigg\{ 
     n_{\bar\sigma,\mathbf{r}}\big( 
              c^\dagger_{\sigma,\mathbf{r}} c_{\sigma,\mathbf{r}+\mathbf{e}_\eta}             
             +   c^\dagger_{\sigma,\mathbf{r}+\mathbf{e}_\eta} c_{\sigma,\mathbf{r}} 
     \big) \\ \nonumber
 &&\;\;\;\;\;\;\;\;\;\;\;\;\;\;\;\;     -
     n_{\bar\sigma,\mathbf{r}+\mathbf{e}_\eta}\big(
              c^\dagger_{\sigma,\mathbf{r}} c_{\sigma,\mathbf{r}+\mathbf{e}_\eta} 
             +  c^\dagger_{\sigma,\mathbf{r}+\mathbf{e}_\eta} c_{\sigma,\mathbf{r}}  
     \big)
     \Bigg\}   \\
 &&= -t U\sum_\sigma
     (n_{\bar\sigma,\mathbf{r}+\mathbf{e}_\eta}-n_{\bar\sigma,\mathbf{r}})\big( 
              c^\dagger_{\sigma,\mathbf{r}} c_{\sigma,\mathbf{r}+\mathbf{e}_\eta}             
             +   c^\dagger_{\sigma,\mathbf{r}+\mathbf{e}_\eta} c_{\sigma,\mathbf{r}} 
     \big) 
\end{eqnarray} 
The terms we get are all assisted hopping terms.

The final expression reads
\begin{eqnarray}
\partial_t j^\eta_\mathbf{r}
 &&=  -t^2  \sum_{\sigma} 
    \Bigg \{
         2 n_{\sigma,\mathbf{r}+\mathbf{e}_\eta}
      -  2 n_{\sigma,\mathbf{r}}
    +\sum_{\mathbf{u}\in\{-\mathbf{e}_{\eta},\mathbf{e}_{\bar\eta},-\mathbf{e}_{\bar\eta}\}}  \Bigg(     
         c^\dagger_{\sigma,\mathbf{r}+\mathbf{u}} c_{\sigma,\mathbf{r}+\mathbf{e}_\eta} 
    -     c^\dagger_{\sigma,\mathbf{r}}  c_{\sigma,\mathbf{r}+\mathbf{e}_\eta-\mathbf{u}}     
    + \mathrm{H.c} \Bigg) \Bigg\}   \\ \nonumber
 &&\;\;\;\;\;
 -t U\sum_\sigma
     (n_{\bar\sigma,\mathbf{r}+\mathbf{e}_\eta}-n_{\bar\sigma,\mathbf{r}})\big( 
              c^\dagger_{\sigma,\mathbf{r}} c_{\sigma,\mathbf{r}+\mathbf{e}_\eta}             
             +   c^\dagger_{\sigma,\mathbf{r}+\mathbf{e}_\eta} c_{\sigma,\mathbf{r}} 
     \big).     
\end{eqnarray}
A straightforward Fourier transformation of both sides yields
\begin{eqnarray} \label{eq:apriori_constitutive}
 &&\partial_t j^\eta_\mathbf{q} = -t^2 \sum_\sigma \Bigg\{ 2(e^{iq_\eta}-1) n_{\sigma,\mathbf{q}}
 + \sum_{\mathbf{u}\in\{-\mathbf{e}_{\eta},\mathbf{e}_{\bar\eta},-\mathbf{e}_{\bar\eta}\}} \Bigg(
   (e^{-i\mathbf{q}\cdot\mathbf{u}}-1) \sum_{\mathbf{k}} e^{i\mathbf{k}\cdot(\mathbf{e}_\eta-\mathbf{u})} 
    c^\dagger_{\sigma,\mathbf{k}+\mathbf{q}} c_{\sigma,\mathbf{k}} 
 + \mathrm{H.c.} \Bigg)  \Bigg\} \\ \nonumber
 &&\;\;\;\;\;\;\;\;\;\;\;\;+t U \sum_\sigma
     \sum_{\mathbf{k},\mathbf{q}'} \Big(e^{-ik_\eta}+e^{i(k_\eta+q_\eta-q'_\eta)}\Big)\Big(1-e^{iq'_\eta}\Big)n_{\bar\sigma,\mathbf{q}'}c^\dagger_{\sigma,\mathbf{k}+\mathbf{q}-\mathbf{q}'} c_{\sigma,\mathbf{k}} .
\end{eqnarray}
It is interesting to consider the limit $\mathbf{q}=0$
\begin{eqnarray} \nonumber
 \partial_t j^\eta_{\mathbf{q}=0} &=& t U \sum_\sigma
     \sum_{\mathbf{k},\mathbf{q}'} \Big(e^{-ik_\eta}+e^{i(k_\eta-q'_\eta)}\Big)\Big(1-e^{iq'_\eta}\Big)n_{\bar\sigma,\mathbf{q}'}c^\dagger_{\sigma,\mathbf{k}-\mathbf{q}'} c_{\sigma,\mathbf{k}} \\
 && =iU \sum_\sigma
     \sum_{\mathbf{k},\mathbf{q}'} \Big(v^\eta_{\mathbf{k}-\mathbf{q}'}-v^\eta_{\mathbf{k}}\Big)n_{\bar\sigma,\mathbf{q}'}c^\dagger_{\sigma,\mathbf{k}-\mathbf{q}'} c_{\sigma,\mathbf{k}}                      
\end{eqnarray}
which clearly shows that scattering events that do not transfer momentum do not contribute to the decay of current - more precisely, only the scattering events that change the velocity of an electron in the direction of the current contribute to the decay of the current.
In principle, this expression can be used to express the optical conductivity via higher-order correlation functions using Eq.\ref{eq:EOM_in_z}. 
\end{widetext}

\section{Equation of motion}
\label{app:eom}

\newcommand{\vc}[1]{{\mathbf{#1}}}
\newcommand{\vck}{\vc{k}}
\newcommand{\braket}[2]{\langle#1|#2\rangle}
\newcommand{\expv}[1]{\langle #1 \rangle}
\newcommand{\corr}[1]{\langle\langle #1 \rangle\rangle}
\newcommand{\bra}[1]{\langle #1 |}
\newcommand{\ket}[1]{| #1 \rangle}
\newcommand{\Tr}{\mathrm{Tr}}
\newcommand{\kor}[1]{\langle\langle #1 \rangle\rangle}
\newcommand{\degg}{^\circ}
\renewcommand{\Im}{\mathrm{Im}\,}
\renewcommand{\Re}{\mathrm{Re}\,}
\newcommand{\dtN}{{\dot N}}
\newcommand{\dtQ}{{\dot Q}}
\newcommand{\beq}[1]{\begin{eqnarray} #1 \end{eqnarray}}
\newcommand{\one}{\mathbf{1}}

\newcommand{\dd}{\mathrm{d}}
\newcommand{\Res}{\mathrm{Res}}
\newcommand{\sign}{\mathrm{sign}}

\newcommand{\Simp}{\mathbf{S}_\mathrm{imp}}
\newcommand{\bfsc}{\mathbf{s}_c}

\newcommand{\Hi}{H_\mathrm{int}}

Here we derive Eq.~\ref{eq:EOM_in_z}, following the procedure from Ref.~\onlinecite{Zitko_unpublished}.

We start with the standard definition of the correlator in real time
\beq{
\corr{A;B}_{t} = i \theta(t) \langle [ A(t), B(0) ]
\rangle,
}
where $A$ and $B$ are bosonic operators, thus we adopt the definition without the minus sign in front.

The equations of motion are obtained by taking time derivatives. The first derivative yields
\begin{eqnarray} \nonumber
\frac{d}{dt} \corr{A;B}_{t} = 
i \theta(t) \langle [ \dot{A}(t), B(0) ] \rangle
+i \delta(t) \langle [ A, B ] \rangle .\\ \label{eq:EOM_first_derivative}
\end{eqnarray}
We now perform the Laplace transform w.r.t. $t$
\beq{ \label{eq:EOM_Laplace_transform}
\corr{A;B}_{z} = \int_{0^+}^\infty dt e^{izt} \corr{A;B}_{t}.
}
Now for $f(t) = \corr{A;B}_t$, we integrate per parts
\begin{eqnarray}
&&\int_{0^+}^\infty dt e^{izt} \frac{d}{dt} f(t) \\ \nonumber
&&=e^{izt}f(t) |^{+\infty}_{0^+} - iz \int dt e^{izt} f(t) \\  \nonumber
&&=-i \langle [A,B] \rangle - iz \corr{A;B}_z. 
\end{eqnarray}
Note that the $\delta$ is not included in the integration domain.
The second term comes directly from the definition of the Laplace transform Eq.~\ref{eq:EOM_Laplace_transform}.
Equating this result with the Laplace transform of the r.h.s. of Eq.~\ref{eq:EOM_first_derivative} we get:
\begin{eqnarray} \label{eq:EOM_usual_form}
-i \langle [A,B] \rangle - iz \corr{A;B}_z = \corr{ i[H,A]; B }_z
\end{eqnarray}
which, up to the factor $i$, is the equation of motion in its usual form.
%

We will want the second derivative to apply to the operator $B$, thus we first perform a time-shift
\beq{
&&\frac{d}{dt} \corr{A;B}_{t} \\ \nonumber
&&=i \theta(t) \langle [ \dot{A}(0), B(-t) ] \rangle
+i \delta(t) \langle [ A, B ] \rangle
}
and only then apply the second time-derivative
\beq{ 
&&\frac{d^2}{dt^2} \corr{A;B}_{t} \\ \nonumber
&&=i \theta(t) \langle [ \dot{A}(0), -\dot{B}(-t) ] \rangle \\ \nonumber
&&\;\;\;+i \delta(t) \langle [ \dot{A}, B ] \rangle
+i \delta'(t) \langle [ A, B ] \rangle.
}
For convenience, we shift back in time
\beq{ \label{eq:EOM_second_derivative}
&&\frac{d^2}{dt^2} \corr{A;B}_{t} \\ \nonumber
&&=-i \theta(t) \langle [ \dot{A}(t), \dot{B}(0) ] \rangle \\ \nonumber
&&\;\;\;+i \delta(t) \langle [ \dot{A}, B ] \rangle
        +i \delta'(t) \langle [ A, B ] \rangle.
}

The integration by parts once more leads to
\begin{eqnarray}
&&\int_{0^+}^\infty dt e^{izt} \frac{d^2}{dt^2} f(t) \\ \nonumber
&&=\frac{d}{dt}\left( e^{izt}\dot{f}(t) \right) |^{+\infty}_{0^+} - iz
\int dt e^{izt} \frac{d}{dt} \dot{f}(t) \\ \nonumber
&&=-i \langle [\dot{A},B] \rangle - iz \left(
-i \langle [A,B] \rangle - iz \corr{A;B}_z
\right) \\ \nonumber
&&=-i \langle [\dot{A},B] \rangle - z \langle [A,B]
\rangle - z^2 \corr{A;B}_z
\end{eqnarray}
where the second and third term come directly from Eq.~\ref{eq:EOM_usual_form}.
Equating this result with the Laplace transform of the RHS of Eq.~\ref{eq:EOM_second_derivative} (there note the minus sign in front of the first term, and note that $ \dot{A}\dot{B} = - [H,A][H,B]$)
\beq{ \nonumber
&&-i \langle [\dot{A},B] \rangle - z \langle [A,B]
\rangle - z^2 \corr{A;B}_z \\
&&= \corr{ [H,A]; [H,B] }_z 
}

We rearange the result
\beq{ 
&&z^2 \corr{A;B}_z =
  \\ \nonumber
&&-i \langle [\dot{A},B] \rangle - z \langle [A,B] \rangle- \corr{ [H,A]; [H,B] }_z 
}
and finally obtain
\beq{ 
&&z^2 \corr{A;B}_z =
  \\ \nonumber
&& -\langle [[A,H],B] \rangle - z \langle [A,B] \rangle+ \corr{ [A,H]; [H,B] }_z 
}
which is the expression used in the derivation in Appendix \ref{app:chi_vs_Lambda}.

\section{Calculation of second-order self-energy}
\label{app:optmal_selfenergy}

The optimal way to compute the second order self-energy Eq.~\ref{eq:Sigma_simple} it is to first evaluate the ``triple density of states'' by a 3D histrogram on a dense energy-grid
\begin{equation}\label{eq:triple_dos}
 \rho_{3,\mathbf{k}}(\varepsilon_1,\varepsilon_2,\varepsilon_3) = \frac{1}{N^2 \Delta\omega^3}\sum_{\mathbf{k'},\mathbf{q}} \delta_{\varepsilon_1,\varepsilon_{\mathbf{k}-\mathbf{q}}}
  \delta_{\varepsilon_2,\varepsilon_{\mathbf{k}'+\mathbf{q}}}
  \delta_{\varepsilon_3,\varepsilon_{\mathbf{k}'}}
\end{equation}
where $\Delta\omega$ is the step in the energy grid.
This calculation only needs to be performed once, for $\mu=0$, and it does not depend on $T$. Then for a given $(\mu,T)$ we accumulate pole amplitudes for the self-energy on the same energy grid as
\begin{eqnarray} \label{eq:Sigma_optimal}
   &&\mathrm{Im}\tilde{\Sigma}_\mathbf{k}(\omega) \\ \nonumber
   &&= -\frac{\pi}{\Delta\omega}\int \mathrm{d}\varepsilon_1 \mathrm{d}\varepsilon_2 \mathrm{d}\varepsilon_3
   \;\rho_{3,\mathbf{k}}(\varepsilon_1,\varepsilon_2,\varepsilon_3)
  \delta_{\omega+\mu-\varepsilon_1-\varepsilon_2+\varepsilon_3}  \\ \nonumber 
    &&\;\;\;\;\;\times \sum_{s=\pm 1}  n_F(s(\varepsilon_1-\mu))  n_F(s(\varepsilon_2-\mu)) n_F(-s(\varepsilon_3-\mu)) .
\end{eqnarray}
The real-part of the self-energy can then be obtained via the standard Kramers-Kronig relation.

In our calculations we consider lattices $L\times L$, up to $L=256$. When calculating histograms, the optimal number of bins is the square root of the number of the data points. Therefore, in Eq.~\ref{eq:triple_dos} the number of energy bins per axis should be equal to $\sqrt{N^2}^{1/3}= L^{2/3}$, but for the sake of numerical simplicity, we take the number of energy bins (per axis) to be $L$. This means that, up to the overhead of evaluating Eq.~\ref{eq:triple_dos} once, we have reduced the complexity of the calculation from $L^4$ (Eq.~\ref{eq:Sigma_simple}) to $L^3$ (Eq.~\ref{eq:Sigma_optimal}), which is a huge speed up.

\begin{figure}[t!]
 \centering{
 \includegraphics[width=3.2in,trim=0cm 0cm 0cm 0cm]{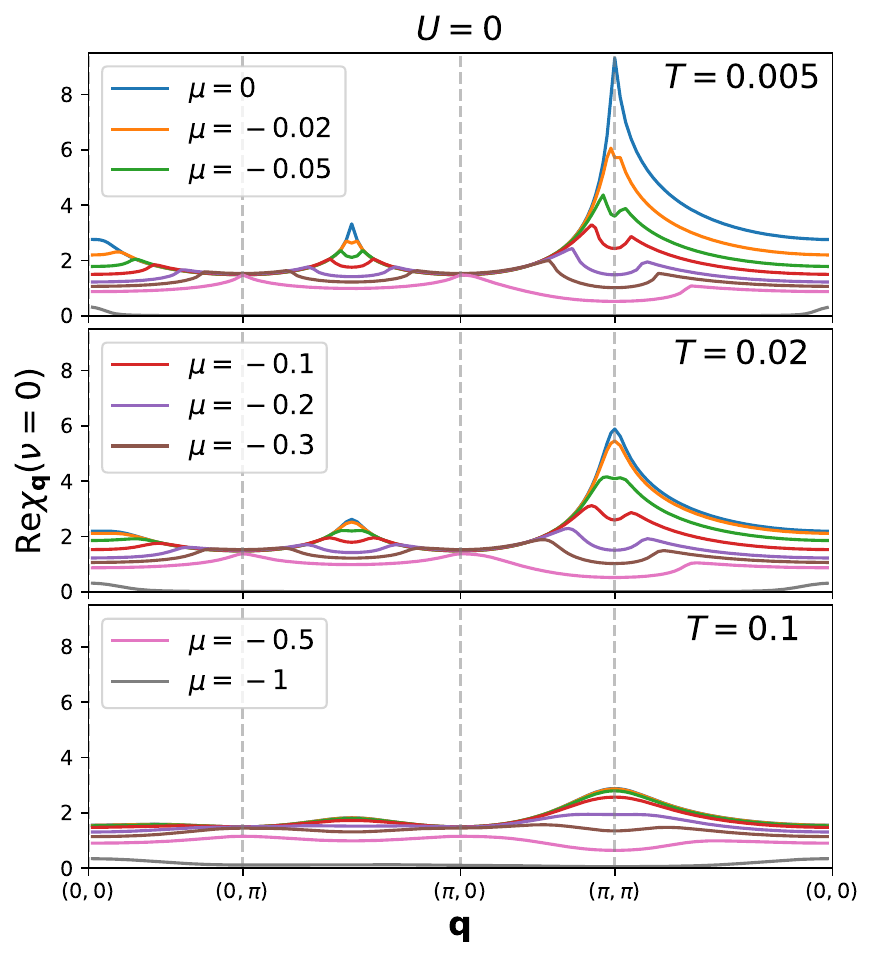}
 }
 \caption{
 Static susceptibility in the non-interacting limit as a function of momentum, at different fillings and temperatures.
}
\label{fig:chic_vs_q}
\end{figure}

\begin{figure}[t!]
 \centering{
 \includegraphics[width=3.2in,trim=0cm 0cm 0cm 0cm]{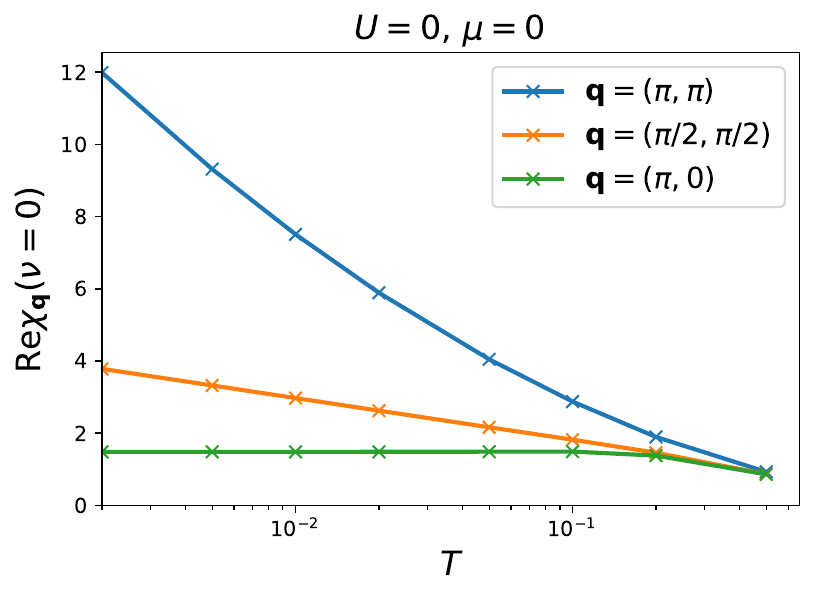}
 }
 \caption{ Static susceptibilities in the non-interacting limit, at half-filling, as a function of temperature. The values at $\mathbf{q}=(\pi,\pi)$ and $\mathbf{q}=(\pi/2,\pi/2)$ appear to diverge as $T\rightarrow 0$. }
\label{fig:chic_vs_T}
\end{figure}

\section{Divergence of static susceptibility at $U=0$, $T\rightarrow 0$}
\label{app:chic}

In this section we show results for the static susceptibility $\mathrm{Re}\chi_\mathbf{q}(\nu=0)$, in the non-interacting limit. The results are obtained by setting $\phi,\varphi=1$, $z=0$ in Eq.~\ref{eq:Qqz_final} and using a $6000\times 6000$ lattice. In Fig.~\ref{fig:chic_vs_q} we observe that there are peaks at $\mathbf{q}=(0,0)$, $\mathbf{q}=(\pi,\pi)$ and $\mathbf{q}=(\pi/2,\pi/2)$, but the highest peak is at $\mathbf{q}=(\pi,\pi)$.
As one dopes away from half-filling, the peaks split. At $\mu=-0.5$, the peaks are at $\mathbf{q}=(0,\pi)$ and the symmetry related $\mathbf{q}=(\pi,0)$. In the nearly empty limit, the only peak is at $\mathbf{q}=(0,0)$.
As expected, the peaks become less pronounced as one increases temperature.
In Fig.~\ref{fig:chic_vs_T} we further observe that the value at $\mathbf{q}=(\pi,\pi)$ and $\mathbf{q}=(\pi/2,\pi/2)$ appears to diverge as $T\rightarrow 0$, indicating competing instabilities towards charge order. It is important to note that in the non-interacting limit, spin and charge susceptibilities are equal. Therefore, any instability towards charge order is accompanied by instability towards spin order. As soon as interaction is turned on, the degeneracy between charge and spin order is lifted.

\bibliography{refs.bib}

\begin{thebibliography}{53}%
\makeatletter
\providecommand \@ifxundefined [1]{%
 \@ifx{#1\undefined}
}%
\providecommand \@ifnum [1]{%
 \ifnum #1\expandafter \@firstoftwo
 \else \expandafter \@secondoftwo
 \fi
}%
\providecommand \@ifx [1]{%
 \ifx #1\expandafter \@firstoftwo
 \else \expandafter \@secondoftwo
 \fi
}%
\providecommand \natexlab [1]{#1}%
\providecommand \enquote  [1]{``#1''}%
\providecommand \bibnamefont  [1]{#1}%
\providecommand \bibfnamefont [1]{#1}%
\providecommand \citenamefont [1]{#1}%
\providecommand \href@noop [0]{\@secondoftwo}%
\providecommand \href [0]{\begingroup \@sanitize@url \@href}%
\providecommand \@href[1]{\@@startlink{#1}\@@href}%
\providecommand \@@href[1]{\endgroup#1\@@endlink}%
\providecommand \@sanitize@url [0]{\catcode `\\12\catcode `\$12\catcode
  `\&12\catcode `\#12\catcode `\^12\catcode `\_12\catcode `\%12\relax}%
\providecommand \@@startlink[1]{}%
\providecommand \@@endlink[0]{}%
\providecommand \url  [0]{\begingroup\@sanitize@url \@url }%
\providecommand \@url [1]{\endgroup\@href {#1}{\urlprefix }}%
\providecommand \urlprefix  [0]{URL }%
\providecommand \Eprint [0]{\href }%
\providecommand \doibase [0]{http://dx.doi.org/}%
\providecommand \selectlanguage [0]{\@gobble}%
\providecommand \bibinfo  [0]{\@secondoftwo}%
\providecommand \bibfield  [0]{\@secondoftwo}%
\providecommand \translation [1]{[#1]}%
\providecommand \BibitemOpen [0]{}%
\providecommand \bibitemStop [0]{}%
\providecommand \bibitemNoStop [0]{.\EOS\space}%
\providecommand \EOS [0]{\spacefactor3000\relax}%
\providecommand \BibitemShut  [1]{\csname bibitem#1\endcsname}%
\let\auto@bib@innerbib\@empty
\bibitem [{\citenamefont {Keimer}\ \emph {et~al.}(2015)\citenamefont {Keimer},
  \citenamefont {Kivelson}, \citenamefont {Norman}, \citenamefont {Uchida},\
  and\ \citenamefont {Zaanen}}]{Keimer2015}%
  \BibitemOpen
  \bibfield  {author} {\bibinfo {author} {\bibfnamefont {B.}~\bibnamefont
  {Keimer}}, \bibinfo {author} {\bibfnamefont {S.~A.}\ \bibnamefont
  {Kivelson}}, \bibinfo {author} {\bibfnamefont {M.~R.}\ \bibnamefont
  {Norman}}, \bibinfo {author} {\bibfnamefont {S.}~\bibnamefont {Uchida}}, \
  and\ \bibinfo {author} {\bibfnamefont {J.}~\bibnamefont {Zaanen}},\ }\href
  {\doibase 10.1038/nature14165} {\bibfield  {journal} {\bibinfo  {journal}
  {Nature}\ }\textbf {\bibinfo {volume} {518}},\ \bibinfo {pages} {179}
  (\bibinfo {year} {2015})}\BibitemShut {NoStop}%
\bibitem [{\citenamefont {Cooper}\ \emph {et~al.}(2009)\citenamefont {Cooper},
  \citenamefont {Wang}, \citenamefont {Vignolle}, \citenamefont {Lipscombe},
  \citenamefont {Hayden}, \citenamefont {Tanabe}, \citenamefont {Adachi},
  \citenamefont {Koike}, \citenamefont {Nohara}, \citenamefont {Takagi},
  \citenamefont {Proust},\ and\ \citenamefont {Hussey}}]{Cooper2009}%
  \BibitemOpen
  \bibfield  {author} {\bibinfo {author} {\bibfnamefont {R.~A.}\ \bibnamefont
  {Cooper}}, \bibinfo {author} {\bibfnamefont {Y.}~\bibnamefont {Wang}},
  \bibinfo {author} {\bibfnamefont {B.}~\bibnamefont {Vignolle}}, \bibinfo
  {author} {\bibfnamefont {O.~J.}\ \bibnamefont {Lipscombe}}, \bibinfo {author}
  {\bibfnamefont {S.~M.}\ \bibnamefont {Hayden}}, \bibinfo {author}
  {\bibfnamefont {Y.}~\bibnamefont {Tanabe}}, \bibinfo {author} {\bibfnamefont
  {T.}~\bibnamefont {Adachi}}, \bibinfo {author} {\bibfnamefont
  {Y.}~\bibnamefont {Koike}}, \bibinfo {author} {\bibfnamefont
  {M.}~\bibnamefont {Nohara}}, \bibinfo {author} {\bibfnamefont
  {H.}~\bibnamefont {Takagi}}, \bibinfo {author} {\bibfnamefont
  {C.}~\bibnamefont {Proust}}, \ and\ \bibinfo {author} {\bibfnamefont {N.~E.}\
  \bibnamefont {Hussey}},\ }\href {\doibase 10.1126/science.1165015} {\bibfield
   {journal} {\bibinfo  {journal} {Science}\ }\textbf {\bibinfo {volume}
  {323}},\ \bibinfo {pages} {603} (\bibinfo {year} {2009})}\BibitemShut
  {NoStop}%
\bibitem [{\citenamefont {Legros}\ \emph {et~al.}(2018)\citenamefont {Legros},
  \citenamefont {Benhabib}, \citenamefont {Tabis}, \citenamefont
  {Lalibert{\'{e}}}, \citenamefont {Dion}, \citenamefont {Lizaire},
  \citenamefont {Vignolle}, \citenamefont {Vignolles}, \citenamefont {Raffy},
  \citenamefont {Li}, \citenamefont {Auban-Senzier}, \citenamefont
  {Doiron-Leyraud}, \citenamefont {Fournier}, \citenamefont {Colson},
  \citenamefont {Taillefer},\ and\ \citenamefont {Proust}}]{Legros2018}%
  \BibitemOpen
  \bibfield  {author} {\bibinfo {author} {\bibfnamefont {A.}~\bibnamefont
  {Legros}}, \bibinfo {author} {\bibfnamefont {S.}~\bibnamefont {Benhabib}},
  \bibinfo {author} {\bibfnamefont {W.}~\bibnamefont {Tabis}}, \bibinfo
  {author} {\bibfnamefont {F.}~\bibnamefont {Lalibert{\'{e}}}}, \bibinfo
  {author} {\bibfnamefont {M.}~\bibnamefont {Dion}}, \bibinfo {author}
  {\bibfnamefont {M.}~\bibnamefont {Lizaire}}, \bibinfo {author} {\bibfnamefont
  {B.}~\bibnamefont {Vignolle}}, \bibinfo {author} {\bibfnamefont
  {D.}~\bibnamefont {Vignolles}}, \bibinfo {author} {\bibfnamefont
  {H.}~\bibnamefont {Raffy}}, \bibinfo {author} {\bibfnamefont {Z.~Z.}\
  \bibnamefont {Li}}, \bibinfo {author} {\bibfnamefont {P.}~\bibnamefont
  {Auban-Senzier}}, \bibinfo {author} {\bibfnamefont {N.}~\bibnamefont
  {Doiron-Leyraud}}, \bibinfo {author} {\bibfnamefont {P.}~\bibnamefont
  {Fournier}}, \bibinfo {author} {\bibfnamefont {D.}~\bibnamefont {Colson}},
  \bibinfo {author} {\bibfnamefont {L.}~\bibnamefont {Taillefer}}, \ and\
  \bibinfo {author} {\bibfnamefont {C.}~\bibnamefont {Proust}},\ }\href
  {\doibase 10.1038/s41567-018-0334-2} {\bibfield  {journal} {\bibinfo
  {journal} {Nat. Phys.}\ }\textbf {\bibinfo {volume} {15}},\ \bibinfo {pages}
  {142} (\bibinfo {year} {2018})}\BibitemShut {NoStop}%
\bibitem [{\citenamefont {Cao}\ \emph {et~al.}(2020)\citenamefont {Cao},
  \citenamefont {Chowdhury}, \citenamefont {Rodan-Legrain}, \citenamefont
  {Rubies-Bigorda}, \citenamefont {Watanabe}, \citenamefont {Taniguchi},
  \citenamefont {Senthil},\ and\ \citenamefont {Jarillo-Herrero}}]{Cao2020}%
  \BibitemOpen
  \bibfield  {author} {\bibinfo {author} {\bibfnamefont {Y.}~\bibnamefont
  {Cao}}, \bibinfo {author} {\bibfnamefont {D.}~\bibnamefont {Chowdhury}},
  \bibinfo {author} {\bibfnamefont {D.}~\bibnamefont {Rodan-Legrain}}, \bibinfo
  {author} {\bibfnamefont {O.}~\bibnamefont {Rubies-Bigorda}}, \bibinfo
  {author} {\bibfnamefont {K.}~\bibnamefont {Watanabe}}, \bibinfo {author}
  {\bibfnamefont {T.}~\bibnamefont {Taniguchi}}, \bibinfo {author}
  {\bibfnamefont {T.}~\bibnamefont {Senthil}}, \ and\ \bibinfo {author}
  {\bibfnamefont {P.}~\bibnamefont {Jarillo-Herrero}},\ }\href {\doibase
  10.1103/physrevlett.124.076801} {\bibfield  {journal} {\bibinfo  {journal}
  {Phys. Rev. Lett}\ }\textbf {\bibinfo {volume} {124}},\ \bibinfo {pages}
  {076801} (\bibinfo {year} {2020})}\BibitemShut {NoStop}%
\bibitem [{\citenamefont {Ayres}\ \emph {et~al.}(2021)\citenamefont {Ayres},
  \citenamefont {Berben}, \citenamefont {{\v{C}}ulo}, \citenamefont {Hsu},
  \citenamefont {van Heumen}, \citenamefont {Huang}, \citenamefont {Zaanen},
  \citenamefont {Kondo}, \citenamefont {Takeuchi}, \citenamefont {Cooper},
  \citenamefont {Putzke}, \citenamefont {Friedemann}, \citenamefont
  {Carrington},\ and\ \citenamefont {Hussey}}]{Ayres2021}%
  \BibitemOpen
  \bibfield  {author} {\bibinfo {author} {\bibfnamefont {J.}~\bibnamefont
  {Ayres}}, \bibinfo {author} {\bibfnamefont {M.}~\bibnamefont {Berben}},
  \bibinfo {author} {\bibfnamefont {M.}~\bibnamefont {{\v{C}}ulo}}, \bibinfo
  {author} {\bibfnamefont {Y.-T.}\ \bibnamefont {Hsu}}, \bibinfo {author}
  {\bibfnamefont {E.}~\bibnamefont {van Heumen}}, \bibinfo {author}
  {\bibfnamefont {Y.}~\bibnamefont {Huang}}, \bibinfo {author} {\bibfnamefont
  {J.}~\bibnamefont {Zaanen}}, \bibinfo {author} {\bibfnamefont
  {T.}~\bibnamefont {Kondo}}, \bibinfo {author} {\bibfnamefont
  {T.}~\bibnamefont {Takeuchi}}, \bibinfo {author} {\bibfnamefont {J.~R.}\
  \bibnamefont {Cooper}}, \bibinfo {author} {\bibfnamefont {C.}~\bibnamefont
  {Putzke}}, \bibinfo {author} {\bibfnamefont {S.}~\bibnamefont {Friedemann}},
  \bibinfo {author} {\bibfnamefont {A.}~\bibnamefont {Carrington}}, \ and\
  \bibinfo {author} {\bibfnamefont {N.~E.}\ \bibnamefont {Hussey}},\ }\href
  {\doibase 10.1038/s41586-021-03622-z} {\bibfield  {journal} {\bibinfo
  {journal} {Nature}\ }\textbf {\bibinfo {volume} {595}},\ \bibinfo {pages}
  {661} (\bibinfo {year} {2021})}\BibitemShut {NoStop}%
\bibitem [{\citenamefont {Grigera}\ \emph {et~al.}(2001)\citenamefont
  {Grigera}, \citenamefont {Perry}, \citenamefont {Schofield}, \citenamefont
  {Chiao}, \citenamefont {Julian}, \citenamefont {Lonzarich}, \citenamefont
  {Ikeda}, \citenamefont {Maeno}, \citenamefont {Millis},\ and\ \citenamefont
  {Mackenzie}}]{Grigera2001}%
  \BibitemOpen
  \bibfield  {author} {\bibinfo {author} {\bibfnamefont {S.~A.}\ \bibnamefont
  {Grigera}}, \bibinfo {author} {\bibfnamefont {R.~S.}\ \bibnamefont {Perry}},
  \bibinfo {author} {\bibfnamefont {A.~J.}\ \bibnamefont {Schofield}}, \bibinfo
  {author} {\bibfnamefont {M.}~\bibnamefont {Chiao}}, \bibinfo {author}
  {\bibfnamefont {S.~R.}\ \bibnamefont {Julian}}, \bibinfo {author}
  {\bibfnamefont {G.~G.}\ \bibnamefont {Lonzarich}}, \bibinfo {author}
  {\bibfnamefont {S.~I.}\ \bibnamefont {Ikeda}}, \bibinfo {author}
  {\bibfnamefont {Y.}~\bibnamefont {Maeno}}, \bibinfo {author} {\bibfnamefont
  {A.~J.}\ \bibnamefont {Millis}}, \ and\ \bibinfo {author} {\bibfnamefont
  {A.~P.}\ \bibnamefont {Mackenzie}},\ }\href {\doibase
  10.1126/science.1063539} {\bibfield  {journal} {\bibinfo  {journal}
  {Science}\ }\textbf {\bibinfo {volume} {294}},\ \bibinfo {pages} {329}
  (\bibinfo {year} {2001})}\BibitemShut {NoStop}%
\bibitem [{\citenamefont {Licciardello}\ \emph {et~al.}(2019)\citenamefont
  {Licciardello}, \citenamefont {Buhot}, \citenamefont {Lu}, \citenamefont
  {Ayres}, \citenamefont {Kasahara}, \citenamefont {Matsuda}, \citenamefont
  {Shibauchi},\ and\ \citenamefont {Hussey}}]{Licciardello2019}%
  \BibitemOpen
  \bibfield  {author} {\bibinfo {author} {\bibfnamefont {S.}~\bibnamefont
  {Licciardello}}, \bibinfo {author} {\bibfnamefont {J.}~\bibnamefont {Buhot}},
  \bibinfo {author} {\bibfnamefont {J.}~\bibnamefont {Lu}}, \bibinfo {author}
  {\bibfnamefont {J.}~\bibnamefont {Ayres}}, \bibinfo {author} {\bibfnamefont
  {S.}~\bibnamefont {Kasahara}}, \bibinfo {author} {\bibfnamefont
  {Y.}~\bibnamefont {Matsuda}}, \bibinfo {author} {\bibfnamefont
  {T.}~\bibnamefont {Shibauchi}}, \ and\ \bibinfo {author} {\bibfnamefont
  {N.~E.}\ \bibnamefont {Hussey}},\ }\href {\doibase 10.1038/s41586-019-0923-y}
  {\bibfield  {journal} {\bibinfo  {journal} {Nature}\ }\textbf {\bibinfo
  {volume} {567}},\ \bibinfo {pages} {213} (\bibinfo {year}
  {2019})}\BibitemShut {NoStop}%
\bibitem [{\citenamefont {Cha}\ \emph {et~al.}(2020)\citenamefont {Cha},
  \citenamefont {Wentzell}, \citenamefont {Parcollet}, \citenamefont
  {Georges},\ and\ \citenamefont {Kim}}]{Cha2020}%
  \BibitemOpen
  \bibfield  {author} {\bibinfo {author} {\bibfnamefont {P.}~\bibnamefont
  {Cha}}, \bibinfo {author} {\bibfnamefont {N.}~\bibnamefont {Wentzell}},
  \bibinfo {author} {\bibfnamefont {O.}~\bibnamefont {Parcollet}}, \bibinfo
  {author} {\bibfnamefont {A.}~\bibnamefont {Georges}}, \ and\ \bibinfo
  {author} {\bibfnamefont {E.-A.}\ \bibnamefont {Kim}},\ }\href {\doibase
  10.1073/pnas.2003179117} {\bibfield  {journal} {\bibinfo  {journal} {PNAS}\
  }\textbf {\bibinfo {volume} {117}},\ \bibinfo {pages} {18341} (\bibinfo
  {year} {2020})}\BibitemShut {NoStop}%
\bibitem [{\citenamefont {Deng}\ \emph {et~al.}(2013)\citenamefont {Deng},
  \citenamefont {Mravlje}, \citenamefont {\ifmmode~\check{Z}\else
  \v{Z}\fi{}itko}, \citenamefont {Ferrero}, \citenamefont {Kotliar},\ and\
  \citenamefont {Georges}}]{Deng2014}%
  \BibitemOpen
  \bibfield  {author} {\bibinfo {author} {\bibfnamefont {X.}~\bibnamefont
  {Deng}}, \bibinfo {author} {\bibfnamefont {J.}~\bibnamefont {Mravlje}},
  \bibinfo {author} {\bibfnamefont {R.}~\bibnamefont {\ifmmode~\check{Z}\else
  \v{Z}\fi{}itko}}, \bibinfo {author} {\bibfnamefont {M.}~\bibnamefont
  {Ferrero}}, \bibinfo {author} {\bibfnamefont {G.}~\bibnamefont {Kotliar}}, \
  and\ \bibinfo {author} {\bibfnamefont {A.}~\bibnamefont {Georges}},\ }\href
  {\doibase 10.1103/PhysRevLett.110.086401} {\bibfield  {journal} {\bibinfo
  {journal} {Phys. Rev. Lett}\ }\textbf {\bibinfo {volume} {110}},\ \bibinfo
  {pages} {086401} (\bibinfo {year} {2013})}\BibitemShut {NoStop}%
\bibitem [{\citenamefont {Vu{\v{c}}i{\v{c}}evi{\'{c}}}\ \emph
  {et~al.}(2015)\citenamefont {Vu{\v{c}}i{\v{c}}evi{\'{c}}}, \citenamefont
  {Tanaskovi{\'{c}}}, \citenamefont {Rozenberg},\ and\ \citenamefont
  {Dobrosavljevi{\'{c}}}}]{VucicevicPRL2015}%
  \BibitemOpen
  \bibfield  {author} {\bibinfo {author} {\bibfnamefont {J.}~\bibnamefont
  {Vu{\v{c}}i{\v{c}}evi{\'{c}}}}, \bibinfo {author} {\bibfnamefont
  {D.}~\bibnamefont {Tanaskovi{\'{c}}}}, \bibinfo {author} {\bibfnamefont
  {M.~J.}\ \bibnamefont {Rozenberg}}, \ and\ \bibinfo {author} {\bibfnamefont
  {V.}~\bibnamefont {Dobrosavljevi{\'{c}}}},\ }\href {\doibase
  10.1103/physrevlett.114.246402} {\bibfield  {journal} {\bibinfo  {journal}
  {Phys. Rev. Lett}\ }\textbf {\bibinfo {volume} {114}},\ \bibinfo {pages}
  {246402} (\bibinfo {year} {2015})}\BibitemShut {NoStop}%
\bibitem [{\citenamefont {Perepelitsky}\ \emph {et~al.}(2016)\citenamefont
  {Perepelitsky}, \citenamefont {Galatas}, \citenamefont {Mravlje},
  \citenamefont {\ifmmode~\check{Z}\else \v{Z}\fi{}itko}, \citenamefont
  {Khatami}, \citenamefont {Shastry},\ and\ \citenamefont
  {Georges}}]{Perepelitsky2016}%
  \BibitemOpen
  \bibfield  {author} {\bibinfo {author} {\bibfnamefont {E.}~\bibnamefont
  {Perepelitsky}}, \bibinfo {author} {\bibfnamefont {A.}~\bibnamefont
  {Galatas}}, \bibinfo {author} {\bibfnamefont {J.}~\bibnamefont {Mravlje}},
  \bibinfo {author} {\bibfnamefont {R.}~\bibnamefont {\ifmmode~\check{Z}\else
  \v{Z}\fi{}itko}}, \bibinfo {author} {\bibfnamefont {E.}~\bibnamefont
  {Khatami}}, \bibinfo {author} {\bibfnamefont {B.~S.}\ \bibnamefont
  {Shastry}}, \ and\ \bibinfo {author} {\bibfnamefont {A.}~\bibnamefont
  {Georges}},\ }\href {\doibase 10.1103/PhysRevB.94.235115} {\bibfield
  {journal} {\bibinfo  {journal} {Phys. Rev. B}\ }\textbf {\bibinfo {volume}
  {94}},\ \bibinfo {pages} {235115} (\bibinfo {year} {2016})}\BibitemShut
  {NoStop}%
\bibitem [{\citenamefont {Kokalj}(2017)}]{Kokalj2017}%
  \BibitemOpen
  \bibfield  {author} {\bibinfo {author} {\bibfnamefont {J.}~\bibnamefont
  {Kokalj}},\ }\href {\doibase 10.1103/PhysRevB.95.041110} {\bibfield
  {journal} {\bibinfo  {journal} {Phys. Rev. B}\ }\textbf {\bibinfo {volume}
  {95}},\ \bibinfo {pages} {041110(R)} (\bibinfo {year} {2017})}\BibitemShut
  {NoStop}%
\bibitem [{\citenamefont {Huang}\ \emph {et~al.}(2019)\citenamefont {Huang},
  \citenamefont {Sheppard}, \citenamefont {Moritz},\ and\ \citenamefont
  {Devereaux}}]{Huang2019}%
  \BibitemOpen
  \bibfield  {author} {\bibinfo {author} {\bibfnamefont {E.~W.}\ \bibnamefont
  {Huang}}, \bibinfo {author} {\bibfnamefont {R.}~\bibnamefont {Sheppard}},
  \bibinfo {author} {\bibfnamefont {B.}~\bibnamefont {Moritz}}, \ and\ \bibinfo
  {author} {\bibfnamefont {T.~P.}\ \bibnamefont {Devereaux}},\ }\href {\doibase
  10.1126/science.aau7063} {\bibfield  {journal} {\bibinfo  {journal}
  {Science}\ }\textbf {\bibinfo {volume} {366}},\ \bibinfo {pages} {987}
  (\bibinfo {year} {2019})}\BibitemShut {NoStop}%
\bibitem [{\citenamefont {Vu\ifmmode \check{c}\else \v{c}\fi{}i\ifmmode
  \check{c}\else \v{c}\fi{}evi\ifmmode~\acute{c}\else \'{c}\fi{}}\ \emph
  {et~al.}(2019)\citenamefont {Vu\ifmmode \check{c}\else \v{c}\fi{}i\ifmmode
  \check{c}\else \v{c}\fi{}evi\ifmmode~\acute{c}\else \'{c}\fi{}},
  \citenamefont {Kokalj}, \citenamefont {\ifmmode~\check{Z}\else
  \v{Z}\fi{}itko}, \citenamefont {Wentzell}, \citenamefont
  {Tanaskovi\ifmmode~\acute{c}\else \'{c}\fi{}},\ and\ \citenamefont
  {Mravlje}}]{VucicevicPRL2019}%
  \BibitemOpen
  \bibfield  {author} {\bibinfo {author} {\bibfnamefont {J.}~\bibnamefont
  {Vu\ifmmode \check{c}\else \v{c}\fi{}i\ifmmode \check{c}\else
  \v{c}\fi{}evi\ifmmode~\acute{c}\else \'{c}\fi{}}}, \bibinfo {author}
  {\bibfnamefont {J.}~\bibnamefont {Kokalj}}, \bibinfo {author} {\bibfnamefont
  {R.}~\bibnamefont {\ifmmode~\check{Z}\else \v{Z}\fi{}itko}}, \bibinfo
  {author} {\bibfnamefont {N.}~\bibnamefont {Wentzell}}, \bibinfo {author}
  {\bibfnamefont {D.}~\bibnamefont {Tanaskovi\ifmmode~\acute{c}\else
  \'{c}\fi{}}}, \ and\ \bibinfo {author} {\bibfnamefont {J.}~\bibnamefont
  {Mravlje}},\ }\href {\doibase 10.1103/PhysRevLett.123.036601} {\bibfield
  {journal} {\bibinfo  {journal} {Phys. Rev. Lett.}\ }\textbf {\bibinfo
  {volume} {123}},\ \bibinfo {pages} {036601} (\bibinfo {year}
  {2019})}\BibitemShut {NoStop}%
\bibitem [{\citenamefont {Vrani\ifmmode~\acute{c}\else \'{c}\fi{}}\ \emph
  {et~al.}(2020)\citenamefont {Vrani\ifmmode~\acute{c}\else \'{c}\fi{}},
  \citenamefont {Vu\ifmmode \check{c}\else \v{c}\fi{}i\ifmmode \check{c}\else
  \v{c}\fi{}evi\ifmmode~\acute{c}\else \'{c}\fi{}}, \citenamefont {Kokalj},
  \citenamefont {Skolimowski}, \citenamefont {\ifmmode~\check{Z}\else
  \v{Z}\fi{}itko}, \citenamefont {Mravlje},\ and\ \citenamefont
  {Tanaskovi\ifmmode~\acute{c}\else \'{c}\fi{}}}]{VranicPRB2020}%
  \BibitemOpen
  \bibfield  {author} {\bibinfo {author} {\bibfnamefont {A.}~\bibnamefont
  {Vrani\ifmmode~\acute{c}\else \'{c}\fi{}}}, \bibinfo {author} {\bibfnamefont
  {J.}~\bibnamefont {Vu\ifmmode \check{c}\else \v{c}\fi{}i\ifmmode
  \check{c}\else \v{c}\fi{}evi\ifmmode~\acute{c}\else \'{c}\fi{}}}, \bibinfo
  {author} {\bibfnamefont {J.}~\bibnamefont {Kokalj}}, \bibinfo {author}
  {\bibfnamefont {J.}~\bibnamefont {Skolimowski}}, \bibinfo {author}
  {\bibfnamefont {R.}~\bibnamefont {\ifmmode~\check{Z}\else \v{Z}\fi{}itko}},
  \bibinfo {author} {\bibfnamefont {J.}~\bibnamefont {Mravlje}}, \ and\
  \bibinfo {author} {\bibfnamefont {D.}~\bibnamefont
  {Tanaskovi\ifmmode~\acute{c}\else \'{c}\fi{}}},\ }\href {\doibase
  10.1103/PhysRevB.102.115142} {\bibfield  {journal} {\bibinfo  {journal}
  {Phys. Rev. B}\ }\textbf {\bibinfo {volume} {102}},\ \bibinfo {pages}
  {115142} (\bibinfo {year} {2020})}\BibitemShut {NoStop}%
\bibitem [{\citenamefont {Kiely}\ and\ \citenamefont
  {Mueller}(2021)}]{Kiely2021}%
  \BibitemOpen
  \bibfield  {author} {\bibinfo {author} {\bibfnamefont {T.~G.}\ \bibnamefont
  {Kiely}}\ and\ \bibinfo {author} {\bibfnamefont {E.~J.}\ \bibnamefont
  {Mueller}},\ }\href {\doibase 10.1103/PhysRevB.104.165143} {\bibfield
  {journal} {\bibinfo  {journal} {Phys. Rev. B}\ }\textbf {\bibinfo {volume}
  {104}},\ \bibinfo {pages} {165143} (\bibinfo {year} {2021})}\BibitemShut
  {NoStop}%
\bibitem [{\citenamefont {Herman}\ \emph {et~al.}(2019)\citenamefont {Herman},
  \citenamefont {Buhmann}, \citenamefont {Fischer},\ and\ \citenamefont
  {Sigrist}}]{Herman2019}%
  \BibitemOpen
  \bibfield  {author} {\bibinfo {author} {\bibfnamefont {F.}~\bibnamefont
  {Herman}}, \bibinfo {author} {\bibfnamefont {J.}~\bibnamefont {Buhmann}},
  \bibinfo {author} {\bibfnamefont {M.~H.}\ \bibnamefont {Fischer}}, \ and\
  \bibinfo {author} {\bibfnamefont {M.}~\bibnamefont {Sigrist}},\ }\href
  {\doibase 10.1103/PhysRevB.99.184107} {\bibfield  {journal} {\bibinfo
  {journal} {Phys. Rev. B}\ }\textbf {\bibinfo {volume} {99}},\ \bibinfo
  {pages} {184107} (\bibinfo {year} {2019})}\BibitemShut {NoStop}%
\bibitem [{\citenamefont {Brown}\ \emph {et~al.}(2019)\citenamefont {Brown},
  \citenamefont {Mitra}, \citenamefont {Guardado-Sanchez}, \citenamefont
  {Nourafkan}, \citenamefont {Reymbaut}, \citenamefont {H{\'{e}}bert},
  \citenamefont {Bergeron}, \citenamefont {Tremblay}, \citenamefont {Kokalj},
  \citenamefont {Huse}, \citenamefont {Schau{\ss}},\ and\ \citenamefont
  {Bakr}}]{Brown2019}%
  \BibitemOpen
  \bibfield  {author} {\bibinfo {author} {\bibfnamefont {P.~T.}\ \bibnamefont
  {Brown}}, \bibinfo {author} {\bibfnamefont {D.}~\bibnamefont {Mitra}},
  \bibinfo {author} {\bibfnamefont {E.}~\bibnamefont {Guardado-Sanchez}},
  \bibinfo {author} {\bibfnamefont {R.}~\bibnamefont {Nourafkan}}, \bibinfo
  {author} {\bibfnamefont {A.}~\bibnamefont {Reymbaut}}, \bibinfo {author}
  {\bibfnamefont {C.-D.}\ \bibnamefont {H{\'{e}}bert}}, \bibinfo {author}
  {\bibfnamefont {S.}~\bibnamefont {Bergeron}}, \bibinfo {author}
  {\bibfnamefont {A.-M.~S.}\ \bibnamefont {Tremblay}}, \bibinfo {author}
  {\bibfnamefont {J.}~\bibnamefont {Kokalj}}, \bibinfo {author} {\bibfnamefont
  {D.~A.}\ \bibnamefont {Huse}}, \bibinfo {author} {\bibfnamefont
  {P.}~\bibnamefont {Schau{\ss}}}, \ and\ \bibinfo {author} {\bibfnamefont
  {W.~S.}\ \bibnamefont {Bakr}},\ }\href {\doibase 10.1126/science.aat4134}
  {\bibfield  {journal} {\bibinfo  {journal} {Science}\ }\textbf {\bibinfo
  {volume} {363}},\ \bibinfo {pages} {379} (\bibinfo {year}
  {2019})}\BibitemShut {NoStop}%
\bibitem [{\citenamefont {Jakli{\v c}}\ and\ \citenamefont {Prelov{\v
  s}ek}(2000)}]{Jaklic2000}%
  \BibitemOpen
  \bibfield  {author} {\bibinfo {author} {\bibfnamefont {J.}~\bibnamefont
  {Jakli{\v c}}}\ and\ \bibinfo {author} {\bibfnamefont {P.}~\bibnamefont
  {Prelov{\v s}ek}},\ }\href@noop {} {\bibfield  {journal} {\bibinfo  {journal}
  {Adv. Phys.}\ }\textbf {\bibinfo {volume} {49}},\ \bibinfo {pages} {1}
  (\bibinfo {year} {2000})}\BibitemShut {NoStop}%
\bibitem [{\citenamefont {Kokalj}\ and\ \citenamefont
  {McKenzie}(2013)}]{Kokalj2013}%
  \BibitemOpen
  \bibfield  {author} {\bibinfo {author} {\bibfnamefont {J.}~\bibnamefont
  {Kokalj}}\ and\ \bibinfo {author} {\bibfnamefont {R.~H.}\ \bibnamefont
  {McKenzie}},\ }\href {\doibase 10.1103/PhysRevLett.110.206402} {\bibfield
  {journal} {\bibinfo  {journal} {Phys. Rev. Lett.}\ }\textbf {\bibinfo
  {volume} {110}},\ \bibinfo {pages} {206402} (\bibinfo {year}
  {2013})}\BibitemShut {NoStop}%
\bibitem [{\citenamefont {Rubtsov}\ \emph {et~al.}(2005)\citenamefont
  {Rubtsov}, \citenamefont {Savkin},\ and\ \citenamefont
  {Lichtenstein}}]{Rubtsov2005}%
  \BibitemOpen
  \bibfield  {author} {\bibinfo {author} {\bibfnamefont {A.~N.}\ \bibnamefont
  {Rubtsov}}, \bibinfo {author} {\bibfnamefont {V.~V.}\ \bibnamefont {Savkin}},
  \ and\ \bibinfo {author} {\bibfnamefont {A.~I.}\ \bibnamefont
  {Lichtenstein}},\ }\href {\doibase 10.1103/PhysRevB.72.035122} {\bibfield
  {journal} {\bibinfo  {journal} {Phys. Rev. B}\ }\textbf {\bibinfo {volume}
  {72}},\ \bibinfo {pages} {035122} (\bibinfo {year} {2005})}\BibitemShut
  {NoStop}%
\bibitem [{\citenamefont {Gull}\ \emph {et~al.}(2011)\citenamefont {Gull},
  \citenamefont {Millis}, \citenamefont {Lichtenstein}, \citenamefont
  {Rubtsov}, \citenamefont {Troyer},\ and\ \citenamefont
  {Werner}}]{GullRMP2011}%
  \BibitemOpen
  \bibfield  {author} {\bibinfo {author} {\bibfnamefont {E.}~\bibnamefont
  {Gull}}, \bibinfo {author} {\bibfnamefont {A.~J.}\ \bibnamefont {Millis}},
  \bibinfo {author} {\bibfnamefont {A.~I.}\ \bibnamefont {Lichtenstein}},
  \bibinfo {author} {\bibfnamefont {A.~N.}\ \bibnamefont {Rubtsov}}, \bibinfo
  {author} {\bibfnamefont {M.}~\bibnamefont {Troyer}}, \ and\ \bibinfo {author}
  {\bibfnamefont {P.}~\bibnamefont {Werner}},\ }\href {\doibase
  10.1103/RevModPhys.83.349} {\bibfield  {journal} {\bibinfo  {journal} {Rev.
  Mod. Phys.}\ }\textbf {\bibinfo {volume} {83}},\ \bibinfo {pages} {349}
  (\bibinfo {year} {2011})}\BibitemShut {NoStop}%
\bibitem [{\citenamefont {Hafermann}\ \emph {et~al.}(2014)\citenamefont
  {Hafermann}, \citenamefont {van Loon}, \citenamefont {Katsnelson},
  \citenamefont {Lichtenstein},\ and\ \citenamefont
  {Parcollet}}]{Hafermann2014}%
  \BibitemOpen
  \bibfield  {author} {\bibinfo {author} {\bibfnamefont {H.}~\bibnamefont
  {Hafermann}}, \bibinfo {author} {\bibfnamefont {E.~G. C.~P.}\ \bibnamefont
  {van Loon}}, \bibinfo {author} {\bibfnamefont {M.~I.}\ \bibnamefont
  {Katsnelson}}, \bibinfo {author} {\bibfnamefont {A.~I.}\ \bibnamefont
  {Lichtenstein}}, \ and\ \bibinfo {author} {\bibfnamefont {O.}~\bibnamefont
  {Parcollet}},\ }\href {\doibase 10.1103/PhysRevB.90.235105} {\bibfield
  {journal} {\bibinfo  {journal} {Phys. Rev. B}\ }\textbf {\bibinfo {volume}
  {90}},\ \bibinfo {pages} {235105} (\bibinfo {year} {2014})}\BibitemShut
  {NoStop}%
\bibitem [{\citenamefont {Forster}(2018)}]{Forster2018}%
  \BibitemOpen
  \bibfield  {author} {\bibinfo {author} {\bibfnamefont {D.}~\bibnamefont
  {Forster}},\ }\href@noop {} {\emph {\bibinfo {title} {Hydrodynamic
  fluctuations, broken symmetry, and correlation functions}}}\ (\bibinfo
  {publisher} {CRC Press},\ \bibinfo {year} {2018})\BibitemShut {NoStop}%
\bibitem [{\citenamefont {Coleman}(2015)}]{Coleman2015}%
  \BibitemOpen
  \bibfield  {author} {\bibinfo {author} {\bibfnamefont {P.}~\bibnamefont
  {Coleman}},\ }\href@noop {} {\emph {\bibinfo {title} {Introduction to
  many-body physics}}}\ (\bibinfo  {publisher} {Cambridge University Press},\
  \bibinfo {address} {Cambridge, England},\ \bibinfo {year} {2015})\BibitemShut
  {NoStop}%
\bibitem [{\citenamefont {Aoki}\ \emph {et~al.}(2014)\citenamefont {Aoki},
  \citenamefont {Tsuji}, \citenamefont {Eckstein}, \citenamefont {Kollar},
  \citenamefont {Oka},\ and\ \citenamefont {Werner}}]{AokiRMP2014}%
  \BibitemOpen
  \bibfield  {author} {\bibinfo {author} {\bibfnamefont {H.}~\bibnamefont
  {Aoki}}, \bibinfo {author} {\bibfnamefont {N.}~\bibnamefont {Tsuji}},
  \bibinfo {author} {\bibfnamefont {M.}~\bibnamefont {Eckstein}}, \bibinfo
  {author} {\bibfnamefont {M.}~\bibnamefont {Kollar}}, \bibinfo {author}
  {\bibfnamefont {T.}~\bibnamefont {Oka}}, \ and\ \bibinfo {author}
  {\bibfnamefont {P.}~\bibnamefont {Werner}},\ }\href {\doibase
  10.1103/RevModPhys.86.779} {\bibfield  {journal} {\bibinfo  {journal} {Rev.
  Mod. Phys.}\ }\textbf {\bibinfo {volume} {86}},\ \bibinfo {pages} {779}
  (\bibinfo {year} {2014})}\BibitemShut {NoStop}%
\bibitem [{\citenamefont {Rohe}\ and\ \citenamefont
  {Honerkamp}(2020)}]{Rohe2020}%
  \BibitemOpen
  \bibfield  {author} {\bibinfo {author} {\bibfnamefont {D.}~\bibnamefont
  {Rohe}}\ and\ \bibinfo {author} {\bibfnamefont {C.}~\bibnamefont
  {Honerkamp}},\ }\href {\doibase 10.21468/SciPostPhys.9.6.084} {\bibfield
  {journal} {\bibinfo  {journal} {SciPost Phys.}\ }\textbf {\bibinfo {volume}
  {9}},\ \bibinfo {pages} {84} (\bibinfo {year} {2020})}\BibitemShut {NoStop}%
\bibitem [{\citenamefont {\ifmmode~\check{S}\else \v{S}\fi{}imkovic}\ \emph
  {et~al.}(2020)\citenamefont {\ifmmode~\check{S}\else \v{S}\fi{}imkovic},
  \citenamefont {LeBlanc}, \citenamefont {Kim}, \citenamefont {Deng},
  \citenamefont {Prokof'ev}, \citenamefont {Svistunov},\ and\ \citenamefont
  {Kozik}}]{Simkovic2020}%
  \BibitemOpen
  \bibfield  {author} {\bibinfo {author} {\bibfnamefont {F.}~\bibnamefont
  {\ifmmode~\check{S}\else \v{S}\fi{}imkovic}}, \bibinfo {author}
  {\bibfnamefont {J.~P.~F.}\ \bibnamefont {LeBlanc}}, \bibinfo {author}
  {\bibfnamefont {A.~J.}\ \bibnamefont {Kim}}, \bibinfo {author} {\bibfnamefont
  {Y.}~\bibnamefont {Deng}}, \bibinfo {author} {\bibfnamefont {N.~V.}\
  \bibnamefont {Prokof'ev}}, \bibinfo {author} {\bibfnamefont {B.~V.}\
  \bibnamefont {Svistunov}}, \ and\ \bibinfo {author} {\bibfnamefont
  {E.}~\bibnamefont {Kozik}},\ }\href {\doibase 10.1103/PhysRevLett.124.017003}
  {\bibfield  {journal} {\bibinfo  {journal} {Phys. Rev. Lett.}\ }\textbf
  {\bibinfo {volume} {124}},\ \bibinfo {pages} {017003} (\bibinfo {year}
  {2020})}\BibitemShut {NoStop}%
\bibitem [{\citenamefont {Kim}\ \emph {et~al.}(2020)\citenamefont {Kim},
  \citenamefont {Simkovic},\ and\ \citenamefont {Kozik}}]{Kim2020}%
  \BibitemOpen
  \bibfield  {author} {\bibinfo {author} {\bibfnamefont {A.~J.}\ \bibnamefont
  {Kim}}, \bibinfo {author} {\bibfnamefont {F.}~\bibnamefont {Simkovic}}, \
  and\ \bibinfo {author} {\bibfnamefont {E.}~\bibnamefont {Kozik}},\ }\href
  {\doibase 10.1103/PhysRevLett.124.117602} {\bibfield  {journal} {\bibinfo
  {journal} {Phys. Rev. Lett.}\ }\textbf {\bibinfo {volume} {124}},\ \bibinfo
  {pages} {117602} (\bibinfo {year} {2020})}\BibitemShut {NoStop}%
\bibitem [{\citenamefont {Klett}\ \emph {et~al.}(2020)\citenamefont {Klett},
  \citenamefont {Wentzell}, \citenamefont {Sch\"afer}, \citenamefont
  {Simkovic}, \citenamefont {Parcollet}, \citenamefont {Andergassen},\ and\
  \citenamefont {Hansmann}}]{Klett2020}%
  \BibitemOpen
  \bibfield  {author} {\bibinfo {author} {\bibfnamefont {M.}~\bibnamefont
  {Klett}}, \bibinfo {author} {\bibfnamefont {N.}~\bibnamefont {Wentzell}},
  \bibinfo {author} {\bibfnamefont {T.}~\bibnamefont {Sch\"afer}}, \bibinfo
  {author} {\bibfnamefont {F.}~\bibnamefont {Simkovic}}, \bibinfo {author}
  {\bibfnamefont {O.}~\bibnamefont {Parcollet}}, \bibinfo {author}
  {\bibfnamefont {S.}~\bibnamefont {Andergassen}}, \ and\ \bibinfo {author}
  {\bibfnamefont {P.}~\bibnamefont {Hansmann}},\ }\href {\doibase
  10.1103/PhysRevResearch.2.033476} {\bibfield  {journal} {\bibinfo  {journal}
  {Phys. Rev. Research}\ }\textbf {\bibinfo {volume} {2}},\ \bibinfo {pages}
  {033476} (\bibinfo {year} {2020})}\BibitemShut {NoStop}%
\bibitem [{\citenamefont {Sch\"afer}\ \emph {et~al.}(2021)\citenamefont
  {Sch\"afer}, \citenamefont {Wentzell}, \citenamefont {\ifmmode~\check{S}\else
  \v{S}\fi{}imkovic}, \citenamefont {He}, \citenamefont {Hille}, \citenamefont
  {Klett}, \citenamefont {Eckhardt}, \citenamefont {Arzhang}, \citenamefont
  {Harkov}, \citenamefont {Le~R\'egent}, \citenamefont {Kirsch}, \citenamefont
  {Wang}, \citenamefont {Kim}, \citenamefont {Kozik}, \citenamefont {Stepanov},
  \citenamefont {Kauch}, \citenamefont {Andergassen}, \citenamefont {Hansmann},
  \citenamefont {Rohe}, \citenamefont {Vilk}, \citenamefont {LeBlanc},
  \citenamefont {Zhang}, \citenamefont {Tremblay}, \citenamefont {Ferrero},
  \citenamefont {Parcollet},\ and\ \citenamefont {Georges}}]{Schaefer2021}%
  \BibitemOpen
  \bibfield  {author} {\bibinfo {author} {\bibfnamefont {T.}~\bibnamefont
  {Sch\"afer}}, \bibinfo {author} {\bibfnamefont {N.}~\bibnamefont {Wentzell}},
  \bibinfo {author} {\bibfnamefont {F.}~\bibnamefont {\ifmmode~\check{S}\else
  \v{S}\fi{}imkovic}}, \bibinfo {author} {\bibfnamefont {Y.-Y.}\ \bibnamefont
  {He}}, \bibinfo {author} {\bibfnamefont {C.}~\bibnamefont {Hille}}, \bibinfo
  {author} {\bibfnamefont {M.}~\bibnamefont {Klett}}, \bibinfo {author}
  {\bibfnamefont {C.~J.}\ \bibnamefont {Eckhardt}}, \bibinfo {author}
  {\bibfnamefont {B.}~\bibnamefont {Arzhang}}, \bibinfo {author} {\bibfnamefont
  {V.}~\bibnamefont {Harkov}}, \bibinfo {author} {\bibfnamefont {F.-M.}\
  \bibnamefont {Le~R\'egent}}, \bibinfo {author} {\bibfnamefont
  {A.}~\bibnamefont {Kirsch}}, \bibinfo {author} {\bibfnamefont
  {Y.}~\bibnamefont {Wang}}, \bibinfo {author} {\bibfnamefont {A.~J.}\
  \bibnamefont {Kim}}, \bibinfo {author} {\bibfnamefont {E.}~\bibnamefont
  {Kozik}}, \bibinfo {author} {\bibfnamefont {E.~A.}\ \bibnamefont {Stepanov}},
  \bibinfo {author} {\bibfnamefont {A.}~\bibnamefont {Kauch}}, \bibinfo
  {author} {\bibfnamefont {S.}~\bibnamefont {Andergassen}}, \bibinfo {author}
  {\bibfnamefont {P.}~\bibnamefont {Hansmann}}, \bibinfo {author}
  {\bibfnamefont {D.}~\bibnamefont {Rohe}}, \bibinfo {author} {\bibfnamefont
  {Y.~M.}\ \bibnamefont {Vilk}}, \bibinfo {author} {\bibfnamefont {J.~P.~F.}\
  \bibnamefont {LeBlanc}}, \bibinfo {author} {\bibfnamefont {S.}~\bibnamefont
  {Zhang}}, \bibinfo {author} {\bibfnamefont {A.-M.~S.}\ \bibnamefont
  {Tremblay}}, \bibinfo {author} {\bibfnamefont {M.}~\bibnamefont {Ferrero}},
  \bibinfo {author} {\bibfnamefont {O.}~\bibnamefont {Parcollet}}, \ and\
  \bibinfo {author} {\bibfnamefont {A.}~\bibnamefont {Georges}},\ }\href
  {\doibase 10.1103/PhysRevX.11.011058} {\bibfield  {journal} {\bibinfo
  {journal} {Phys. Rev. X}\ }\textbf {\bibinfo {volume} {11}},\ \bibinfo
  {pages} {011058} (\bibinfo {year} {2021})}\BibitemShut {NoStop}%
\bibitem [{\citenamefont {Vu\ifmmode \check{c}\else \v{c}\fi{}i\ifmmode
  \check{c}\else \v{c}\fi{}evi\ifmmode~\acute{c}\else \'{c}\fi{}}\ and\
  \citenamefont {\ifmmode~\check{Z}\else
  \v{Z}\fi{}itko}(2021)}]{VucicevicPRB2021}%
  \BibitemOpen
  \bibfield  {author} {\bibinfo {author} {\bibfnamefont {J.}~\bibnamefont
  {Vu\ifmmode \check{c}\else \v{c}\fi{}i\ifmmode \check{c}\else
  \v{c}\fi{}evi\ifmmode~\acute{c}\else \'{c}\fi{}}}\ and\ \bibinfo {author}
  {\bibfnamefont {R.}~\bibnamefont {\ifmmode~\check{Z}\else \v{Z}\fi{}itko}},\
  }\href {\doibase 10.1103/PhysRevB.104.205101} {\bibfield  {journal} {\bibinfo
   {journal} {Phys. Rev. B}\ }\textbf {\bibinfo {volume} {104}},\ \bibinfo
  {pages} {205101} (\bibinfo {year} {2021})}\BibitemShut {NoStop}%
\bibitem [{\citenamefont {Vu\ifmmode \check{c}\else \v{c}\fi{}i\ifmmode
  \check{c}\else \v{c}\fi{}evi\ifmmode~\acute{c}\else \'{c}\fi{}}\ \emph
  {et~al.}(2013)\citenamefont {Vu\ifmmode \check{c}\else \v{c}\fi{}i\ifmmode
  \check{c}\else \v{c}\fi{}evi\ifmmode~\acute{c}\else \'{c}\fi{}},
  \citenamefont {Terletska}, \citenamefont {Tanaskovi\ifmmode~\acute{c}\else
  \'{c}\fi{}},\ and\ \citenamefont {Dobrosavljevi\ifmmode~\acute{c}\else
  \'{c}\fi{}}}]{VucicevicPRB2013}%
  \BibitemOpen
  \bibfield  {author} {\bibinfo {author} {\bibfnamefont {J.}~\bibnamefont
  {Vu\ifmmode \check{c}\else \v{c}\fi{}i\ifmmode \check{c}\else
  \v{c}\fi{}evi\ifmmode~\acute{c}\else \'{c}\fi{}}}, \bibinfo {author}
  {\bibfnamefont {H.}~\bibnamefont {Terletska}}, \bibinfo {author}
  {\bibfnamefont {D.}~\bibnamefont {Tanaskovi\ifmmode~\acute{c}\else
  \'{c}\fi{}}}, \ and\ \bibinfo {author} {\bibfnamefont {V.}~\bibnamefont
  {Dobrosavljevi\ifmmode~\acute{c}\else \'{c}\fi{}}},\ }\href {\doibase
  10.1103/PhysRevB.88.075143} {\bibfield  {journal} {\bibinfo  {journal} {Phys.
  Rev. B}\ }\textbf {\bibinfo {volume} {88}},\ \bibinfo {pages} {075143}
  (\bibinfo {year} {2013})}\BibitemShut {NoStop}%
\bibitem [{\citenamefont {Georges}\ \emph {et~al.}(1996)\citenamefont
  {Georges}, \citenamefont {Kotliar}, \citenamefont {Krauth},\ and\
  \citenamefont {Rozenberg}}]{Georges1996}%
  \BibitemOpen
  \bibfield  {author} {\bibinfo {author} {\bibfnamefont {A.}~\bibnamefont
  {Georges}}, \bibinfo {author} {\bibfnamefont {G.}~\bibnamefont {Kotliar}},
  \bibinfo {author} {\bibfnamefont {W.}~\bibnamefont {Krauth}}, \ and\ \bibinfo
  {author} {\bibfnamefont {M.~J.}\ \bibnamefont {Rozenberg}},\ }\href {\doibase
  10.1103/RevModPhys.68.13} {\bibfield  {journal} {\bibinfo  {journal} {Rev.
  Mod. Phys.}\ }\textbf {\bibinfo {volume} {68}},\ \bibinfo {pages} {13}
  (\bibinfo {year} {1996})}\BibitemShut {NoStop}%
\bibitem [{\citenamefont {Wilson}(1975)}]{wilson1975}%
  \BibitemOpen
  \bibfield  {author} {\bibinfo {author} {\bibfnamefont {K.~G.}\ \bibnamefont
  {Wilson}},\ }\href {\doibase 10.1103/RevModPhys.47.773} {\bibfield  {journal}
  {\bibinfo  {journal} {Rev. Mod. Phys.}\ }\textbf {\bibinfo {volume} {47}},\
  \bibinfo {pages} {773} (\bibinfo {year} {1975})}\BibitemShut {NoStop}%
\bibitem [{\citenamefont {Krishna-murthy}\ \emph {et~al.}(1980)\citenamefont
  {Krishna-murthy}, \citenamefont {Wilkins},\ and\ \citenamefont
  {Wilson}}]{krishna1980a}%
  \BibitemOpen
  \bibfield  {author} {\bibinfo {author} {\bibfnamefont {H.~R.}\ \bibnamefont
  {Krishna-murthy}}, \bibinfo {author} {\bibfnamefont {J.~W.}\ \bibnamefont
  {Wilkins}}, \ and\ \bibinfo {author} {\bibfnamefont {K.~G.}\ \bibnamefont
  {Wilson}},\ }\href {\doibase 10.1103/PhysRevB.21.1003} {\bibfield  {journal}
  {\bibinfo  {journal} {Phys. Rev. B}\ }\textbf {\bibinfo {volume} {21}},\
  \bibinfo {pages} {1003} (\bibinfo {year} {1980})}\BibitemShut {NoStop}%
\bibitem [{\citenamefont {Bulla}\ \emph {et~al.}(2008)\citenamefont {Bulla},
  \citenamefont {Costi},\ and\ \citenamefont {Pruschke}}]{bulla2008}%
  \BibitemOpen
  \bibfield  {author} {\bibinfo {author} {\bibfnamefont {R.}~\bibnamefont
  {Bulla}}, \bibinfo {author} {\bibfnamefont {T.~A.}\ \bibnamefont {Costi}}, \
  and\ \bibinfo {author} {\bibfnamefont {T.}~\bibnamefont {Pruschke}},\ }\href
  {\doibase 10.1103/RevModPhys.80.395} {\bibfield  {journal} {\bibinfo
  {journal} {Rev. Mod. Phys.}\ }\textbf {\bibinfo {volume} {80}},\ \bibinfo
  {pages} {395} (\bibinfo {year} {2008})}\BibitemShut {NoStop}%
\bibitem [{\citenamefont {\ifmmode~\check{Z}\else \v{Z}\fi{}itko}\ and\
  \citenamefont {Pruschke}(2009)}]{resolution}%
  \BibitemOpen
  \bibfield  {author} {\bibinfo {author} {\bibfnamefont {R.}~\bibnamefont
  {\ifmmode~\check{Z}\else \v{Z}\fi{}itko}}\ and\ \bibinfo {author}
  {\bibfnamefont {T.}~\bibnamefont {Pruschke}},\ }\href {\doibase
  10.1103/PhysRevB.79.085106} {\bibfield  {journal} {\bibinfo  {journal} {Phys.
  Rev. B}\ }\textbf {\bibinfo {volume} {79}},\ \bibinfo {pages} {085106}
  (\bibinfo {year} {2009})}\BibitemShut {NoStop}%
\bibitem [{\citenamefont {\v{S}imkovic IV}\ \emph {et~al.}(2022)\citenamefont
  {\v{S}imkovic IV}, \citenamefont {Rossi},\ and\ \citenamefont
  {Ferrero}}]{Simkovic2021}%
  \BibitemOpen
  \bibfield  {author} {\bibinfo {author} {\bibfnamefont {F.}~\bibnamefont
  {\v{S}imkovic IV}}, \bibinfo {author} {\bibfnamefont {R.}~\bibnamefont
  {Rossi}}, \ and\ \bibinfo {author} {\bibfnamefont {M.}~\bibnamefont
  {Ferrero}},\ }\href {\doibase 10.1103/PhysRevResearch.4.043201} {\bibfield
  {journal} {\bibinfo  {journal} {Phys. Rev. Res.}\ }\textbf {\bibinfo {volume}
  {4}},\ \bibinfo {pages} {043201} (\bibinfo {year} {2022})}\BibitemShut
  {NoStop}%
\bibitem [{\citenamefont {Rohringer}\ \emph {et~al.}(2018)\citenamefont
  {Rohringer}, \citenamefont {Hafermann}, \citenamefont {Toschi}, \citenamefont
  {Katanin}, \citenamefont {Antipov}, \citenamefont {Katsnelson}, \citenamefont
  {Lichtenstein}, \citenamefont {Rubtsov},\ and\ \citenamefont
  {Held}}]{RohringerRMP2018}%
  \BibitemOpen
  \bibfield  {author} {\bibinfo {author} {\bibfnamefont {G.}~\bibnamefont
  {Rohringer}}, \bibinfo {author} {\bibfnamefont {H.}~\bibnamefont
  {Hafermann}}, \bibinfo {author} {\bibfnamefont {A.}~\bibnamefont {Toschi}},
  \bibinfo {author} {\bibfnamefont {A.~A.}\ \bibnamefont {Katanin}}, \bibinfo
  {author} {\bibfnamefont {A.~E.}\ \bibnamefont {Antipov}}, \bibinfo {author}
  {\bibfnamefont {M.~I.}\ \bibnamefont {Katsnelson}}, \bibinfo {author}
  {\bibfnamefont {A.~I.}\ \bibnamefont {Lichtenstein}}, \bibinfo {author}
  {\bibfnamefont {A.~N.}\ \bibnamefont {Rubtsov}}, \ and\ \bibinfo {author}
  {\bibfnamefont {K.}~\bibnamefont {Held}},\ }\href {\doibase
  10.1103/RevModPhys.90.025003} {\bibfield  {journal} {\bibinfo  {journal}
  {Rev. Mod. Phys.}\ }\textbf {\bibinfo {volume} {90}},\ \bibinfo {pages}
  {025003} (\bibinfo {year} {2018})}\BibitemShut {NoStop}%
\bibitem [{\citenamefont {Taheridehkordi}\ \emph {et~al.}(2019)\citenamefont
  {Taheridehkordi}, \citenamefont {Curnoe},\ and\ \citenamefont
  {LeBlanc}}]{Taheridehkordi2019}%
  \BibitemOpen
  \bibfield  {author} {\bibinfo {author} {\bibfnamefont {A.}~\bibnamefont
  {Taheridehkordi}}, \bibinfo {author} {\bibfnamefont {S.~H.}\ \bibnamefont
  {Curnoe}}, \ and\ \bibinfo {author} {\bibfnamefont {J.~P.~F.}\ \bibnamefont
  {LeBlanc}},\ }\href {\doibase 10.1103/PhysRevB.99.035120} {\bibfield
  {journal} {\bibinfo  {journal} {Phys. Rev. B}\ }\textbf {\bibinfo {volume}
  {99}},\ \bibinfo {pages} {035120} (\bibinfo {year} {2019})}\BibitemShut
  {NoStop}%
\bibitem [{\citenamefont {Vu\ifmmode \check{c}\else \v{c}\fi{}i\ifmmode
  \check{c}\else \v{c}\fi{}evi\ifmmode~\acute{c}\else \'{c}\fi{}}\ and\
  \citenamefont {Ferrero}(2020)}]{VucicevicPRB2020}%
  \BibitemOpen
  \bibfield  {author} {\bibinfo {author} {\bibfnamefont {J.}~\bibnamefont
  {Vu\ifmmode \check{c}\else \v{c}\fi{}i\ifmmode \check{c}\else
  \v{c}\fi{}evi\ifmmode~\acute{c}\else \'{c}\fi{}}}\ and\ \bibinfo {author}
  {\bibfnamefont {M.}~\bibnamefont {Ferrero}},\ }\href {\doibase
  10.1103/PhysRevB.101.075113} {\bibfield  {journal} {\bibinfo  {journal}
  {Phys. Rev. B}\ }\textbf {\bibinfo {volume} {101}},\ \bibinfo {pages}
  {075113} (\bibinfo {year} {2020})}\BibitemShut {NoStop}%
\bibitem [{\citenamefont {Taheridehkordi}\ \emph
  {et~al.}(2020{\natexlab{a}})\citenamefont {Taheridehkordi}, \citenamefont
  {Curnoe},\ and\ \citenamefont {LeBlanc}}]{Taheridehkordi2020}%
  \BibitemOpen
  \bibfield  {author} {\bibinfo {author} {\bibfnamefont {A.}~\bibnamefont
  {Taheridehkordi}}, \bibinfo {author} {\bibfnamefont {S.~H.}\ \bibnamefont
  {Curnoe}}, \ and\ \bibinfo {author} {\bibfnamefont {J.~P.~F.}\ \bibnamefont
  {LeBlanc}},\ }\href {\doibase 10.1103/PhysRevB.101.125109} {\bibfield
  {journal} {\bibinfo  {journal} {Phys. Rev. B}\ }\textbf {\bibinfo {volume}
  {101}},\ \bibinfo {pages} {125109} (\bibinfo {year}
  {2020}{\natexlab{a}})}\BibitemShut {NoStop}%
\bibitem [{\citenamefont {Taheridehkordi}\ \emph
  {et~al.}(2020{\natexlab{b}})\citenamefont {Taheridehkordi}, \citenamefont
  {Curnoe},\ and\ \citenamefont {LeBlanc}}]{Taheridehkordi2020b}%
  \BibitemOpen
  \bibfield  {author} {\bibinfo {author} {\bibfnamefont {A.}~\bibnamefont
  {Taheridehkordi}}, \bibinfo {author} {\bibfnamefont {S.~H.}\ \bibnamefont
  {Curnoe}}, \ and\ \bibinfo {author} {\bibfnamefont {J.~P.~F.}\ \bibnamefont
  {LeBlanc}},\ }\href {\doibase 10.1103/PhysRevB.102.045115} {\bibfield
  {journal} {\bibinfo  {journal} {Phys. Rev. B}\ }\textbf {\bibinfo {volume}
  {102}},\ \bibinfo {pages} {045115} (\bibinfo {year}
  {2020}{\natexlab{b}})}\BibitemShut {NoStop}%
\bibitem [{\citenamefont {Vu\ifmmode \check{c}\else \v{c}\fi{}i\ifmmode
  \check{c}\else \v{c}\fi{}evi\ifmmode~\acute{c}\else \'{c}\fi{}}\ \emph
  {et~al.}(2021)\citenamefont {Vu\ifmmode \check{c}\else \v{c}\fi{}i\ifmmode
  \check{c}\else \v{c}\fi{}evi\ifmmode~\acute{c}\else \'{c}\fi{}},
  \citenamefont {Stipsi\ifmmode~\acute{c}\else \'{c}\fi{}},\ and\ \citenamefont
  {Ferrero}}]{VucicevicPRR2021}%
  \BibitemOpen
  \bibfield  {author} {\bibinfo {author} {\bibfnamefont {J.}~\bibnamefont
  {Vu\ifmmode \check{c}\else \v{c}\fi{}i\ifmmode \check{c}\else
  \v{c}\fi{}evi\ifmmode~\acute{c}\else \'{c}\fi{}}}, \bibinfo {author}
  {\bibfnamefont {P.}~\bibnamefont {Stipsi\ifmmode~\acute{c}\else \'{c}\fi{}}},
  \ and\ \bibinfo {author} {\bibfnamefont {M.}~\bibnamefont {Ferrero}},\ }\href
  {\doibase 10.1103/PhysRevResearch.3.023082} {\bibfield  {journal} {\bibinfo
  {journal} {Phys. Rev. Research}\ }\textbf {\bibinfo {volume} {3}},\ \bibinfo
  {pages} {023082} (\bibinfo {year} {2021})}\BibitemShut {NoStop}%
\bibitem [{\citenamefont {McNiven}\ \emph {et~al.}(2022)\citenamefont
  {McNiven}, \citenamefont {Terletska}, \citenamefont {Andrews},\ and\
  \citenamefont {LeBlanc}}]{McNiven2022}%
  \BibitemOpen
  \bibfield  {author} {\bibinfo {author} {\bibfnamefont {B.~D.~E.}\
  \bibnamefont {McNiven}}, \bibinfo {author} {\bibfnamefont {H.}~\bibnamefont
  {Terletska}}, \bibinfo {author} {\bibfnamefont {G.~T.}\ \bibnamefont
  {Andrews}}, \ and\ \bibinfo {author} {\bibfnamefont {J.~P.~F.}\ \bibnamefont
  {LeBlanc}},\ }\href {\doibase 10.1103/PhysRevB.106.035145} {\bibfield
  {journal} {\bibinfo  {journal} {Phys. Rev. B}\ }\textbf {\bibinfo {volume}
  {106}},\ \bibinfo {pages} {035145} (\bibinfo {year} {2022})}\BibitemShut
  {NoStop}%
\bibitem [{\citenamefont {Wu}\ \emph {et~al.}(2017)\citenamefont {Wu},
  \citenamefont {Ferrero}, \citenamefont {Georges},\ and\ \citenamefont
  {Kozik}}]{Wu2017}%
  \BibitemOpen
  \bibfield  {author} {\bibinfo {author} {\bibfnamefont {W.}~\bibnamefont
  {Wu}}, \bibinfo {author} {\bibfnamefont {M.}~\bibnamefont {Ferrero}},
  \bibinfo {author} {\bibfnamefont {A.}~\bibnamefont {Georges}}, \ and\
  \bibinfo {author} {\bibfnamefont {E.}~\bibnamefont {Kozik}},\ }\href
  {\doibase 10.1103/PhysRevB.96.041105} {\bibfield  {journal} {\bibinfo
  {journal} {Phys. Rev. B}\ }\textbf {\bibinfo {volume} {96}},\ \bibinfo
  {pages} {041105} (\bibinfo {year} {2017})}\BibitemShut {NoStop}%
\bibitem [{\citenamefont {\ifmmode~\check{S}\else \v{S}\fi{}imkovic}\ and\
  \citenamefont {Kozik}(2019)}]{Simkovic2019}%
  \BibitemOpen
  \bibfield  {author} {\bibinfo {author} {\bibfnamefont {F.}~\bibnamefont
  {\ifmmode~\check{S}\else \v{S}\fi{}imkovic}}\ and\ \bibinfo {author}
  {\bibfnamefont {E.}~\bibnamefont {Kozik}},\ }\href {\doibase
  10.1103/PhysRevB.100.121102} {\bibfield  {journal} {\bibinfo  {journal}
  {Phys. Rev. B}\ }\textbf {\bibinfo {volume} {100}},\ \bibinfo {pages}
  {121102(R)} (\bibinfo {year} {2019})}\BibitemShut {NoStop}%
\bibitem [{\citenamefont {Rossi}\ \emph {et~al.}(2020)\citenamefont {Rossi},
  \citenamefont {{\v{S}}imkovic},\ and\ \citenamefont {Ferrero}}]{Rossi2020}%
  \BibitemOpen
  \bibfield  {author} {\bibinfo {author} {\bibfnamefont {R.}~\bibnamefont
  {Rossi}}, \bibinfo {author} {\bibfnamefont {F.}~\bibnamefont
  {{\v{S}}imkovic}}, \ and\ \bibinfo {author} {\bibfnamefont {M.}~\bibnamefont
  {Ferrero}},\ }\href {\doibase 10.1209/0295-5075/132/11001} {\bibfield
  {journal} {\bibinfo  {journal} {{EPL}}\ }\textbf {\bibinfo {volume} {132}},\
  \bibinfo {pages} {11001} (\bibinfo {year} {2020})}\BibitemShut {NoStop}%
\bibitem [{\citenamefont {Kim}\ \emph {et~al.}(2021)\citenamefont {Kim},
  \citenamefont {Prokof'ev}, \citenamefont {Svistunov},\ and\ \citenamefont
  {Kozik}}]{Kim2021}%
  \BibitemOpen
  \bibfield  {author} {\bibinfo {author} {\bibfnamefont {A.~J.}\ \bibnamefont
  {Kim}}, \bibinfo {author} {\bibfnamefont {N.~V.}\ \bibnamefont {Prokof'ev}},
  \bibinfo {author} {\bibfnamefont {B.~V.}\ \bibnamefont {Svistunov}}, \ and\
  \bibinfo {author} {\bibfnamefont {E.}~\bibnamefont {Kozik}},\ }\href
  {\doibase 10.1103/PhysRevLett.126.257001} {\bibfield  {journal} {\bibinfo
  {journal} {Phys. Rev. Lett}\ }\textbf {\bibinfo {volume} {126}},\ \bibinfo
  {pages} {257001} (\bibinfo {year} {2021})}\BibitemShut {NoStop}%
\bibitem [{\citenamefont {Xu}\ \emph {et~al.}(2022)\citenamefont {Xu},
  \citenamefont {Shi}, \citenamefont {Vitali}, \citenamefont {Qin},\ and\
  \citenamefont {Zhang}}]{Xu2022}%
  \BibitemOpen
  \bibfield  {author} {\bibinfo {author} {\bibfnamefont {H.}~\bibnamefont
  {Xu}}, \bibinfo {author} {\bibfnamefont {H.}~\bibnamefont {Shi}}, \bibinfo
  {author} {\bibfnamefont {E.}~\bibnamefont {Vitali}}, \bibinfo {author}
  {\bibfnamefont {M.}~\bibnamefont {Qin}}, \ and\ \bibinfo {author}
  {\bibfnamefont {S.}~\bibnamefont {Zhang}},\ }\href {\doibase
  10.1103/physrevresearch.4.013239} {\bibfield  {journal} {\bibinfo  {journal}
  {Phys. Rev. Research}\ }\textbf {\bibinfo {volume} {4}},\ \bibinfo {pages}
  {013239} (\bibinfo {year} {2022})}\BibitemShut {NoStop}%
\bibitem [{\citenamefont {Parcollet}\ \emph {et~al.}(2015)\citenamefont
  {Parcollet}, \citenamefont {Ferrero}, \citenamefont {Ayral}, \citenamefont
  {Hafermann}, \citenamefont {Krivenko}, \citenamefont {Messio},\ and\
  \citenamefont {Seth}}]{Parcollet2015}%
  \BibitemOpen
  \bibfield  {author} {\bibinfo {author} {\bibfnamefont {O.}~\bibnamefont
  {Parcollet}}, \bibinfo {author} {\bibfnamefont {M.}~\bibnamefont {Ferrero}},
  \bibinfo {author} {\bibfnamefont {T.}~\bibnamefont {Ayral}}, \bibinfo
  {author} {\bibfnamefont {H.}~\bibnamefont {Hafermann}}, \bibinfo {author}
  {\bibfnamefont {I.}~\bibnamefont {Krivenko}}, \bibinfo {author}
  {\bibfnamefont {L.}~\bibnamefont {Messio}}, \ and\ \bibinfo {author}
  {\bibfnamefont {P.}~\bibnamefont {Seth}},\ }\href {\doibase
  10.1016/j.cpc.2015.04.023} {\bibfield  {journal} {\bibinfo  {journal}
  {Comput. Phys. Commun.}\ }\textbf {\bibinfo {volume} {196}},\ \bibinfo
  {pages} {398} (\bibinfo {year} {2015})}\BibitemShut {NoStop}%
\bibitem [{\citenamefont {{\v Z}itko}()}]{Zitko_unpublished}%
  \BibitemOpen
  \bibfield  {author} {\bibinfo {author} {\bibfnamefont {R.}~\bibnamefont {{\v
  Z}itko}},\ }\href@noop {} {}\Eprint {http://arxiv.org/abs/unpublished}
  {unpublished} \BibitemShut {NoStop}%
\end{thebibliography}%
\bibliographystyle{apsrev4-1}

\end{document}